\documentclass[11pt]{article}

\usepackage[margin=0.95in]{geometry}
\usepackage[T1]{fontenc}
\usepackage[utf8]{inputenc}
\usepackage{lmodern}
\usepackage{microtype}
\usepackage{amsmath,amsthm,amssymb,amsfonts,bm,amsbsy,bbm,mathtools}
\usepackage{latexsym}
\usepackage{epsfig}
\usepackage{graphicx,rotating,lscape}
\usepackage{verbatim}
\usepackage[round]{natbib}
\usepackage{multirow,color}
\usepackage{subfigure}
\usepackage{float}
\usepackage{physics}
\usepackage{booktabs}
\usepackage{array}
\usepackage{tikz}
\usetikzlibrary{positioning,shapes.geometric}
\usepackage{setspace}
\usepackage[colorlinks=true,citecolor=blue,linkcolor=blue,urlcolor=blue,bookmarks=false]{hyperref}

\newenvironment{keywords}{\par\small\noindent\textbf{\textit{Keywords:}} }{\par}

\numberwithin{equation}{section}
\newtheorem{Lem}{Lemma}
\newtheorem{Th}{Theorem}

\newtheorem{prop}{Proposition}
\newtheorem{example}{Example}

\newtheorem{Rem}{Remark}
\newtheorem{assump}{Assumption}

\let\oldRem\Rem
\let\endoldRem\endRem
\renewenvironment{Rem}{\oldRem\normalfont}{\endoldRem}

\renewcommand{\thefootnote}{\arabic{footnote}}

\def\A{{\bf A}}
\def\a{{\bf a}}

\def\b{{\bf b}}

\def\c{{\bf c}}
\def\f{{\bf f}}

\def\G{{\bf G}}
\def\h{{\bf h}}
\def\I{{\bf I}}

\def\S{{\bf S}}

\def\M{{\bf M}}

\def\Q{{\bf Q}}
\def\U{{\bf U}}
\def\V{{\bf V}}
\def\v{{\bf v}}

\def\X{{\bf X}}
\def\x{{\bf x}}
\def\Y{{\bf Y}}

\def\W{{\bf W}}
\def\w{{\bf w}}
\def\D{{\bf D}}

\def\u{{\bf u}}
\def\Z{{\bf Z}}

\def\g{{\bf g}}
\def\H{{\bf H}}
\def\p{{\bf p}}
\def\P{{\bf P}}
\def\q{{\bf q}}

\def\calH{{\cal H}}

\def\calT{{\cal T}}

\def\mR{\mathbb{R}}

\def\ba{{\boldsymbol\alpha}}

\def\bt{{\boldsymbol\theta}}

\def\bpi{\boldsymbol\pi}

\def\bb{{\boldsymbol\beta}}

\def\bpsi{\boldsymbol\psi}

\def\bg{\boldsymbol\gamma}

\def\0{{\bf 0}}
\def\trans{^{\rm T}}
\def\pr{\hbox{pr}}
\def\wh{\widehat}
\def\wt{\widetilde}

\def\var{\hbox{var}}

\def\eff{_{\rm eff}}

\def\log{{\rm log}}

\def\squarebox#1{\hbox to #1{\hfill\vbox to #1{\vfill}}}

\def\expit{\mathrm{expit}}

\def\fx{f_{\X}}

\def\fone{f_1}
\def\fzero{f_0}

\def\sumi{\sum_{i=1}^n}

\def\sumIP1{\sum_{i=1, i\in P_1}^N}

\def\sumK{\sum_{k=1}^N}

\def\boxit#1{\vbox{\hrule\hbox{\vrule\kern6pt\vbox{\kern6pt#1\kern6pt}\kern6pt\vrule}\hrule}}
\newcommand{\macomment}[1]{}
\newcommand{\jiwei}[1]{}
\newcommand{\qt}[1]{}

\newcommand{\jiweizhao}[1]{}

\usetikzlibrary{positioning,shapes.geometric}

\newcommand{\rmv}[1]{}

\makeatletter
\newcommand*{\ind}{%
	\mathbin{%
		\mathpalette{\@ind}{}%
	}%
}
\newcommand*{\nind}{%
	\mathbin{% % The final symbol is a binary math operator
		\mathpalette{\@ind}{\not}% \mathpalette helps for the adaptation
	}%
}
\newcommand*{\@ind}[2]{%
	\sbox0{$#1\perp\m@th$}% box 0 contains \perp symbol
	\sbox2{$#1=$}% box 2 for the height of =
	\sbox4{$#1\vcenter{}$}% box 4 for the height of the math axis
	\rlap{\copy0}% first \perp
	\dimen@=\dimexpr\ht2-\ht4-.2pt\relax
	\kern\dimen@
	{#2}%
	\kern\dimen@
	\copy0 % second \perp
}
\makeatother

\DeclareMathOperator*{\argmin}{arg\,min}

\newcommand{\E}{\mathrm{E}}
\newcommand{\Eone}{\mathrm{E}_1}
\newcommand{\Ezero}{\mathrm{E}_0}

\def\wh{\widehat}
\def\wt{\widetilde}

\def\pr{\hbox{pr}}

\def\trans{^{\rm T}}
\def\arbone{_{\rm arb,1}}
\def\arbtwo{_{\rm arb,2}}
\def\eff{_{\rm eff}}

\def\effone{_{\rm eff,1}}
\def\efftwo{_{\rm eff,2}}
\def\effthree{_{\rm eff,3}}

\def\np{_{\rm np}}
\def\nvone{_{\rm nv1}}
\def\nvtwo{_{\rm nv2}}
\def\proone{_{\rm alt}}

\def\bb{{\boldsymbol\beta}}
\def\bt{{\boldsymbol\theta}}
\def\bg{{\boldsymbol\gamma}}

\def\bpi{\boldsymbol\pi}
\def\bpsi{\boldsymbol\psi}

\def\ba{{\boldsymbol\alpha}}

\def\bpsi{{\boldsymbol\psi}}

\def\boeta{{\boldsymbol\eta}}

\def\0{{\bf 0}}

\def\X{{\bf X}}
\def\x{{\bf x}}
\def\Z{{\bf Z}}

\def\c{{\bf c}}
\def\W{{\bf W}}
\def\w{{\bf w}}
\def\U{{\bf U}}
\def\u{{\bf u}}
\def\V{{\bf V}}
\def\v{{\bf v}}
\def\S{{\bf S}}

\def\b{{\bf b}}
\def\A{{\bf A}}
\def\D{{\bf D}}
\def\a{{\bf a}}

\def\f{{\bf f}}
\def\g{{\bf g}}
\def\h{{\bf h}}

\def\G{{\bf G}}
\def\H{{\bf H}}
\def\Q{{\bf Q}}
\def\I{{\bf I}}
\def\M{{\bf M}}

\def\J{{\bf J}}
\def\F{{\bf F}}

\def\calH{{\mathcal H}}
\def\calB{{\mathcal B}}

\def\calF{{\mathcal F}}
\def\calT{{\mathcal T}}

\def\expit{{\mbox{expit}}}

\def\ind{\mathbbm{1}}

\def\H{{\bf H}}

\def\mR{\mathbb{R}}

\def\fone{f_1}
\def\fzero{f_0}
\def\fx{f_{\X}}

\def\var{\hbox{var}}
\def\sd{\hbox{sd}}
\def\Est{\hbox{Est}}
\def\Bias{\hbox{Bias}}
\def\logistic{\hbox{logistic}}
\def\rforest{\hbox{rforest}}
\def\xgboost{\hbox{xgboost}}

\newcommand{\tg}[1]{}

\allowdisplaybreaks[2]

\begin{document}

\title{Efficient Estimation of Average Treatment Effect on the Treated under Endogenous Treatment Assignment}

\setcounter{footnote}{1}
\author{Trinetri Ghosh\thanks{US Food \& Drug Administration. This work was completed while Trinetri Ghosh was affiliated with the University of Wisconsin-Madison.}~,~
Jiawei Shan\thanks{Departments of Statistics and of Biostatistics \& Medical Informatics, University of Wisconsin-Madison.}~,~
Menggang Yu\thanks{Department of Biostatistics, University of Michigan.}~,~
Jiwei Zhao\thanks{Departments of Statistics and of Biostatistics \& Medical Informatics, University of Wisconsin-Madison; e-mail: \texttt{jiwei.zhao@wisc.edu}. }\\[0.4em]
}

\begingroup
\renewcommand{\thefootnote}{*}
\footnotetext[1]{The authors are listed in alphabetical order.}
\endgroup
\maketitle
	
	\begin{abstract}%  
		In this paper, we consider estimation of average treatment effect on the treated (ATT), an interpretable and relevant causal estimand to policy makers when treatment assignment is endogenous.
		By considering shadow variables that are unrelated to the treatment assignment but related to the outcomes of interest, we establish identification of the ATT.
		Then we focus on efficient estimation of the ATT by characterizing the geometric structure of the likelihood, deriving the semiparametric efficiency bound for ATT estimation and proposing an estimator that can achieve this bound. We rigorously establish the theoretical results of the proposed estimator.
		The finite sample performance of the proposed estimator is studied through comprehensive simulation studies as well as an application to our motivating study.
	\end{abstract}
	
	\begin{keywords}
		Causal inference, endogeneity, identifiability, semiparametric efficiency, influence function
	\end{keywords}
	
	\section{Introduction}\label{sec:intro}

	Among many causal estimands, the most well studied is the average treatment effect (ATE), which captures the average difference in potential outcomes between treatment and control for all study subjects \citep{imbens2015causal, hernan_robins_2022, Morgan_Winship_2014}. Another popular estimand is the average treatment effect on the treated (ATT) \citep{imbens2015causal, hernan_robins_2022}. \cite{heckmansmith1995} and \cite{heckman2001policy} argued that, in many economics and program evaluation studies, ATT might be the implicit quantity sought after, hence more relevant and more interpretable to policymakers.
	For example, when evaluating whether a smoking cessation campaign intervention implemented by the government can decrease the smoking prevalence in a city, researchers and policymakers might be more interested in the effect of the intervention among those who actually received the campaign (i.e., ATT) but not on those who never received \citep{wang2017g}.
	Similar situations also arise in the medical field where a major interest is to assess the treatment effect on actual participants.
	For instance, patients who agreed to take a new drug with a recommended higher dose, instead of patients who did not, would wonder about the benefit of the higher dose.
	It is also the case in our motivating study, with details presented in Section~\ref{sec:real_data}, where the goal is to assess the effect of the stereotactic body radiation therapy (SBRT) among patients with non-small-cell lung cancer (NSCLC). Radiation oncologists would be more interested in the treatment effect on those who actually received SBRT, but not on those for whom the SBRT was never intended.
	
	A key assumption for identifying ATT is strong ignorability of treatment assignment \citep{rosenbaum1983central}, which means that the treatment assignment mechanism is independent of the potential outcomes given the observed covariates or confounders.
	This assumption is also known as unconfoundedness, selection on observables, or exogeneity \citep{imbens2004nonparametric, imbens2015causal}.
	Under this assumption, ATT can be identified and estimated based on propensity scores \citep{rosenbaum1983central, hahn1998role, dehejia1999causal, lechner1999earnings, hirano2003efficient, mccaffrey2004propensity, hainmueller2012entropy, hartman2015sample, pirracchio2016propensity, zhao2017entropy, moodie2018doubly}, regression \citep{heckman1997matching, heckman1998matching, imbens2005mean},
	matching \citep{cochran1973controlling, rubin1973matching, rubin1973use, abadie2006large, abadie2011bias},
	or combinations of these methods \citep{abadie2002simple, pirracchio2018balance}. However, in observational studies, the  unconfoundedness assumption may be frequently violated \citep{frolich2013identification, mebane2013causal}. That is, the differences in outcome measures between treated and untreated may be due to unobserved differences in selecting treatment status that also predict the outcome. Consequently, ATT estimators that rely on unconfoundedness can be substantially biased, and the corresponding estimand may not even be identifiable at all.

	To make progress in the absence of exogeneity, researchers have turned to `helper' variables, in particular, instrumental variables (IVs) that satisfy three key requirements: independence (from unmeasured confounders), exclusion (from dependence with the potential outcomes), and relevance (to the treatment received) \citep{IVSIMreview2014, clarke2012instrumental, ChenJCE2011}.
	When such an IV exists,  ATT may be properly estimated in the presence of unmeasured confounding, with further modeling assumptions either on the outcome or on the IVs \citep{wright1928tariff, robins1994correcting, clarke2012instrumental, liu2020identification, imbens2015causal, hernan_robins_2022}. It is important to emphasize, however, that full nonparametric identification of ATT generally remains impossible even with valid IVs \citep{PearlBiometrika95, Pearl_2009}. Moreover, credible IVs are often difficult to find \citep{newhouse1998econometrics, angrist2001instrumental, hernan2006instruments, martens2006instrumental}, because there may be many unmeasured confounders and it can be challenging to rule out the IV's dependence on all of them \citep{ChenJCE2011}.
	In addition, negative controls (NCs) provide another class of helper variables for identification under unmeasured confounding. This approach typically requires a pair of NCs: a negative control exposure and a negative control outcome, which together can help establish identifiability of causal effects \citep{MiaoGengTchetgenTchetgen2018,TchetgenTchetgenYingCuiShiEtAl2024}.

	\label{rev:intro_SV}
	In this paper, we consider an alternative kind of `helper' variables, denoted as $\Z$ throughout, that are (i) independent of the treatment assignment conditional on the (potentially) untreated outcome and other observed covariates, and (ii) related to the
	(potentially) untreated outcome, possibly through unmeasured confounders. We call such variables shadow variables (SVs) due to their link with the nonresponse instrument in the literature of missing data analysis
	\citep{wang2014instrumental, zhao2015semiparametric, miao2016identifiability, zhao2022versatile}.
	SVs may be more readily available in the settings where treatment assignment is more proximal or nearer in time to the measurement of SVs. In such settings, we might have more information about the treatment assignment to rule out dependence of treatment assignment on the SVs. For example, in our motivating study for newly diagnosed NSCLC patients, a new radiation therapy, SBRT, is compared to the standard surgery procedure, lobectomy.
    According to our collaborating radiation oncologists, lobectomy is the standard treatment for most patients and SBRT has been more commonly given to elderly patients who may be at higher risk of surgical complications.  In addition, SBRT is among the most cost-effective treatment options for early-stage NSCLC, requiring fewer visits and generally being covered by most insurance types \citep{ShahHahnStetsonFriedbergEtAl2013,SBRT2022}. Therefore, some variables related to social determinants of health (SDOH) should not drive the treatment assignment and can be considered as $\Z$, such as	income level (binary, income$<$\$48,000 versus income$\geq$\$48,000) and insurance level (binary, private versus non-private).

	\label{rev:intro_contribution}
	Our main contributions are threefold. 
	First, leveraging a valid SV, we establish the nonparametric identifiability of ATT, encoded as $\Delta$ throughout. 
    SVs have been studied in the missing data literature, but, to our best knowledge, this paper provides the first result for the identification of a causal estimand with the aid of an SV.
	Second, under a parametric treatment assignment model indexed by $\bt$, we develop a comprehensive semiparametric theory for inference on both $\bt$ and $\Delta$. In particular, we derive the nuisance tangent space and its orthogonal complement, and obtain the efficient influence functions \citep{bickel1993efficient, tsiatis2006semiparametric}. This complete geometric characterization clarifies how nuisance components are intertwined, thereby guiding the construction of flexible estimators and the derivation of efficient procedures. We also propose an alternative estimator $\wh\Delta\proone$ which, albeit not as efficient as $\wh\Delta\eff$ in theory, is substantially easier to implement numerically. Moreover, we elucidate the relationship between $\wh\Delta\eff$ and $\wh\Delta\proone$ by connecting their influence functions via the projection technique \citep{tsiatis2006semiparametric}.
	Third, the derived semiparametric theory enables the construction of Neyman orthogonal scores, which in turn permits the use of modern machine learning methods for nuisance estimation while retaining valid inference for the target parameters. Simulations and a real-data application demonstrate the effectiveness, scalability, and practical value of our proposed methods.

	The remainder of the paper is organized as follows. Section~\ref{sec:complexity} introduces the basic framework and provides sufficient conditions for identifying the likelihood and the parameter of interest. Section~\ref{sec:est.delta} develops the semiparametric theory and the corresponding efficient estimators; it also introduces a simpler alternative estimator and clarifies its connection to the efficient one. Section~\ref{sec:tech} describes implementation details and establishes the asymptotic properties of the proposed estimators. Section~\ref{sec:simu} presents simulation studies evaluating finite-sample performance and comparing our methods with existing approaches. Section~\ref{sec:real_data} applies the proposed method to the NCDB data. Additional technical details are deferred to the Appendix.

\section{Preliminaries}\label{sec:complexity}

We adopt the potential outcome framework, which implicitly requires the stable unit treatment value assumption (SUTVA) \citep{neyman1923applications, rubin1974estimating, rubin1990comment} throughout this article. Let $T$ denote the treatment where $T=1$ means treatment/intervention and $T=0$ control. The observed outcome is then $Y=TY_1+(1-T)Y_0$, where $Y_t$ denotes the potential outcome under $T=t$, $t\in\{0,1\}$. In addition, we observe a vector of pre-treatment covariates $\X$.

The traditional strong ignorability assumption requires $T \perp\!\!\!\perp (Y_1,Y_0) \mid \X$ \citep{rosenbaum1983central}. Consequently, $\pr(T=1\mid y_1,y_0,\x)=\pr(T=1\mid \x)$, so in observational studies the non-randomized treatment assignment mechanism can be modeled based on observed covariates alone, say $w(\x)\equiv\pr(T=1\mid \x)$. When the ignorability assumption fails, one might have to directly model $\pr(T=1\mid y_1,y_0,\x)$, which involves the potential outcomes. In this article, we instead propose to model the conditional distribution of $T$ given $\X$ and the potential untreated outcome $Y_0$, denoted as
\begin{eqnarray*}
\pi(y_0,\x)\equiv \pr(T=1\mid y_0,\x)=\int \pr(T=1\mid y_0, y_1, \x)f(y_1\mid y_0,\x) {\rm d} y_1,
\end{eqnarray*}
where $f(y_1\mid y_0,\x)$ is the conditional density of $Y_1$ given $Y_0$ and $\X$.
In particular, we allow $\pr(T=1\mid y_0, y_1, \x)\neq \pr(T=1\mid y_0,\x)\neq \pr(T=1\mid\x)$.

\subsection{Identification}\label{ssec:identification}

Because for each subject, only one of the potential outcomes is observed, violation of the strong ignorability assumption implies that our parameter of interest is not directly identifiable from the likelihood of the observed data, without imposing extra assumptions.
Throughout the ATT is defined as
\begin{align}
 \Delta =&~ \E(Y_1-Y_0\mid T=1) \nonumber \\
=&~ \frac{\iint y_1 w(\x)\fone(y_1\mid \x,1)g(\x)\dd y_1 \dd\x - \iint y_0
	\frac{\pi(y_0,\x)}{1-\pi(y_0,\x)}
	\{1-w(\x)\}\fzero(y_0\mid \x,0)g(\x)\dd y_0\dd\x }{\int w(\x)g(\x)\dd \x},\label{eq:att}
\end{align}
where $\fone(y_1\mid\x,1)$ is the conditional density of $Y_1$ given $\X$ and $T=1$, $\fzero(y_0\mid \x,0)$ the conditional density of $Y_0$ given $\X$ and $T=0$, and
$g(\x)$ is the marginal distribution of $\X$.

Note that the likelihood function from one single observation in this model is
\begin{align}
&\{\fone(y_1\mid\x,1)w(\x)g(\x)\}^t
\left[\fzero(y_0\mid \x,0)\{1-w(\x)\}g(\x)\right]^{1-t}\nonumber\\
=&~ w(\x)^t \{1-w(\x)\}^{1-t} \fone(y_1\mid\x,1)^t \fzero(y_0\mid \x,0)^{1-t} g(\x).\label{eq:likelihood1}
\end{align}
Therefore one can identify $\fone(y_1\mid\x,1)$, $\fzero(y_0\mid \x,0)$, $g(\x)$ and $w(\x)$ from  (\ref{eq:likelihood1}). However $\pi(y_0,\x)$ cannot be identified, which plays a critical role in the numerator of $\Delta$ from \eqref{eq:att}.
In other words the parameter of interest $\Delta$ cannot be identified from (\ref{eq:likelihood1}) without extra assumptions.

To identify $\pi(y_0,\x)$, we write the following  transformation of the identifiable $w(\x)$ as
\begin{align}\label{eq:iden_pi}
\{1-w(\x)\}^{-1} = \int \frac{\fzero(y_0\mid \x,0)}{1-\pi(y_0,\x)} \dd y_0.
\end{align}
Then if for any two different functions $\pi_1(y_0,\x)$ and $\pi_2(y_0,\x)$ such that
\[
\int \frac{\fzero(y_0\mid \x,0)}{1-\pi_1(y_0,\x)} \dd y_0 = \int \frac{\fzero(y_0\mid \x,0)}{1-\pi_2(y_0,\x)}\dd y_0 \mbox{ implies } \pi_1(y_0,\x)=\pi_2(y_0,\x),
\]
we have hope to identify  $\pi(y_0,\x)$. In particular, we assume that $\X$ can be partitioned into two subsets of variables $\U$ and $\Z$, i.e. $\X=(\U\trans,\Z\trans)\trans$, such that the following set of sufficient conditions for identification hold:
\begin{itemize}
	\item[(i)]{\em Conditional completeness of the potential outcome distribution}:
	\begin{eqnarray} \Ezero\{h(Y_0,\U)\mid \X,0\}  = 0 \mbox{ implies } h(Y_0,\U)=0 \mbox{ almost surely},\label{eq:assumecompleteness}\end{eqnarray}
	
	\item[(ii)]{\em Exclusion restriction from treatment assignment}:
	\begin{eqnarray} \pr(T=1\mid y_0,\x) = \pr(T=1\mid y_0,\u).\label{eq:assumeshadow}
	\end{eqnarray}
\end{itemize}

The assumption (\ref{eq:assumecompleteness}) is the well-known completeness condition, popularly used in missing data analysis, instrumental variable models, and econometrics; see, e.g., \cite{newey2003instrumental, hu2018nonparametric, zhao2022versatile}. It is a condition imposed on the distribution $\fzero(y_0\mid \x,0)$. By writing $\Ezero\{h(Y_0,\U)\mid \X,0\} = \Ezero\{h(Y_0,\U)\mid \U, \Z,0\}$, 
the condition essentially requires the shadow variable $\Z$ to be correlated with $Y_0$ and to exhibit sufficient variation relative to $Y_0$.
For instance, in the case of categorical $Y_0$ and $\Z$, it requires $\Z$  to have at least as many categories as $Y_0$.
In our motivating example where $Y_0$ is binary, we choose $\Z$ to consist of two binary covariates (hence to have four categories) to ensure that this assumption holds.
The potential outcome $Y_0$ can also be continuous. In that case, several commonly used models for $f_0(y_0\mid \x,0)$, such as the full exponential families as illustrated in Example \ref{exa:exp_family} below, satisfy the completeness condition \citep{newey2003instrumental,d2011completeness}. \label{resp:completeness}

\begin{example}
	[Full exponential family]
	% , Theorem 2.2 of \cite{newey2003instrumental}]
	\label{exa:exp_family}
	If the conditional distribution of $y_0$ given $(\x,T=0)$ is absolutely continuous and $f_0(y_0\mid \x,0) = s(y_0,\u)t(\x) \exp\{ \mu(\x)\trans\tau(y_0,\u)\}$, where 
    $s(y_0,\u)>0$, $\tau(y_0,\u)$ is one-to-one in $y_0$, and the support of $\mu(\x)$ given $\u$ is an open set, then the completeness condition \eqref{eq:assumecompleteness} holds.
\end{example}

The assumption \eqref{eq:assumeshadow} requires $\Z$ to be independent of treatment assignment conditional on the untreated potential outcome $y_0$ and the other covariates $\u$. When a medical intervention or treatment can be conveniently delivered to patients, then a patient's access to transportation or distance to the health care center (e.g. rural vs urban) should not be related to the treatment assignment. So such variables can be part of $\Z$.
A cost-effective treatment may be given to patients purely based on toxicity and efficacy trade-off; hence, some socioeconomic variables or educational background variables can be part of $\Z$ as well.

Under assumptions (\ref{eq:assumecompleteness}) and (\ref{eq:assumeshadow}), the model likelihood (\ref{eq:likelihood1}) becomes
\begin{eqnarray}\label{eq:likelihood2edn}
\left[\int\frac{\fzero(y_0\mid \x,0)}{1-\pi(y_0,\u)}\dd y_0-1\right]^{t}
\left[\int\frac{\fzero(y_0\mid \x,0)}{1-\pi(y_0,\u)}\dd y_0\right]^{-1}
\fone(y_1\mid\x,1)^t \fzero(y_0\mid \x,0)^{1-t} g(\x),
\end{eqnarray}
and we have the following identifiability result with its proof contained in Appendix~\ref{sec:lemmaidenproof}.
\begin{Lem}\label{lemmaiden}
	Under assumptions (\ref{eq:assumecompleteness}) and (\ref{eq:assumeshadow}), the model likelihood (\ref{eq:likelihood2edn}) is fully identifiable, hence the parameter $\Delta$ is identifiable.
\end{Lem}

\begin{Rem}\label{rem:comp_missing}
	Identification of $\Delta$ using shadow variables is conceptually similar to approaches in the missing-data literature \citep{wang2014instrumental,zhao2015semiparametric}. 
    %Causal inference can be naturally framed as a missing-data problem, where the treatment assignment $T$ serves as a missingness indicator. 
    When ATT is the primary parameter of interest, $\E(Y_1\mid T=1)$ is inherently identifiable, and $\E(Y_0\mid T=1)$ is the key quantity to be identified. Thus, it suffices to identify the missingness propensity for $Y_0$, i.e., $\pr(T=1\mid y_0,\x)$, instead of $\pr(T=1\mid y_0,y_1,\x)$, as discussed above. 
	It is worth noting that although $\E(Y_0\mid T=1)$ is identifiable, its efficient influence function (EIF) differs from that under ignorability \citep{hahn1998role}, because the likelihood in the current setting, given by \eqref{eq:likelihood2}, differs from that under ignorability. We derive the corresponding EIFs in Section~\ref{sec:est.delta}.
\end{Rem}

\begin{Rem}\label{rem:comp_IV_SV}
	Instrumental variables (IVs) are another commonly used class of auxiliary variables in the causal inference literature for identifying causal effects in the presence of unmeasured confounding \citep{AngristImbensRubin1996,WangTchetgenTchetgen2018a}. A valid IV is typically required to be exogenous, relevant to the treatment, and to affect the outcome only through the treatment. These conditions often imply that an IV is a pre-treatment, externally generated variable that does not directly influence the outcome. In contrast, a shadow variable (SV) is assumed not to directly affect treatment assignment while being predictive of the outcome. Consequently, an SV can be post-treatment or measured contemporaneously with the treatment, as long as it does not drive the treatment assignment.
\end{Rem}

The assumption (\ref{eq:assumeshadow}) indicates that we can model $T$ based on $y_0$ and $\u$. 
Note that identifying $\pi(\cdot)$ from \eqref{eq:iden_pi} amounts to solving a Fredholm integral equation of the first kind \citep{Kress_LinearIntegralEquations}. Such problems are notoriously known to be ill-posed, which makes nonparametric modeling challenging.
In the rest of the paper, we therefore adopt a parametric model for $T$:
\begin{eqnarray}\label{eq:assumepar}
\pr(T=1\mid y_0,\x)=\pr(T=1\mid y_0,\u)=\pi(y_0,\u;\bt),
\end{eqnarray}
where the form of $\pi(\cdot)$ is known and $\bt$ is an unknown $d$-dimensional parameter. We focus on the corresponding efficient estimation of both $\bt$ and $\Delta$, and defer a discussion of the purely nonparametric approach to Appendix \ref{sec:nonparametric}. \label{resp:nonp1}
In our motivating example, we adopt the logistic regression model
\begin{eqnarray*}
\pi(y_0,\u;\bt)=\expit(\theta_1+\theta_2 y_0+\bt_3\trans\u),
\end{eqnarray*}
and the estimates for $\theta_2$ are highly significant (see Table~\ref{tab:propensity_comparison_cdcc} of Section~\ref{sec:real_data}).
Indeed, this dependence of the treatment assignment on the potential untreated outcome $Y_0$ indicates that the traditional strong ignorability assumption may not be sensible in this study.

\section{Estimation}\label{sec:est.delta}

Throughout the paper, when there is no ambiguity, we abbreviate $\pi(y_0,\u;\bt)$ as $\pi(\bt)$ and $\pi(y_{0i},\u_i;\bt)$ as $\pi_i(\bt)$.
We denote $\partial \pi(\bt)/\partial \bt$ by $\dot\bpi(\bt)$.
We denote the joint density of $Y_1,Y_0$ given $\X$ by $f(y_1,y_0\mid\x)$; therefore, $\fone(y_1\mid\x)=\int f(y_1,y_0\mid\x)\dd y_0$ and $\fzero(y_0\mid\x)=\int f(y_1,y_0\mid\x)\dd y_1$.
We write the conditional expectation with respect to $\fzero(y_0\mid \x)$ as $\Ezero(\cdot\mid \x)$ and that with respect to $\fzero(y_0\mid \x,0)$ as $\Ezero(\cdot\mid \x,0)$. Similar notation also applies to $\Eone(\cdot\mid \x)$ and $\Eone(\cdot\mid \x,1)$.
We write $\W=(T,Y,\X\trans)\trans$, $p=\pr(T=1)$, and specify the true value of unknown parameters with superscript $^0$ whenever needed, such as $\bt^0$ and $\Delta^0$.
For clarity of presentation, we also adopt the simplified notations summarized in Table~\ref{tab:notations}. These notations leverage the identity $\Ezero\{\a(Y_0,\x)\mid \x\} = \{1-w(\x)\}\Ezero[\{1-\pi(\bt^0)\}^{-1}\a(Y_0,\x)\mid\x,0]$ for any $\a(y_0,\x)$, which allows us to identify quantities of the form $\Ezero(\cdot\mid \x)$. \label{rev:notation}

\label{rev:EIF}
In this section, our main goal is to derive the efficient influence functions for $(\bt^0,\Delta^0)$ under model assumptions \eqref{eq:assumecompleteness}--\eqref{eq:assumeshadow}, which in turn motivate an efficient estimation approach for the parameters of interest. In semiparametric theory, an estimator $\wh{\bb}=\wh{\bb}(\w_1,\dots,\w_n)$ of $\bb^0$ is called regular and asymptotically linear (RAL) if
\begin{align*}
\sqrt{n}(\wh{\bb}-\bb^0)=n^{-1/2} \sum_{i=1}^n \bpsi\left(\w_i\right)+o_p(1),
\end{align*}
where $\bpsi(\cdot)$ is a mean-zero function, referred to as the influence function of $\wh{\bb}$. Under mild regularity conditions, the central limit theorem implies that $\sqrt{n}(\wh{\bb}-\bb)\xrightarrow{d} N(\0,\E(\bpsi \bpsi\trans))$, provided that $\E(\bpsi\bpsi\trans)$ is finite and nonsingular. Among all RAL estimators of $\bb^0$, the one with the smallest asymptotic variance has the influence function that is the efficient influence function (EIF), denoted as $\bpsi\eff$, and the corresponding semiparametric efficiency bound is $\E\left(\bpsi\eff \bpsi\eff\trans\right)$ \citep{bickel1993efficient,tsiatis2006semiparametric}.

\begin{table}[htbp]
\centering
\caption{Simplified notations used throughout the paper}
\label{tab:notations}
\begin{tabular}{cl}
\toprule
Notation & \multicolumn{1}{c}{Definition} \\
\midrule
$w(\x)$ & $1-\Ezero[ \{1-\pi(\bt^0)\}^{-1}\mid\x,0]^{-1}$\\
$\A(\x)$ & $ \Ezero [\{1-\pi(\bt^0)\}^{-1}\dot\bpi(\bt^0)\mid \x] = \{1-w(\x)\}\Ezero [\{1-\pi(\bt^0)\}^{-2}\dot\bpi(\bt^0)\mid \x,0]$ \\
$B(\x)$ & $\Ezero [\{1-\pi(\bt^0)\}^{-1}\pi(\bt^0)\mid \x]
= \{1-w(\x)\}\Ezero [\{1-\pi(\bt^0)\}^{-2}\pi(\bt^0)\mid \x,0]$ \\
$C(\x)$ & $\Ezero [\{1-\pi(\bt^0)\}^{-1}\pi(\bt^0)^2Y_0\mid \x] = \{1-w(\x)\}\Ezero [\{1-\pi(\bt^0)\}^{-2}\pi(\bt^0)^2Y_0\mid \x,0]$ \\
$V(\x)$
		& $p \E\left\{\phi\proone(\w;\bt^0,\Delta^0)\mid\x\right\} + \Ezero [Y_0 \{1-\pi(\bt^0)\}^{-1}\pi(\bt^0)\mid \x]$ \\
		& $= w(\x)\{\Eone(Y_1\mid \x,1)-\Delta^0\} + \{1-w(\x)\}\Ezero [Y_0 \{1-\pi(\bt^0)\}^{-2}\pi(\bt^0)^2\mid \x,0]$ \\
$\M$ & $\E\{B(\x)^{-1}\A(\x)\A(\x)\trans\}$ \\
$\Q$ & $\E [Y_0 \{1-\pi(\bt^0)\}^{-1}\dot\bpi(\bt^0)]$ \\
$\D$ & $\E\{B(\x)^{-1} V(\x) \A(\x)\}$ \\
\bottomrule
\multicolumn{2}{l}{\footnotesize $^* \phi\proone(\w;\bt,\Delta)$ is defined in (\ref{eq:IF.pro1}).}
\end{tabular}
\end{table}

\subsection{Efficient Estimation of $\Delta$}\label{subsec:eff_delta}

Under the assumptions (\ref{eq:assumecompleteness}) and (\ref{eq:assumepar}), the model likelihood is
\begin{eqnarray}\label{eq:likelihood2}
\left[\int\frac{\fzero(y_0\mid \x,0)}{1-\pi(y_0,\u;\bt)}\dd y_0-1\right]^{t}
\left[\int\frac{\fzero(y_0\mid \x,0)}{1-\pi(y_0,\u;\bt)}\dd y_0\right]^{-1}
\fone(y_1\mid\x,1)^t \fzero(y_0\mid \x,0)^{1-t} g(\x),
\end{eqnarray}
which contains three nonparametric nuisance parts $\fone(y_1\mid\x,1)$, $\fzero(y_0\mid\x,0)$, $g(\x)$, and one $d$-dimensional parameter $\bt$.
We consider the Hilbert space $\calH$ of all $d$-dimensional zero-mean measurable functions of the observed data with finite variance, equipped with the inner product $\langle \h_1,\h_2\rangle=\E\{\h_1(\cdot)\trans \h_2(\cdot)\}$, where $\h_1, \h_2\in \calH$.
We first derive the nuisance tangent space and its orthogonal complement, where the nuisance tangent space is defined as the mean squared closure of the nuisance tangent spaces of parametric submodels spanned by the nuisance score vectors.
We have  the following result whose proof can be found in Appendix~\ref{sec:derivelambdaperp}.
\begin{prop}\label{prop:Hdecomp}
	The space $\calH$ can be decomposed as
	\begin{eqnarray*}
	\calH = \Lambda_0 \oplus \Lambda_1 \oplus \Lambda_2 \oplus \Lambda^\perp,
	\end{eqnarray*}
	where
	\begin{align*}
	\Lambda_0 =&~  \left[
	(1-t)\c(y_0,\x)+\frac{t-w(\x)}{w(\x)}\Ezero\{\c(y_0,\x)\mid\x\} : \c(y_0,\x)\in\mR^{d}, \Ezero\{\c(y_0,\x)\mid\x,0\}=\0
	\right],\\
	\Lambda_1 =&~  \left[ t \b(y_1,\x)\in\mR^{d}: \Eone\{\b(y_1,\x)\mid\x,1\}=\0 \right],\\
	\Lambda_2 =&~  \left[ \a(\x)\in\mR^{d}: \E\{\a(\x)\}=\0 \right],
	\end{align*}
	are the nuisance tangent spaces with respect to $\fzero(y_0\mid\x,0)$, $\fone(y_1\mid\x,1)$, and $g(\x)$, respectively, and the orthogonal complement is
	\begin{eqnarray*}
	\Lambda^\perp
	=  \left\{\frac{t-\pi(\bt^0)}{1-\pi(\bt^0)}\g(\x): \g(\x)\in \mR^{d} \right\}.
	\end{eqnarray*}
	The notation $\oplus$ represents the direct sum of two spaces that are orthogonal to each other.
\end{prop}
The tangent spaces presented in Proposition~\ref{prop:Hdecomp} depend on the parameterization of $\pi(y_0,\u)$ and would differ under a nonparametric specification. 
\label{rev:eif_nonp}
Proposition~\ref{prop:Hdecomp} indicates that, any influence function for estimating $\Delta$ can be written as the summation of four pairwise orthogonal elements  in spaces $\Lambda_2$, $\Lambda_1$, $\Lambda_0$ and $\Lambda^\perp$, respectively.
In particular we study the efficient influence function,
which must be of the form
\begin{align}
\label{eq:eif.att}
\phi\eff(\w;\bt^0,\Delta^0) =&~ \LaTeXunderbrace{a^*(\x)}_{\in\Lambda_2} + \LaTeXunderbrace{t b^*(y_1,\x)}_{\in\Lambda_1} + \LaTeXunderbrace{(1-t)c^*(y_0,\x)+\frac{t-w(\x)}{w(\x)}\Ezero(c^*\mid \x)}_{\in\Lambda_0} + \LaTeXunderbrace{\H\trans\S\eff(\w;\bt^0)}_{\in\Lambda^\perp},\nonumber\\
\equiv&~ \phi(\w;\bt^0,\Delta^0) + \H\trans\S\eff(\w;\bt^0),
\end{align}
where $\E(a^*)=0$, $\Eone(b^*\mid \x,1)=0$, $\Ezero(c^*\mid \x,0)=0$, $\H$ is a $d \times 1$ vector and $\S\eff(\w;\bt)$ is the efficient score of $\bt$, to be delineated later in Section~\ref{subsec:est.theta}.
The explicit formulas of $\phi(\w;\bt,\Delta)$ and $\H$ are summarized below, with the proof contained in Appendix~\ref{sec:eifdelta}.
\begin{prop}\label{prop:Hphiprotwo}
	We can derive that $\H = p^{-1}\M^{-1}(\D-\Q)$ and $\phi(\w;\bt,\Delta)=$
	% {\small{
	% 		\begin{eqnarray}\label{eq:IF.pro2}
	% 		&& 
	% 		%\phi(\w;\bt,\Delta) \n\\
	% 		%&=&
	% 		p^{-1} \left\{\frac{t-\pi(\bt)}{1-\pi(\bt)}Y - \frac{t-\pi(\bt)}{1-\pi(\bt)} \frac1{B(\x)}\left(w(\x)\Eone(Y_1\mid\x,1)+\{1-w(\x)\}\Ezero\left[\frac{\pi(\bt^0)^2}{\{1-\pi(\bt^0)\}^2}Y_0\mid\x,0\right]\right)\right. \n\\
	% 		&&\ \ \ \left.-\Delta\left\{t-\frac{t-\pi(\bt)}{1-\pi(\bt)}\frac{w(\x)}{B(\x)}\right\}
	% 		\right\}.
	% 		\end{eqnarray}
	% }}
	\begin{align}\label{eq:IF.pro2}
		p^{-1} \left[\frac{t-\pi(\bt)}{1-\pi(\bt)}Y - \frac{t-\pi(\bt)}{1-\pi(\bt)} \frac1{B(\x)}\left\{w(\x)\Eone(Y_1\mid\x,1)+C(\x)\right\} 
		-\Delta\left\{t-\frac{t-\pi(\bt)}{1-\pi(\bt)}\frac{w(\x)}{B(\x)}\right\}
		\right],
	\end{align}
	where the simplified notations $\{\M,\Q,\D,B(\x),C(\x)\}$ are defined in Table~\ref{tab:notations}.
\end{prop}
It is clear that any estimator of $\Delta$, $\wh\Delta$, depends on a generic estimator of $\bt$, $\wh\bt$.
This is also reflected in the efficient influence function $\phi\eff(\w;\bt,\Delta)$ in (\ref{eq:eif.att}). For a given $\bt$, the efficient estimator of $\Delta$ can be obtained from solving $\frac1n\sumi \phi(\w_i;\bt,\Delta)=0$; i.e., 
\begin{align*}
\wh\Delta\eff(\bt) =
\frac{
	\frac1n\sumi \left[ \frac{t_i-\pi_i(\bt)}{1-\pi_i(\bt)}y_i - \frac{t_i-\pi_i(\bt)}{1-\pi_i(\bt)}
	\frac1{\wh B(\x_i)}
	\left\{\wh w(\x_i)\wh\E_1(Y_1\mid \x_i,1)+
	% \{1-\wh w(\x_i)\}\wh\E_0\left[\frac{\pi^2(\bt)Y_0}{\{1-\pi(\bt)\}^2}\mid \x_i,0\right]
	\wh C(\x_i)
	\right\}
	\right]
}{
	\frac1n \sumi \left\{ t_i - \frac{t_i-\pi_i(\bt)}{1-\pi_i(\bt)}\frac{\wh w(\x_i)}{\wh B(\x_i)}\right\}
}.
\end{align*}

The estimator $\wh\Delta\eff(\bt)$ literally stands for a family of estimators.
If the truth value of $\bt$, $\bt^0$, is plugged in, one would obtain $\wh\Delta\eff(\bt^0)$. Because $\bt$ is unknown, a natural way is to plug in the efficient estimator of $\bt$, $\wh\bt\eff$, that solves $\frac1n\sumi \S\eff(\w_i;\bt)=\0$. Then one would obtain $\wh\Delta\eff(\wh\bt\eff)$, which is also shorthanded as $\wh\Delta\eff$ throughout the paper.
Its implementation needs modeling of the nuisance parts $\Ezero(\cdot\mid\x,0)$ and $\Eone(\cdot\mid \x,1)$.
The property of the estimator $\wh\Delta\eff(\bt)$ certainly depends on how well the nuisance models $\wh \E_0(\cdot\mid \x,0)$ and $\wh\E_1(\cdot\mid \x,1)$ work.
It can also be easily checked that $\wh\Delta\eff(\bt)$ satisfies the Neyman  orthogonality condition \citep{neyman1959optimal, neyman1979c, chernozhukov2018double}.
In Section \ref{sec:tech}, we will show with details that under some regularity conditions, formalized later in Assumptions~\ref{assump:consistency.para}-\ref{assump:quality.nuisance.est},
\begin{align}
\sqrt{n}\left\{\wh\Delta\eff(\bt^0)-\Delta^0\right\} =&~ n^{-1/2} \sumi \phi(\w_i;\bt^0,\Delta^0) + o_p(1) \xrightarrow{d} N \left[0, \E\{\phi^2 (\w;\bt^0,\Delta^0)\}\right],\nonumber\\
\sqrt{n}\left\{\wh\Delta\eff(\wh\bt\eff)-\Delta^0\right\} =&~ n^{-1/2} \sumi \phi\eff(\w_i;\bt^0,\Delta^0) + o_p(1) \xrightarrow{d} N \left[0, \E\{\phi\eff^2 (\w;\bt^0,\Delta^0)\}\right].\label{eq:deltaprotwo}
\end{align}
Note that the estimator $\wh\Delta\eff(\bt^0)$ is not of practical interest since it is infeasible, whereas the estimator $\wh\Delta\eff(\wh\bt\eff)$ achieves the semiparametric efficiency bound $\E\left\{\phi\eff^2(\w;\bt^0,\Delta^0)\right\}$.
Because
\begin{eqnarray*}
\E\left\{\phi\eff^2(\w;\bt^0,\Delta^0)\right\} = \E\left\{\phi^2(\w;\bt^0,\Delta^0)\right\} + p^{-2}(\D-\Q)\trans\M^{-1}(\D-\Q),
\end{eqnarray*}
the component $p^{-2}(\D-\Q)\trans\M^{-1}(\D-\Q)$, or equivalently the component $\H\trans\S\eff(\w;\bt)$ in (\ref{eq:eif.att}), amounts to the estimation variability brought by $\wh\bt\eff$.

\subsection{Efficient Estimation of $\bt$}\label{subsec:est.theta}

In this section, we propose an efficient estimator of $\bt$ by characterizing its efficient influence function. Theorem~\ref{th:theta.ATT2} establishes that the nuisance tangent space admits the decomposition $\Lambda=\Lambda_0 \oplus \Lambda_1 \oplus \Lambda_2$. It follows that
any element in its perpendicular $\Lambda^\perp$ with an arbitrary $d$-dimensional nondegenerate function $\g(\x)$ yields a consistent estimator of $\bt$; i.e., the solution of
\begin{eqnarray}\label{eq:solvethetausingg}
\frac1n\sumi \frac{t_i-\pi_i(\bt)}{1-\pi_i(\bt)}\g(\x_i)=\0.
\end{eqnarray}
The optimal choice of $\g(\x)$ that provides the minimum estimation variability will be given by the efficient score function $\S\eff$, which is defined as the projection of the score function
\begin{align*}
\S_{\bt} =&~  \frac{\{t-w(\x)\}\{1-w(\x)\}}{w(\x)} \Ezero [\{1-\pi(\bt)\}^{-2}\dot\bpi(\bt)\mid \x,0]\\
=&~  \frac{t-w(\x)}{w(\x)} \Ezero\left[ \{1-\pi(\bt^0)\}^{-1}\dot\bpi(\bt^0)\mid\x \right]
\end{align*}
derived in Appendix \ref{sec:eifdelta} onto the space $\Lambda^\perp$ \citep[Theorem 3.5]{tsiatis2006semiparametric}.
It can be easily seen that $\S_\bt \perp \Lambda_2$ and $\S_\bt \perp \Lambda_1$.
Therefore, we can decompose $\S_\bt$ as
\begin{align*}
& \frac{t-w(\x)}{w(\x)} \Ezero\left[ \{1-\pi(\bt^0)\}^{-1}\dot\bpi(\bt^0)\mid\x \right]\\
=&~ \LaTeXunderbrace{(1-t)\c_1(y_0,\x)+\frac{t-w(\x)}{w(\x)}\Ezero\{\c_1(y_0,\x)\mid\x\}}_{\in\Lambda_0}
+ \LaTeXunderbrace{\frac{t-\pi(\bt^0)}{1-\pi(\bt^0)}\g_1(\x)}_{\in\Lambda^\perp},
\end{align*}
where $\Ezero(\c_1\mid\x,0)=\0$.
Taking $t=1$ and $t=0$ respectively of the above expression, it is straightforward to obtain
\begin{align*}
&\c_1(y_0,\x)=\left\{\frac{\pi(\bt^0)}{1-\pi(\bt^0)} - \frac{w(\x)}{1-w(\x)}\right\}\g_1(\x),
\\
\mbox{and}~~&\g_1(\x)=
\frac{\Ezero [\{1-\pi(\bt^0)\}^{-1}\dot\bpi(\bt^0)\mid \x]}{\Ezero [\{1-\pi(\bt^0)\}^{-1}\pi(\bt^0)\mid \x]}
=\frac{\A(\x)}{B(\x)}.
\end{align*}
Hence, the efficient score function is
\begin{align}
\S\eff (\w;\bt) 
=\frac{t-\pi(\bt)}{1-\pi(\bt)} \frac{\A(\x)}{B(\x)}.
\label{eq:Seff}
\end{align}
This implies that, once the nuisance functions have been estimated, the efficient estimator $\wh\bt\eff$ can be obtained by solving $\frac1n\sumi \S\eff(\w_i;\bt) = \0$, which is
\begin{eqnarray*}
\frac1n \sumi \frac{t_i-\pi_i(\bt)}{1-\pi_i(\bt)} 
\frac{\wh \A(\x_i)}{\wh B(\x_i)} = \0.
\end{eqnarray*}
Under suitable regularity conditions, we can show that, the estimator $\wh\bt\eff$ achieves the semiparametric efficiency bound. That is,
\begin{eqnarray}\label{eq:bteff}
\sqrt{n}(\wh\bt\eff-\bt^0) = n^{-1/2}\sumi \M^{-1}\S\eff(\w_i;\bt^0) + o_p(1) \xrightarrow{d} N \left(\0, \M^{-1}\right),
\end{eqnarray}
where $\M = \var(\S\eff) = \E(\S\eff\S\eff\trans) = -\E\left(\partial{\S\eff}/{\partial\bt\trans}\right) =
\E[B(\x)^{-1}\A(\x)\A(\x)\trans]$ and is assumed to be invertible.
It can be easily checked that the efficient score $\S\eff(\w;\bt)$ also satisfies the Neyman  orthogonality condition.
We will defer the proof of this result to Section \ref{sec:tech}.

Last but not least, when $\wh\bt\eff$ is obtained for the model $\pi(y_0,\u;\bt)=\expit(\theta_1+\theta_2 y_0+\bt_3\trans\u)$, one can easily develop a hypothesis testing procedure for
\begin{eqnarray*}
H_0: \theta_2=0 \mbox{ versus } H_1: \theta_2\neq 0,
\end{eqnarray*}
based on the corresponding Wald-type statistics, and evaluate whether the treatment assignment indeed depends on the potential untreated outcome.

\subsection{A Simpler Estimator for $\Delta$}\label{subsec:att.alter}

Though less efficient, a simpler
estimator of $\Delta$ exists.
According to the definition of $\Delta$, i.e., $\Delta = p^{-1}\{\E(T Y_1)-\E(T Y_0)\} = p^{-1}\left[\E(T Y_1)-\E\left\{(1-T)\frac{\pi(\bt)}{1-\pi(\bt)}Y_0\right\}\right]$,
an immediate estimator of $\Delta$ is
\begin{eqnarray*}
\wh\Delta\proone(\bt) = \frac{\frac1n\sumi \left\{t_i y_{1i} - (1-t_i)\frac{\pi_i(\bt)}{1-\pi_i(\bt)}y_{0i}\right\}}{\frac1n\sumi t_i},
\end{eqnarray*}
whose corresponding influence function is
\begin{eqnarray}\label{eq:IF.pro1}
\phi\proone (\w;\bt,\Delta) = p^{-1}\left\{t y_1 - (1-t)\frac{\pi(\bt)}{1-\pi(\bt)}y_0 - t\Delta\right\}.
\end{eqnarray}
It can be easily verified that,
\begin{equation}\label{eq:delta_alt}
	\begin{aligned}
	\sqrt{n}\{\wh\Delta\proone(\bt^0)-\Delta^0\} =&~ n^{-1/2}\sumi \phi\proone (\w_i;\bt^0,\Delta^0) + o_p(1) \xrightarrow{d} N \left[0, \E\{\phi\proone^2 (\w;\bt^0,\Delta^0)\}\right],\\
	\sqrt{n}\{\wh\Delta\proone(\wh\bt\eff)-\Delta^0\} =&~ n^{-1/2}\sumi \left\{\phi\proone (\w_i;\bt^0,\Delta^0)-p^{-1}\Q \M^{-1}\S\eff(\w_i;\bt^0)\right\} + o_p(1), 
	\end{aligned}
\end{equation}
under standard regularity conditions (e.g.\ those in Theorems 2.6 and 3.4 of \cite{NeweyMcFadden1994}).

Unfortunately, neither the estimator $\wh\Delta\proone(\bt^0)$ nor the estimator $\wh\Delta\proone(\wh\bt\eff)$ is efficient.
This can be seen from the fact that $\phi\proone (\w;\bt^0,\Delta^0)\notin \Lambda$.
Indeed, we have the following projection result
\begin{eqnarray}\label{eq:decom}
\phi\proone(\w;\bt^0,\Delta^0) = \LaTeXunderbrace{\phi(\w;\bt^0,\Delta^0)}_{\Pi\{\phi\proone(\w;\bt^0,\Delta^0)\mid \Lambda\}} + \LaTeXunderbrace{\frac{t-\pi(\bt^0)}{1-\pi(\bt^0)}p^{-1}B(\x)^{-1}V(\x)}_{\Pi\{\phi\proone(\w;\bt^0,\Delta^0)\mid \Lambda^\perp\}}.
\end{eqnarray}
Immediately, we have
\begin{eqnarray*}
\E\{\phi\proone^2(\w;\bt^0,\Delta^0)\} = \E\{\phi^2(\w;\bt^0,\Delta^0)\} + \E\left[\Pi\{\phi\proone(\w;\bt^0,\Delta^0)\mid \Lambda^\perp\}^2\right] > \E\{\phi^2(\w;\bt^0,\Delta^0)\}.
\end{eqnarray*}
Thus, under regularity conditions that justify the first-order linear representations in \eqref{eq:deltaprotwo} and \eqref{eq:delta_alt}, we have the following comparison.
\begin{Lem}\label{lem:comparison}
Under Assumptions~\ref{assump:consistency.para}--\ref{assump:quality.nuisance.est} presented later,
	when $\bt^0$ is known, the estimator $\wh\Delta\proone(\bt^0)$ is less efficient than $\wh\Delta\eff(\bt^0)$.
\end{Lem}
To make the comparison result complete, we also summarize the following result when $\bt$ is estimated by the efficient estimator $\wh\bt\eff$. Its efficiency comparison is guaranteed since $\wh\Delta\eff(\wh\bt\eff)$ is semiparametrically efficient among all RAL estimators.
\begin{Lem}\label{lem:comparison2}
Under Assumptions~\ref{assump:consistency.para}--\ref{assump:quality.nuisance.est} presented later,
	the estimator $\wh\Delta\proone(\wh\bt\eff)$ is less efficient than $\wh\Delta\eff(\wh\bt\eff)$.
\end{Lem}

\section{Implementation and Asymptotic Properties}\label{sec:tech}

In this section, we provide implementation details for using machine learning tools to construct the efficient estimators $\wh\bt\eff$ and $\wh\Delta\eff(\wh\bt\eff)$
and study their asymptotic properties.
As mentioned in Section~\ref{sec:est.delta}, jointly obtaining $\wh\bt\eff$ and $\wh\Delta\eff(\wh\bt\eff)$ is equivalently to solving
\begin{eqnarray*}
\begin{cases}
	&n^{-1}\sumi \S\eff(\w_i;\bt)=\0
	\\
	&n^{-1}\sumi \phi(\w_i;\bt,\Delta)=0,
\end{cases}
\end{eqnarray*}
which involves the nuisance components \label{rev:nuisance_def}
\begin{align*}
\boeta = \left\{
	\begin{array}{c}
		%  \Eone(Y_1\mid \x,1),~
		%  \Ezero\left[\frac{1}{1-\pi(\bt^0)}\mid \x,0\right], ~
		% \Ezero\left[\frac{\pi(\bt^0)}{\{1-\pi(\bt^0)\}^2}\mid \x,0\right],~\\
		% \Ezero\left[\frac{\pi(\bt^0)^2Y_0}{\{1-\pi(\bt^0)\}^2}\mid \x,0\right],~
		% \Ezero\left[\frac{\dot\bpi(\bt^0)}{\{1-\pi(\bt^0)\}^2}\mid \x,0\right]
		 \Eone(Y_1\mid \x,1),~
		 \Ezero[\{1-\pi(\bt^0)\}^{-1}\mid \x,0], ~
		 \Ezero[\{1-\pi(\bt^0)\}^{-2}\pi(\bt^0)\mid \x,0],\\
		 \Ezero[Y_0\{1-\pi(\bt^0)\}^{-2}\pi(\bt^0)^2\mid \x,0],~
		 \Ezero[\{1-\pi(\bt^0)\}^{-2}\dot\bpi(\bt^0)\mid \x,0]
	\end{array}
\right\}\in\calT,
\end{align*}
where $\calT$ denotes the nuisance function space. 
As a special case, when the potential outcomes $(Y_0,Y_1)$ are binary, the evaluation of $\boeta$ reduces to evaluating $\boeta=\{f_0(1\mid\x,0),f_1(1\mid\x,1)\}=\{\Ezero(Y_0\mid \x,0),~\Eone(Y_1\mid \x,1)\}$.
To emphasize the role of the nuisance components in the technical development, we put both parameters together as
$\bpsi=\{\bt\trans,\Delta\}\trans$, 
and denote
\begin{eqnarray*}
\G(\W;\bpsi,\boeta)=\{\S\eff(\W;\bpsi,\boeta)\trans,\phi(\W;\bpsi,\boeta)\}\trans.
\end{eqnarray*}
In Appendix~\ref{sec:proof.theta.ATT2}, we show in Lemma~\ref{lem:Neyman_orth} that the score function $\G$ satisfies the Neyman orthogonality condition \citep{chernozhukov2018double}, in the sense that the pathwise (Gateaux) derivative
\begin{eqnarray*}
\partial_r \E\left[ \G\{\W;\bpsi^0,\boeta^0+r(\boeta-\boeta^0)\}\right] |_{r=0}=\0, ~~ \mbox{for any }\boeta\in\calT.
\end{eqnarray*}
This ensures that small perturbations in the estimation of nuisance functions have only second-order effects on the target parameter. \label{resp:gateaux}
In addition, we apply the sample splitting technique \citep{bickel1982adaptive, schick1986asymptotically, chernozhukov2018double}, widely used to eliminate the first-order overfitting bias induced by the nuisance estimation.
Let $K$ be a fixed positive integer that does not change with the sample size $n$. We take $K$-fold random partition $(I_k)_{k=1}^K$ of the observational indices, $[n]=\{1,2,\ldots,n\}$, such that the size of each fold, $I_k$, is $\vert I_k \vert = n^*=n/K$.
For each $k\in[K]=\{1,2,\ldots,K\}$, let us define $I_k^c=[n] \backslash I_k$ and construct a machine learning (ML) estimator $\wh\boeta^{[-k]} = \wh\boeta\{(\W_i)_{i\in I_k^c}\}\in\calT_{n}$, where $\calT_{n}\subset\calT$ is a nuisance realization set that is a suitably shrinking neighborhood of $\bpsi$. 
Then we obtain $\wh\bpsi$ by solving the cross-fitting estimating equation
\begin{eqnarray}\label{eq:cross.seff.ATT2}
\frac1n \sumK\sum_{i\in I_k} \G(\W_i; \bpsi,\wh\boeta^{[-k]})=\0.
\end{eqnarray}

\begin{Rem}\label{rem:implementation}
	Note that all nuisance functions in $\boeta$ take the form $\E_1(\cdot \mid \x,1)$ and $\E_0(\cdot \mid \x,0)$. Therefore, to obtain $\wh\boeta$, we can rely on off-the-shelf machine learning methods such as XGBoost or random forests to fit regression-style learners for these conditional means. Because some nuisance functions depend on the true parameter value $\bt^0$, we can simply replace it with a consistent, not necessarily efficient, estimator $\wh\bt$ solving \eqref{eq:solvethetausingg} when constructing the regression target. In the special case where $(Y_0, Y_1)$ are binary, this training step simplifies further: it suffices to learn $\E_0(Y_0 \mid \x,0)$ and $\E_1(Y_1 \mid \x,1)$.
\end{Rem}

Let $o(\alpha_n)$, $O(\alpha_n)$, $\omega(\alpha_n)$, $\Omega(\alpha_n)$ and $\Theta(\alpha_n)$ represent sequences that growing at a smaller, equal or smaller, larger, equal or larger and equal rate of $\alpha_n$, respectively.
Further, for any square-integrable vector-valued function $\f(\cdot)=(f_1(\cdot),\dots,f_p(\cdot))\trans$, random variable $\W$, and probability measure $P$, define $\Vert \f(\cdot)\Vert_{P,2} = \Vert \f(\W)\Vert_{P,2}= \{\int\sum_{j=1}^{p}\vert f_j(\w)\vert^2 \dd P(\w)\}^{1/2}$. We use $\partial_\x f$ to denote $(\partial/\partial\x) f$ and $\partial_\x f(\x_0)$ for $(\partial/\partial\x) f(\x)\mid_{\x=\x_0}$.

The asymptotic properties for the estimators $\wh\bt\eff$ and $\wh\Delta\eff(\wh\bt\eff)$ mainly follow Theorem 3.3 of \cite{chernozhukov2018double}, where they presented the general results of the debiased ML estimators for the nonlinear Neyman orthogonal score.
Nevertheless, we first provide the necessary assumptions in our own setting.

\begin{assump}\label{assump:consistency.para}
	%The influence function $\G(\W;\bpsi^0,\boeta)$ obeys Neyman orthogonality and $\E\{\G(\W;\bpsi^0,\boeta^0)\}=\0$.
	There exists $\delta_n=\Omega(n^{-1/2}\log n)$ such that parameter $\bt$ belongs to a compact set $\calB_1\in \mR^{d}$, which contains a ball of radius $c_1\delta_n$ centered at the true parameter $\bt^0$, where $c_1$ is some finite positive constant. Also, parameter $\Delta$ belongs to compact sets $\calB_2\in \mR$ and $(\Delta^0-c_1\delta_n,\Delta^0+c_1\delta_n)\subseteq\calB_2$. In other words, $\calB\in\mR^{d+1}$ contains a ball of radius $\Omega(n^{-1/2}\log n)$ centered at $\bpsi^0$.
	The treatment assignment satisfies $c<\pi(\bt)<1-c$, where $0<c\leq 1/2$. Further, $\pi(\bt)$ is thrice differentiable with respect to $\bt$ and the derivatives are continuous and bounded.
	In addition, $Y_t=\Theta(1)$ and $\E_t(Y_t^2\mid \x,t)=\Theta(1)$ for $t=0,1$.
	%$\vert Y_t \vert \leq c_2$ and $\Vert \E(Y_t^2\mid \X,t)\Vert_\infty \leq c_3$ for $t=0,1$, where $c_2$ and $c_3$ are finite positive constants.
\end{assump}

\begin{assump}\label{assump:quality.nuisance.est}
	The nuisance estimates achieve an $o_p(n^{-1/4})$ convergence rate, i.e.,
	$\Vert \wh \boeta^{[-k]}  - \boeta\Vert_{P,2} = o_p(n^{-1/4}),$
	for $ k\in\{1,2,\ldots,K\}$.
\end{assump}

Assumption~\ref{assump:consistency.para} is on regularity conditions.
Assumption~\ref{assump:quality.nuisance.est} is about the quality of ML nuisance estimates, which ensures that the second-order residuals of the DML estimator are negligible. This rate is achievable for many ML methods under suitable structural assumptions on the nuisance functions; see, for example, \cite{BickelRitovTsybakov2009} for $\ell_1$-penalized methods in sparse models, \cite{WagerAthey2018} for regression trees and random forests, and \cite{FarrellLiangMisra2021} for neural networks.
Now we establish the large sample property of the estimator $\wh\bpsi$ in the following theorem.

\begin{Th}\label{th:theta.ATT2}
	Under the Assumptions~\ref{assump:consistency.para}--\ref{assump:quality.nuisance.est}, we have
	\begin{eqnarray*}
	&&	\sqrt{n}(\wh\bpsi-\bpsi^0) = n^{-1/2}\sumi\J^{-1}\G(\W_i;\bpsi^0,\boeta^0) + o_p(1) \xrightarrow{d} N(\0, \V), \mbox{ where }\\
	&& \J = \partial_{\bpsi\trans} \E\{\G(\W;\bpsi,\boeta^0)\} \mid_{\bpsi=\bpsi^0} = \left(\begin{array}{cc}
		-\M & \0\\
		p^{-1}(\D-\Q)\trans & -1
	\end{array}\right), \mbox{ and }\\
	&& \V = \left(\begin{array}{cc}
		\M^{-1} & p^{-1}\M^{-1}(\D-\Q)\\
		p^{-1}(\D-\Q)\trans\M^{-1} & p^{-2}(\D-\Q)\trans\M^{-1}(\D-\Q) + \E\{\phi^2(\w;\bt^0,\Delta^0)\}
	\end{array}\right),
	\end{eqnarray*}
	and the estimator $\wh\bpsi$ achieves the semiparametric efficiency bound.
\end{Th}
Note that here $\wh\bpsi=(\wh\bt\eff\trans, \wh\Delta\eff(\wh\bt\eff))\trans$; hence, the results in (\ref{eq:bteff}) and (\ref{eq:deltaprotwo}) follow immediately.
The proof of Theorem~\ref{th:theta.ATT2} is provided in Appendix~\ref{sec:proof.theta.ATT2}.
In addition, the asymptotic variance $\V$ does not involve any unknown functions beyond $\boeta$ and therefore can be estimated using a simple plug-in estimator; see Theorem 3.2 of \cite{chernozhukov2018double} for a general variance estimation formula and a proof of consistency. \label{rev:var_est}

\section{Simulation Studies}\label{sec:simu}

In this section, we present some numerical results.
We implement the estimates of $\bt$ and $\Delta$ introduced in Section~\ref{sec:est.delta}.
We also compare the finite-sample performance of estimators of $\Delta$ with two naive estimators that assume strong ignorability \citep{hahn1998role}. The first naive estimator is
\begin{eqnarray*}
\wh\Delta\nvone = \frac{n^{-1}\sumi \left\{t_iy_{1i}-(1-t_i)\frac{\wh{w}(\x_i)}{1-\wh{w}(\x_i)}y_{0i}\right\}}{n^{-1}\sumi t_i},
\end{eqnarray*}
whose corresponding influence function is $p^{-1} \left\{T Y_1 - (1-T)\frac{w(\x)}{1-w(\x)}Y_0 - T \Delta\right\}$.
The second naive estimator is
\begin{eqnarray*}
\wh\Delta\nvtwo = \frac{n^{-1}\sumi \left\{t_iy_{1i}-(1-t_i)\frac{\wh{w}(\x_i)}{1-\wh{w}(\x_i)}y_{0i} - \frac{t_i-\wh{w}(\x_i)}{1-\wh{w}(\x_i)}\wh\E_0(Y_{0i}\mid\x_i)\right\}}{n^{-1}\sumi t_i},
\end{eqnarray*}
which is based on the EIF 
$p^{-1}\left[T\{Y_1 - \Ezero(Y_0\mid \X) - \Delta\} - (1-T)\frac{w(\x)}{1-w(\x)}\{Y_0 - \Ezero(Y_0\mid \X)\}\right]$.
To implement the estimator $\wh\Delta\nvone$, one needs to correctly specify the propensity score model $w(\x)$.
The estimator $\wh\Delta\nvtwo$ enjoys the so-called double robustness property in that one only needs to correctly specify either $w(\x)$ or $\Ezero(Y_0\mid \x)$, both of which are essentially conditional expectations.

In the simulations, we consider a random sample with $600$ observations with the following data generating process.
We first generate
$\X\sim N_2(\0,\I_2)$.
For the potential outcomes,
$Y_1$ follows a Bernoulli distribution with success probability $\expit(x_1)$, whereas $Y_0$ follows a Bernoulli distribution with probability of success $\expit(x_2)$.
We generate a binary treatment $T$ according to
\begin{eqnarray}\label{eq:trtmodelinsim}
\pr(T=1\mid y_0,\x)=\pr(T=1\mid y_0,x_1)=\expit(\theta_1 + \theta_2 y_0 + \theta_3 x_1),
\end{eqnarray}
where $\bt=(0.3,-0.3,-0.25)\trans$.
It can be computed that the true value of ATT is around 0.012.

We solve (\ref{eq:Seff}) to obtain the estimator of $\bt$, and solve (\ref{eq:IF.pro1}) and (\ref{eq:IF.pro2}) to obtain estimates for $\Delta$ respectively.
We also fit random forest (rforest, via R package \verb"ranger") and XG-boost (xgboost, via R package \verb"xgboost") models using all simulated data for the nuisance models $\E_0(Y_0\mid \X)$ and $\E_1(Y_1\mid \X)$.
The results based on 500 simulation replications are reported in Table~\ref{tab:ATT_comparison_simu2}.
In this table, $\wh\sd_p$ is the estimated standard deviation (SD) based on 200 perturbation samples \citep{jin2001simple}.
The conclusions are quite clear.
For $\bt$, we can see that its estimate yields very small bias, and the estimated SD $\wh\sd_p$ is close to the empirical SD.
For $\Delta$, the naive estimators are biased and should not be used in practice;
on the contrary, the proposed estimators all yield almost zero bias.
Regarding different machine learning models for estimating the nuisances, we note that the performances of the corresponding estimators of $\bt$ and $\Delta$, albeit slightly vary, are overall very similar and reach the same conclusion.

\begin{table}[!htbp]
	\centering
	\caption{Simulation: results for $\bt$ and ATT. The notation $\sd$, $\wh\sd_p$ and CP refer to the empirical standard deviation, the standard deviation based on 200 perturbation samples, and $95\%$ coverage probability, respectively.
		The notation $\wh\theta\effone$, $\wh\theta\efftwo$ and $\wh\theta\effthree$ refer to the three coordinates of the parameter estimate $\wh\bt\eff$, respectively.
	}
	\begin{tabular}{l | clccrrrr}
		\hline
		&&$w(\X)$&$\Ezero(Y_0\mid\X)$& $\Eone(Y_1\mid\X)$ & $\Bias$ & $\sd$ &  $\wh\sd_p$&CP\\
		\hline
		\multirow{6}{*}{Naive}
		&\multirow{3}{*}{ $\wh\Delta\nvone$} &$\logistic$&-&-&-0.063&0.044&0.041&0.636\\
		&&$\rforest$&-&-&0.228&0.035&0.038&0.000\\
		&&$\xgboost$&-&-&-0.004&0.047&0.043&0.946\\\cline{2-9}
		&\multirow{3}{*}{$\wh\Delta\nvtwo$}
		%&$\logistic$&correct&-&-0.0627&0.0441&0.0412&0.650\\
		%&&$\rforest$&correct&-&-0.0439&0.0348&0.0306&0.676\\
		%&&$\xgboost$&correct&-&-0.0591&0.0448&0.0397&0.640\\
		%&
		&$\logistic$& $\logistic$&  -&-0.063&0.044&0.041&0.646\\
		%&&$\logistic$& $\rforest$ &-&-0.2353&0.0169 &0.0000&0.0163\\
		&&$\rforest$&$\rforest$&-&-0.038&0.039&0.038&0.818\\
		%&&$\logistic$&$\xgboost$&-&-0.2429&0.0178&0.0000&0.0179\\
		&&$\xgboost$&$\xgboost$&-&-0.055&0.043&0.040&0.698\\\cline{2-9}
		\hline
		&&$\pi(Y_0,\U;\bt)$&$\Ezero(Y_0\mid\X)$& $\Eone(Y_1\mid\X)$ & $\Bias$ & $\sd$& $\wh\sd_p$&CP\\
		\hline
		\multirow{15}{*}{Proposed}
		%        &\multirow{1}{*}{$\wh\theta\effone$}&$\logistic$&correct&-&-0.0163&0.2260&0.2577&0.962\\
		%        &\multirow{1}{*}{$\wh\theta\efftwo$}&$\logistic$&correct&-&0.0004&0.4508 &0.5879&0.980\\
		%        &\multirow{1}{*}{$\wh\theta\effthree$}&$\logistic$&correct&-&-0.0079&0.0882&0.0899&0.962\\
		%        & $\wh\Delta\proone(\wh\bt\eff)$ &$\logistic$&correct&-&-0.0030&0.1063&0.1077&0.936\\
		%        & $\wh\Delta\eff(\wh\bt\eff)$ &$\logistic$&correct&correct&-0.0026&0.1061&0.1091&0.952\\
		%        \cline{2-9}
		&\multirow{1}{*}{$\wh\theta\effone$}&\multirow{5}{*}{model (\ref{eq:trtmodelinsim})}&$\logistic$&-&-0.017&0.226&0.253&0.968\\
		&\multirow{1}{*}{$\wh\theta\efftwo$}&&$\logistic$&-&0.001&0.452 &0.442&0.986\\
		&\multirow{1}{*}{$\wh\theta\effthree$}&&$\logistic$&-&-0.008&0.088&0.092&0.964\\
		& $\wh\Delta\proone(\wh\bt\eff)$ &&$\logistic$ &-&-0.003&0.106&0.104&0.936\\
		& $\wh\Delta\eff(\wh\bt\eff)$ &&$\logistic$ &$\logistic$&-0.003&0.106&0.111&0.946\\
		\cline{2-9}
		&\multirow{1}{*}{$\wh\theta\effone$}&\multirow{5}{*}{model (\ref{eq:trtmodelinsim})}&$\rforest$&-&0.002&0.136&0.144&0.960\\
		&\multirow{1}{*}{$\wh\theta\efftwo$}&&$\rforest$&-&0.001&0.222 &0.239&0.964\\
		&\multirow{1}{*}{$\wh\theta\effthree$}&&$\rforest$&-&-0.008&0.088&0.089&0.958\\
		& $\wh\Delta\proone(\wh\bt\eff)$ && $\rforest$ &-&-0.001&0.052&0.052&0.944\\
		& $\wh\Delta\eff(\wh\bt\eff)$ && $\rforest$ &$\rforest$&-0.001&0.051&0.052&0.956\\
		\cline{2-9}
		&\multirow{1}{*}{$\wh\theta\effone$}&\multirow{5}{*}{model (\ref{eq:trtmodelinsim})}&$\xgboost$&-&-0.008&0.179&0.172&0.944\\
		&\multirow{1}{*}{$\wh\theta\efftwo$}&&$\xgboost$&-&0.007&0.328&0.312&0.946\\
		&\multirow{1}{*}{$\wh\theta\effthree$}&&$\xgboost$&-&-0.0067&0.088&0.089&0.956\\
		& $\wh\Delta\proone(\wh\bt\eff)$  &&$\xgboost$&-&-0.005&0.078&0.071&0.924\\
		& $\wh\Delta\eff(\wh\bt\eff)$ &&$\xgboost$&$\xgboost$&-0.004&0.078&0.082&0.932\\
		\hline
	\end{tabular}
	\label{tab:ATT_comparison_simu2}
\end{table}

To further investigate the efficiency gain of the estimator $\wh\Delta\eff(\bt)$ over $\wh\Delta\proone(\bt)$ for a particular $\bt$, we use the following four choices of $\bt$: the efficient estimate $\wh\bt\eff$, the true value $\bt^0$, $\wh\bt\arbone$ corresponding to the solution of (\ref{eq:solvethetausingg}) with
$\g(\x)=\{w(\x),w(\x)x_1,w(\x)x_1^2\}\trans$, and $\wh\bt\arbtwo$ corresponding to the solution of (\ref{eq:solvethetausingg}) with $\g(\x)=\{w(\x),\expit(\theta_1+\theta_2+\theta_3 x_1)\Ezero(Y_0\mid\x),w(\x)x_1\}\trans$.
The results are summarized in Table~\ref{tab:ATT_comparison_simu2_RE}.
When using the efficient estimate $\wh\bt\eff$, the estimator $\wh\Delta\eff(\wh\bt\eff)$ is still advantageous, but the efficiency gain is quite minimum.
On the other hand, when using the true value $\bt^0$, the estimator $\wh\Delta\eff(\bt^0)$ is a lot more efficient compared to $\wh\Delta\proone(\bt^0)$.
This indicates, because of the projection result (\ref{eq:decom}), the estimator based on the influence function $\phi(\w;\bt,\Delta)$ does improve the estimation efficiency compared to the one using $\phi\proone(\w;\bt,\Delta)$, but this gain can be largely mitigated by efficiently estimating $\wh\bt\eff$.
In general, when we arbitrarily choose $\g(\x)$ functions in (\ref{eq:solvethetausingg}) which lead to estimators $\wh\bt\arbone$ and $\wh\bt\arbtwo$ in Table~\ref{tab:ATT_comparison_simu2_RE}, the efficiency gain could be somewhere between 15\% and 70\%.

\begin{table}[!htbp]
	\centering
	\caption{
		Simulation: results for the relative efficiency (RE) of $\wh\Delta\eff(\bt)$ over $\wh\Delta\proone(\bt)$ with different estimates of $\bt$. Here RE is defined as the mean squared error (MSE) of the estimator $\wh\Delta\proone(\bt)$ divided by that of $\wh\Delta\eff(\bt)$. RE$>$1 indicates that $\wh\Delta\eff(\bt)$ is more efficient. The estimator $\wh\bt\arbone$ corresponds to the choice of $\g(\x)=\{w(\x),w(\x)x_1,w(\x)x_1^2\}\trans$ and the estimator $\wh\bt\arbtwo$ corresponds to the choice of $\g(\x)=\{w(\x),\expit(\theta_1+\theta_2+\theta_3 x_1)\Ezero(Y_0\mid\x),w(\x)x_1\}\trans$, in (\ref{eq:solvethetausingg}).
	}
	{\small{
			\begin{tabular}{clccrrrr}
				\hline
				&$\pi(Y_0,\U;\bt)$&$\Ezero(Y_0\mid\X)$& $\Eone(Y_1\mid\X)$ & $\Bias$ & $\sd$& MSE&RE\\
				\hline
				$\wh\Delta\proone(\wh\bt\eff)$ & \multirow{2}{*}{model (\ref{eq:trtmodelinsim})}&logistic &-&-0.003&0.106&0.011&\\
				$\wh\Delta\eff(\wh\bt\eff)$ &&logistic &logistic&-0.003&0.106& 0.011\\
				\cline{2-8}
				&&&&&&&1.004\\
				\hline
				$\wh\Delta\proone(\wh\bt\eff)$ & \multirow{2}{*}{model (\ref{eq:trtmodelinsim})}&rforest &-&-0.001&0.052&0.003&\\
				$\wh\Delta\eff(\wh\bt\eff)$ &&rforest &rforest&-0.001&0.051& 0.003\\
				\cline{2-8}
				&&&&&&&1.059\\
				\hline
				$\wh\Delta\proone(\wh\bt\eff)$ & \multirow{2}{*}{model (\ref{eq:trtmodelinsim})}&xgboost &-&-0.005&0.078&0.006&\\
				$\wh\Delta\eff(\wh\bt\eff)$ &&xgboost &xgboost&-0.004&0.078& 0.006\\
				\cline{2-8}
				&&&&&&&1.014\\
				\hline\hline
				$\wh\Delta\proone(\bt^0)$ & \multirow{2}{*}{model (\ref{eq:trtmodelinsim})}&logistic &-&-0.000&0.060&0.004&\\
				$\wh\Delta\eff(\bt^0)$ &&logistic &logistic&-0.000&0.043& 0.002\\
				\cline{2-8}
				&&&&&&&1.908\\
				\hline
				$\wh\Delta\proone(\bt^0)$ & \multirow{2}{*}{model (\ref{eq:trtmodelinsim})}&rforest &-&-0.000&0.060&0.004&\\
				$\wh\Delta\eff(\bt^0)$ &&rforest &rforest&-0.000&0.037& 0.001\\
				\cline{2-8}
				&&&&&&&2.658\\
				\hline
				$\wh\Delta\proone(\bt^0)$ & \multirow{2}{*}{model (\ref{eq:trtmodelinsim})}&xgboost &-&-0.000&0.060&0.004&\\
				$\wh\Delta\eff(\bt^0)$ &&xgboost &xgboost&-0.000&0.042& 0.002\\
				\cline{2-8}
				&&&&&&&2.040\\
				\hline\hline
				$\wh\Delta\proone(\wh\bt\arbone)$ & \multirow{2}{*}{model (\ref{eq:trtmodelinsim})}&logistic &-&-0.019&0.152&0.023&\\
				$\wh\Delta\eff(\wh\bt\arbone)$ &&logistic &logistic&-0.017&0.141& 0.020\\
				\cline{2-8}
				&&&&&&&1.171\\
				\hline
				$\wh\Delta\proone(\wh\bt\arbone)$ & \multirow{2}{*}{model (\ref{eq:trtmodelinsim})}&rforest &-&-0.014&0.145&0.021&\\
				$\wh\Delta\eff(\wh\bt\arbone)$ &&rforest &rforest&-0.011&0.123&0.015\\
				\cline{2-8}
				&&&&&&&1.407\\
				\hline
				$\wh\Delta\proone(\wh\bt\arbone)$ & \multirow{2}{*}{model (\ref{eq:trtmodelinsim})}&xgboost &-&-0.019&0.147&0.022&\\
				$\wh\Delta\eff(\wh\bt\arbone)$ &&xgboost &xgboost&-0.016&0.131&0.017\\
				\cline{2-8}
				&&&&&&&1.267\\
				\hline\hline
				$\wh\Delta\proone(\wh\bt\arbtwo)$ & \multirow{2}{*}{model (\ref{eq:trtmodelinsim})}&logistic &-&-0.002&0.047&0.002&\\
				$\wh\Delta\eff(\wh\bt\arbtwo)$ &&logistic &logistic&-0.002&0.044& 0.002\\
				\cline{2-8}
				&&&&&&&1.146\\
				\hline
				$\wh\Delta\proone(\wh\bt\arbtwo)$ & \multirow{2}{*}{model (\ref{eq:trtmodelinsim})}&rforest &-&-0.002&0.048&0.002&\\
				$\wh\Delta\eff(\wh\bt\arbtwo)$ &&rforest &rforest&-0.001&0.037& 0.001\\
				\cline{2-8}
				&&&&&&&1.672\\
				\hline
				$\wh\Delta\proone(\wh\bt\arbtwo)$ & \multirow{2}{*}{model (\ref{eq:trtmodelinsim})}&xgboost &-&-0.002&0.047&0.002&\\
				$\wh\Delta\eff(\wh\bt\arbtwo)$ &&xgboost &xgboost&-0.002&0.043& 0.002\\
				\cline{2-8}
				&&&&&&&1.238\\
				\hline
			\end{tabular}
	}}
	\label{tab:ATT_comparison_simu2_RE}
\end{table}

\section{Application to NCDB}\label{sec:real_data}

In this section, we analyze the National Cancer Database (NCDB), a clinical oncology database sourced from hospital registry data that are collected in more than 1,500 Commission on Cancer (CoC)-accredited facilities.

In our application, we focus on 7,233 stage T2 lung cancer patients out of whom 6,124 were treated with stereotactic body radiation therapy (SBRT, $T=1$) and 1,109 were treated with lobectomy surgery ($T=0$).
The estimand of interest is the ATT on 3-year survival rates; that is, the binary outcome $Y$ equals 1 when a patient was recorded as alive at least 3 years after being diagnosed with lung cancer and 0 otherwise.
There are five $\X$ variables in this database. Out of them, three are continuous---age, size of tumor (logarithmic scale), year of diagnosis, and the other two are categorical---income level (binary, income$<$\$48,000 versus income$\geq$\$48,000), and insurance level (binary, private versus non-private). An important prognostic variable, Charlson-Deyo comorbidity index score (CDCI), is also recorded in the data set.
We conduct our analyses stratified on CDCI=0, and CDCI$>$0.
We standardize all the continuous variables throughout the analysis.

In our analysis, we consider two naive estimators for $\Delta$: $\wh\Delta\nvone$ and $\wh\Delta\nvtwo$ which involve both $w(\x)$ and $\Ezero(Y_0\mid \x)$.
We consider three methods to fit each of them: logistic regression (logistic), random forest (rforest), and XG-boost (xgboost).
The logistic regression model for $w(\x)$ is comparable to the modeling of the propensity score in our proposed method.
Also note that the efficient influence function under strong ignorability was first derived in \cite{hahn1998role}, and can be regarded as a special case of the efficient influence function derived in this paper.

The strong ignorability assumption assumes that the choice of treatment does not depend on potential outcomes $Y_1$ or $Y_0$ but on all other covariates.
This assumption might not be realistic for applications.
In our application, when the physician assigned the patients to either radiation therapy or surgery, their decision should not depend on the patients' income or insurance.
On the other hand, the decision might potentially depend on $Y_1$ and $Y_0$. As discussed in the introduction section, $\Z$ consists of income and insurance levels.
Therefore, to apply our method, we assume
\begin{eqnarray}\label{eq:assinapp}
\pr(T=1\mid Y_0, \X) = \pi(y_0,\u;\bt) = \expit(\theta_1 + \theta_2y_0+\xi\trans\u),
\end{eqnarray}
where $\bt=(\theta_1, \theta_2, \xi\trans)\trans$ and   $\u$ includes year of diagnosis, age, and size of tumor.
We first implement the proposed method $\wh\Delta\proone$, relying on $\wh\bt\eff$ that is obtained from solving (\ref{eq:Seff}) which only needs a model for $\Ezero(\cdot\mid \x,0)$.
We then implement the proposed method $\wh\Delta\eff$ according to the efficient influence function of $\Delta$.
The implementation of $\wh\Delta\eff$ relies on $\wh\bt\eff$, $\Ezero(\cdot\mid \x, 0)$ and $\Eone(\cdot\mid \x, 1)$.
For comparison, we also fit logistic regression (logistic), random forest (rforest), and XG-boost (xgboost) methods to estimate $\Ezero(\cdot\mid \x,0)$ or $\Eone(\cdot\mid \x,1)$.

The results of $\wh\bt\eff$ for estimating $\bt$ in model (\ref{eq:assinapp}) are contained in Table~\ref{tab:propensity_comparison_cdcc}.
Similar to our simulation studies, different machine learning methods all reach the same conclusion.
It is clear that the effect of $Y_0$ is statistically significant, no matter which model is used for estimating $\E_0(Y_0\mid \x,0)$, and regardless of CDCI=0 or CDCI$>$0.
This indicates, in this application, the assumption (\ref{eq:assinapp}) that the treatment assignment mechanism does depend on the untreated potential outcome $Y_0$ is sensible.
This analytical result also echoes our earlier arguments that are solely based on clinical experience.

\begin{table}[h]
	\centering
	\caption{
		Real Data Application: estimation results $\wh\bt\eff$ by fitting logistic regression models for the propensity score by solving (\ref{eq:Seff}), when the data are stratified based on CDCI=0 and CDCI$>$0. For comparison purpose, the model $\Ezero(Y_0\mid \x,0)$ is fitted using logistic regression, random forest and XG-boost, respectively.
	}
	\begin{tabular}{l|lrrrrr}
		\hline
		$\Ezero(Y_0\mid\x,0)$&&Intercept & $Y_0$ & Year of diagnosis & Age & Size of tumor \\
		\hline
		\multicolumn{7}{c}{CDCI$=$0}\\
		\hline
		\multirow{3}{*}{$\logistic$}
		& ${\rm Est}$ &-1.369&-0.848&0.595&0.568&-0.053\\
		& $\wh \sd_p$ &0.054&0.006&0.049&0.045&0.047\\
		& p-value &0.000&0.000&0.000&0.000&0.262\\
		\hline
		\multirow{3}{*}{$\rforest$}
		& ${\rm Est}$ &-1.396&-0.723&0.593&0.570&-0.051\\
		& $\wh \sd_p$ &0.081&0.146&0.049&0.042&0.044\\
		& p-value &0.000&0.000&0.000&0.000&0.243\\
		\hline
		\multirow{3}{*}{$\xgboost$}
		& ${\rm Est}$ &-1.366&-0.840&0.594&0.567&-0.052\\
		& $\wh \sd_p$ &0.057&0.012&0.051&0.046&0.049\\
		& p-value &0.000&0.000&0.000&0.000&0.284\\
		\hline\hline
		\multicolumn{7}{c}{CDCI$>$0}\\
		\hline
		\multirow{3}{*}{$\logistic$}
		& ${\rm Est}$ &-1.830&-0.876&0.833&0.476&-0.147\\
		& $\wh \sd_p$ &0.067&0.005&0.070&0.053&0.064\\
		& p-value &0.000&0.000&0.000&0.000&0.021\\
		\hline
		\multirow{3}{*}{$\rforest$}
		& ${\rm Est}$ & -1.846&-0.765&0.823&0.477&-0.144\\
		& $\wh \sd_p$ &0.066&0.064&0.068&0.055&0.059\\
		& p-value &0.000&0.000&0.000&0.000&0.016\\
		\hline
		\multirow{3}{*}{$\xgboost$}
		& ${\rm Est}$ &-1.826&-0.873&0.830&0.477&-0.148\\
		& $\wh \sd_p$ &0.067&0.006&0.066&0.055&0.063\\
		& p-value &0.000&0.000&0.000&0.000&0.019\\
		\hline
	\end{tabular}
	\label{tab:propensity_comparison_cdcc}
\end{table}

Different ATT estimation results using naive methods and proposed methods are summarized in Table~\ref{tab:comparison_ATT_cdcc}.
From Table~\ref{tab:comparison_ATT_cdcc}, among treated patients with CDCI=0, the naive methods suggest that the 3-year survival rate would have been 5\%--20\% higher had these patients received $T=0$ instead. In contrast, the proposed methods suggest little to no difference in 3-year survival between the two treatments for this subgroup.
Among treated patients with CDCI>0, the naive methods suggest a larger increase of 15\%--31\% in 3-year survival under $T=0$, whereas the proposed methods indicate a more modest but still statistically significant increase of about 7\%--11\%.

To put it in a nutshell, if the patients in $T=1$ group were assigned to $T=0$, naive methods that assume strong ignorability show that the average 3-year survival rate could increase 20-30\%.
This seems unrealistic.
On the contrary, the proposed methods indicate that the conclusion varies according to the level of CDCI: the 3-year survival rate does increase for CDCI$>$0 patients but only about 7-11\%, whereas it almost stays the same for CDCI=0 patients.

\begin{table}[h]
	\centering
	\caption{Real Data Application: comparison of estimated ATTs by naive methods (assuming strong ignorability) and proposed methods, using different estimation procedures for nuisance models, after stratifying the data based on CDCI$=$0 and CDCI$>$0.}
	\begin{tabular}{l | ccccrrr}
		\hline
		\multicolumn{8}{c}{CDCI$=$0}\\
		\hline
		&&$w(\X)$&$\Ezero(Y_0\mid\X)$& $\Eone(Y_1\mid\X)$ & $\Est$ & $\wh\sd_p$ & p-value\\
		\hline
		\multirow{6}{*}{Naive}
		&\multirow{3}{*}{ $\wh\Delta\nvone$} &$\logistic$&-&-&-0.1908&0.0222&0.0000\\
		&&$\rforest$&-&-&-0.0539&0.0209 &0.0101\\
		&&$\xgboost$&-&-&-0.2025&0.0274&0.0000\\\cline{2-8}
		&\multirow{3}{*}{$\wh\Delta\nvtwo$}&$\logistic$& $\logistic$&  -&  -0.1942& 0.0235& 0.0000\\
		%&&$\logistic$& $\rforest$ &-&-0.2043&0.0231&0.0000&0.0229\\
		&&$\rforest$&$\rforest$&-&-0.2045&0.0218&0.0000\\
		%&&$\logistic$&$\xgboost$&-&-0.1912&0.0235&0.0000&0.0241\\
		&&$\xgboost$&$\xgboost$&-&-0.1882&0.0256&0.0000\\\cline{2-8}
		\hline
		&&$\pi(Y_0,\U;\bt)$&$\Ezero(Y_0\mid\X,0)$& $\Eone(Y_1\mid\X,1)$ & $\Est$ & $\wh\sd_p$ & p-value\\
		\hline
		\multirow{6}{*}{Proposed}
		&\multirow{3}{*}{$\wh\Delta\proone(\wh\bt\eff)$}&\multirow{3}{*}{model (\ref{eq:trtmodelinsim})}& $\logistic$&-&0.0118& 0.0232&0.6095\\
		&&& $\rforest$ &-&-0.0270&0.0396&0.4948\\
		&&&$\xgboost$&-&0.0084&0.0222&0.7044\\\cline{2-8}
		&\multirow{3}{*}{$\wh\Delta\eff(\wh\bt\eff)$}&\multirow{3}{*}{model (\ref{eq:trtmodelinsim})}& $\logistic$&$\logistic$&0.0091& 0.0229&0.6900\\
		&&& $\rforest$ &$\rforest$&-0.0465&0.0321&0.1475\\
		&&&$\xgboost$&$\xgboost$&-0.0103&0.0224&0.6440\\\cline{2-8}
		\hline\hline
		\multicolumn{8}{c}{CDCI$>$0}\\
		\hline
		&&$w(\X)$&$\Ezero(Y_0\mid\X)$& $\Eone(Y_1\mid\X)$ & $\Est$ & $\wh\sd_p$ & p-value \\
		\hline
		\multirow{6}{*}{Naive}
		&\multirow{3}{*}{ $\wh\Delta\nvone$} &$\logistic$&-&-&-0.2841&0.0249&0.0000\\
		&&$\rforest$&-&-&-0.1554&0.0239&0.0000\\
		&&$\xgboost$&-&-&-0.3154&0.0307&0.0000\\\cline{2-8}
		&\multirow{3}{*}{$\wh\Delta\nvtwo$}&$\logistic$& $\logistic$&  -&  -0.2838& 0.0236& 0.0000\\
		%&&$\logistic$& $\rforest$ &-&-0.2893&0.0253&0.0000&0.0257\\
		&&$\rforest$&$\rforest$&-&-0.2872&0.0248&0.0000\\
		%&&$\logistic$&$\xgboost$&-&-0.2868&0.0251&0.0000&0.0260\\
		&&$\xgboost$&$\xgboost$&-&-0.2892&0.0275&0.0000\\\cline{2-8}
		\hline
		&&$\pi(Y_0,\U,\bt)$&$\Ezero(Y_0\mid\X,0)$& $\Eone(Y_1\mid\X,1)$ & $\Est$ & $\wh\sd_p$ & p-value\\
		\hline
		\multirow{6}{*}{Proposed}
		&\multirow{3}{*}{$\wh\Delta\proone(\wh\bt\eff)$}&\multirow{3}{*}{model (\ref{eq:trtmodelinsim})}& $\logistic$&-&  -0.0726& 0.0231&0.0017\\
		&&& $\rforest$ &-&-0.1050&0.0294&0.0004\\
		&&&$\xgboost$&-&-0.0747&0.0241&0.0020\\\cline{2-8}
		&\multirow{3}{*}{$\wh\Delta\eff(\wh\bt\eff)$}&\multirow{3}{*}{model (\ref{eq:trtmodelinsim})}& $\logistic$&$\logistic$&  -0.0750& 0.0240&0.0018\\
		&&& $\rforest$ &$\rforest$&-0.1072&0.0226&0.0000\\
		&&&$\xgboost$&$\xgboost$&-0.1125&0.0262&0.0000\\
		\hline
	\end{tabular}
	\label{tab:comparison_ATT_cdcc}
\end{table}

\begin{figure}[h]
	\begin{center}
		\includegraphics[width=0.32\textwidth,height=0.18\textheight]{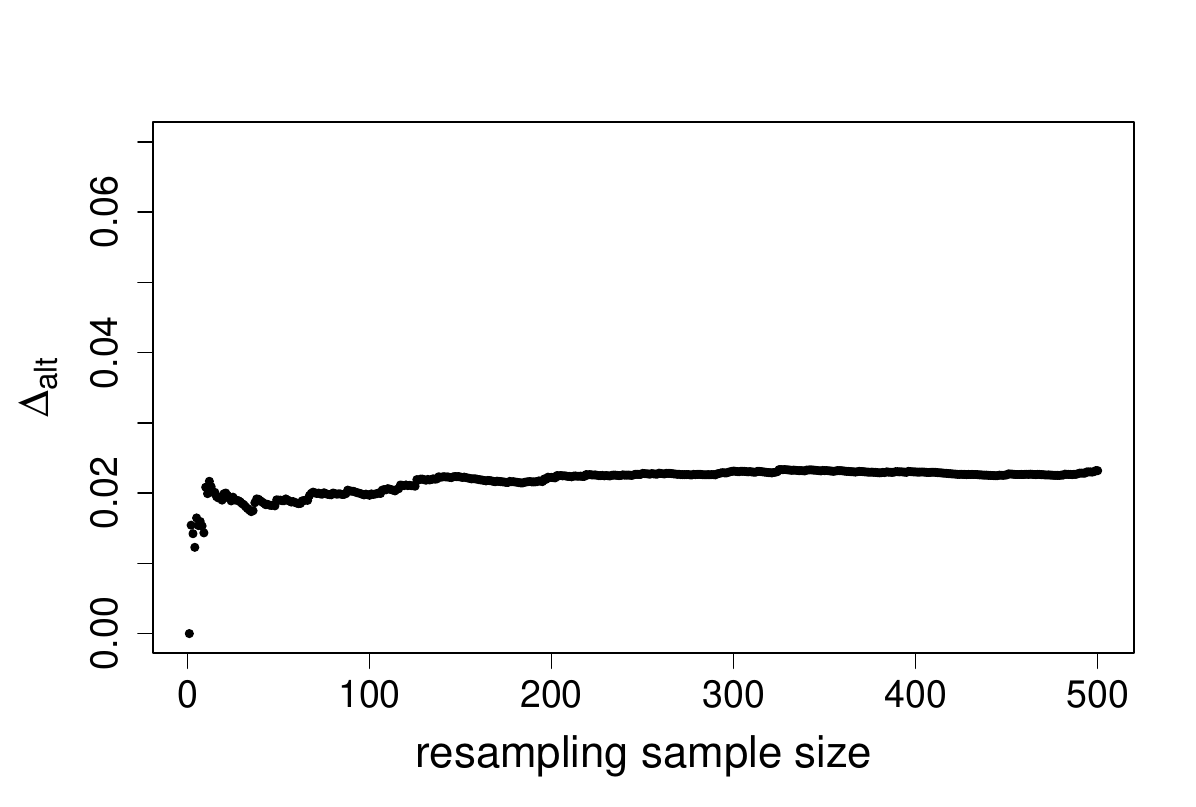}
		\includegraphics[width=0.32\textwidth,height=0.18\textheight]{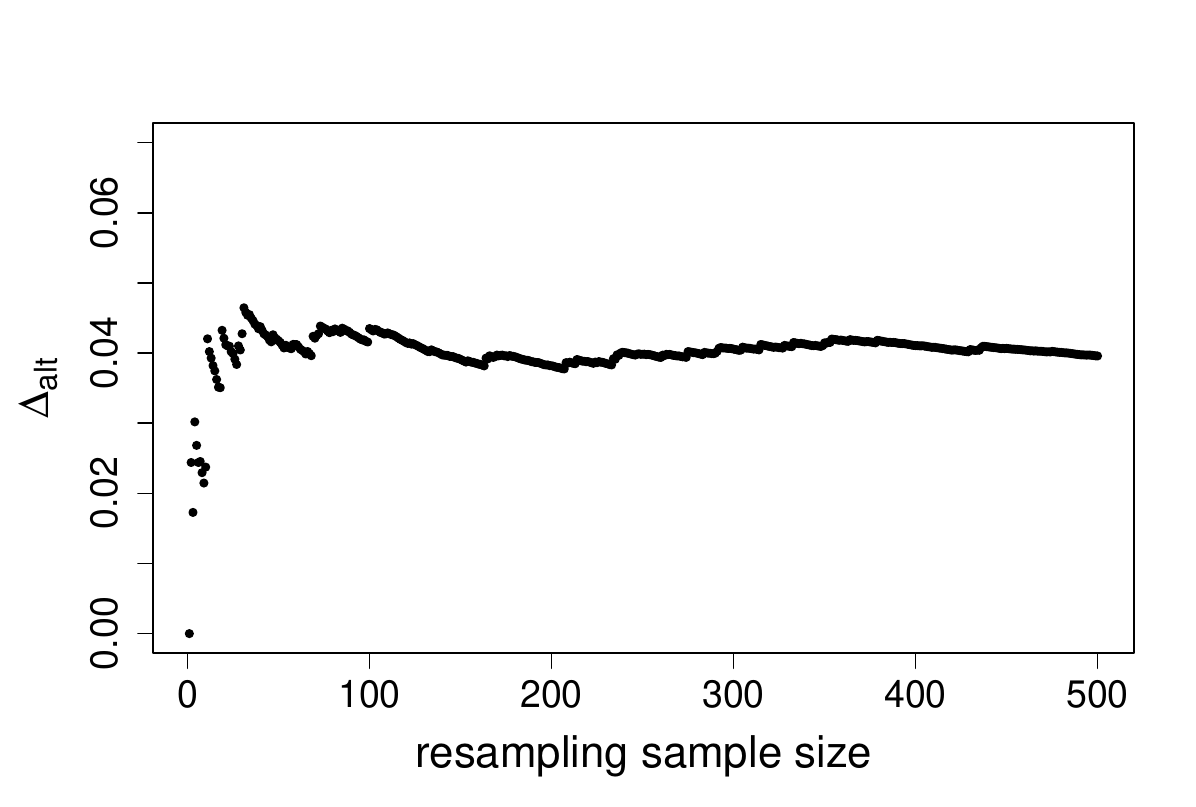}
		\includegraphics[width=0.32\textwidth,height=0.18\textheight]{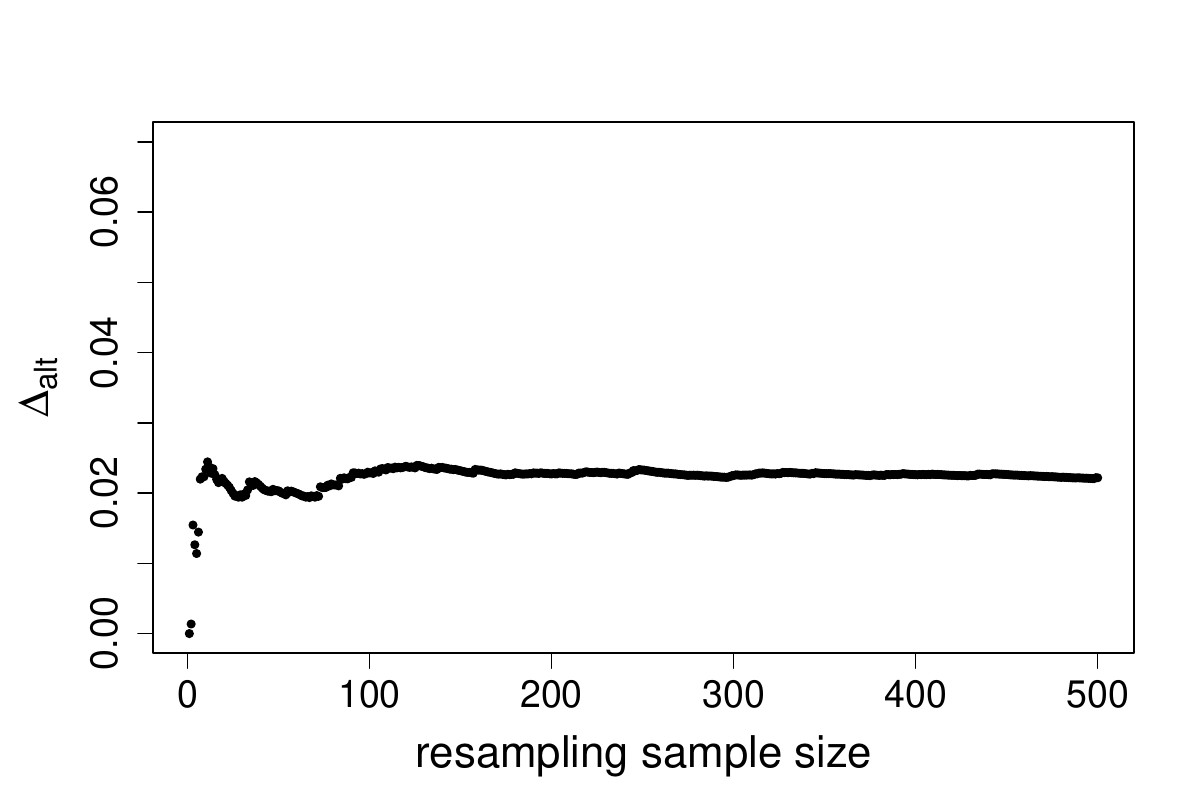}\\
		\includegraphics[width=0.32\textwidth,height=0.18\textheight]{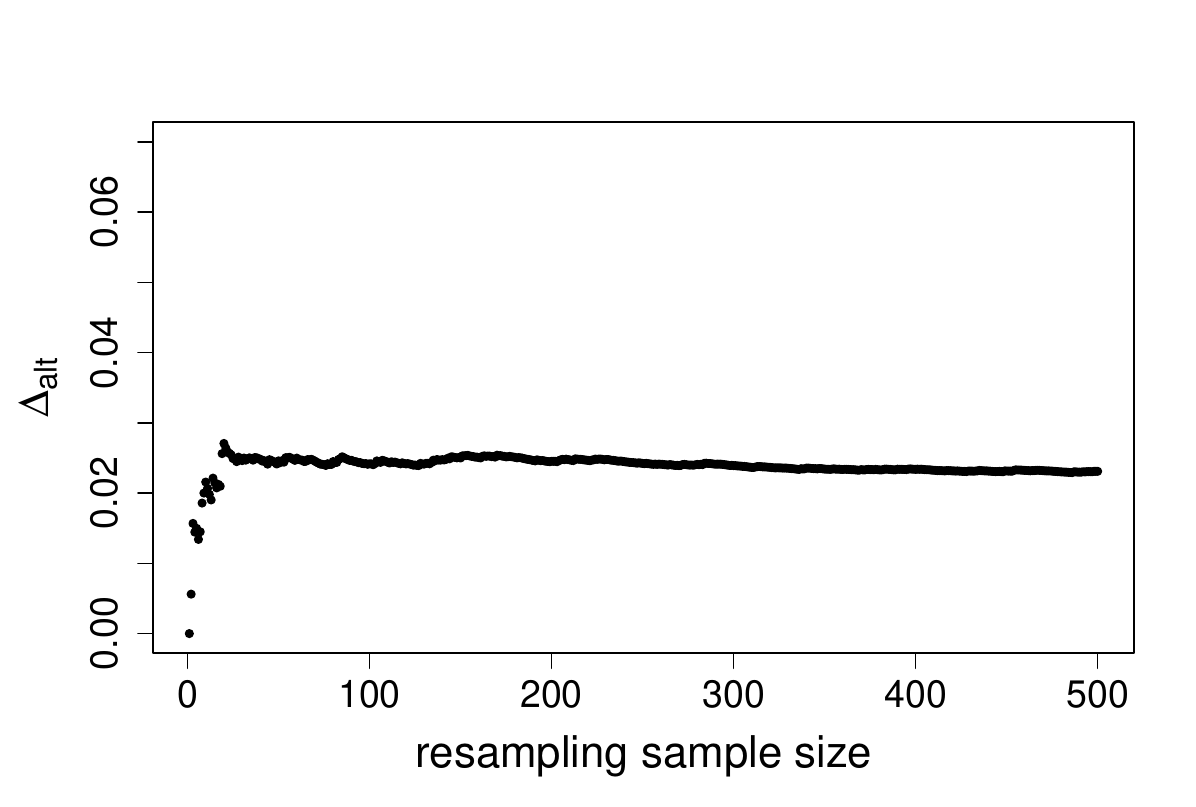}
		\includegraphics[width=0.32\textwidth,height=0.18\textheight]{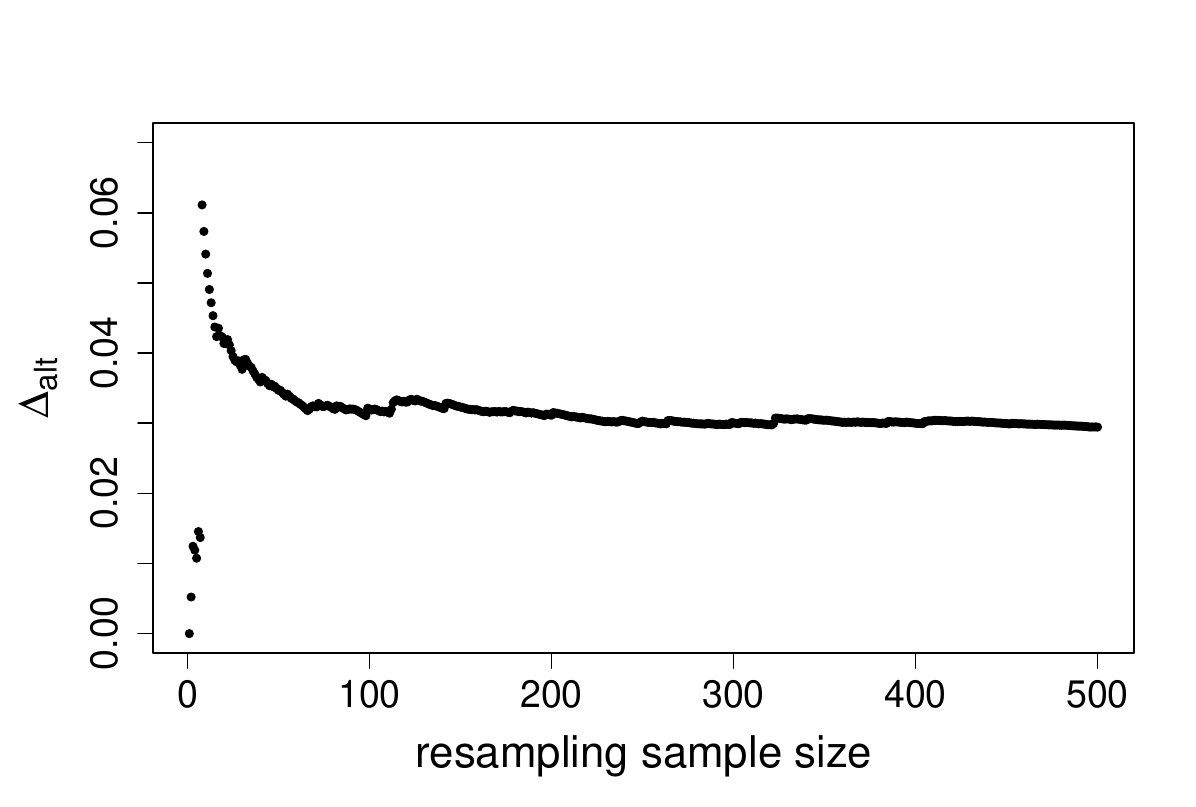}
		\includegraphics[width=0.32\textwidth,height=0.18\textheight]{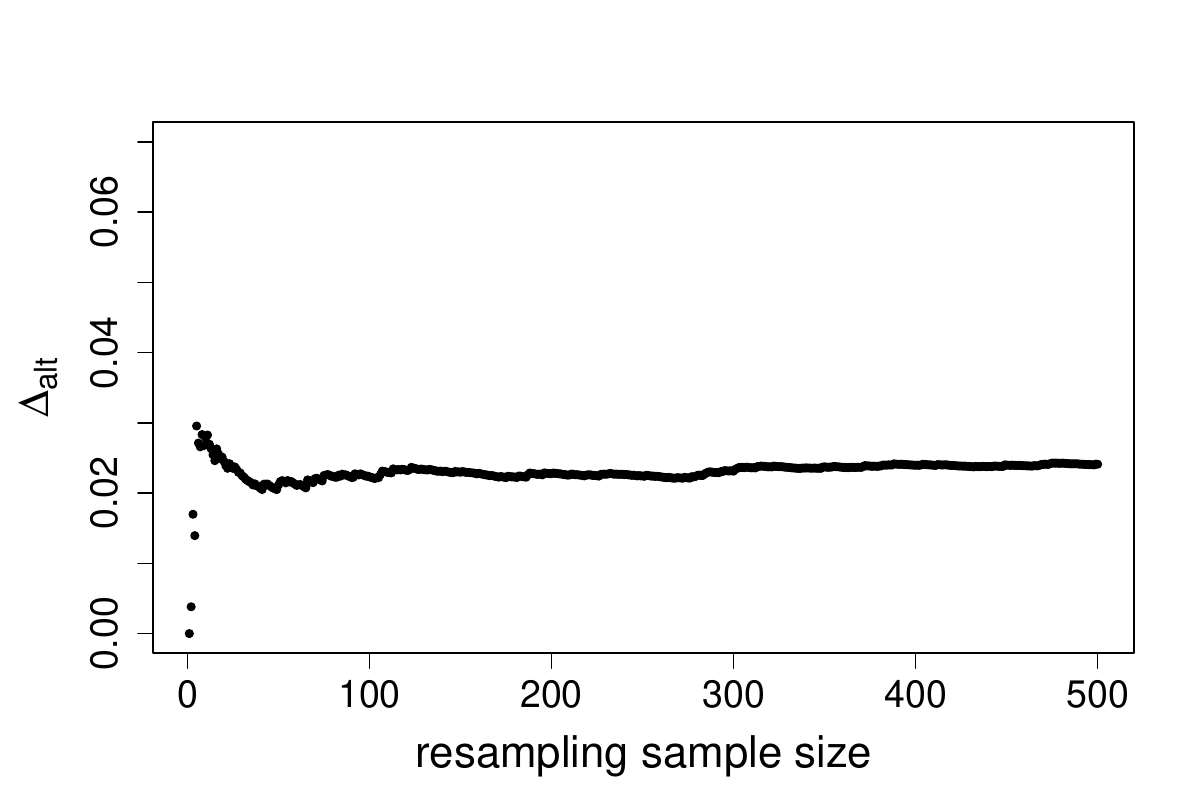}
		\caption{Real Data Application: standard deviations for $\wh\Delta\proone(\wh\bt\eff)$ based on 500 perturbed resamplings. The first row illustrates the standard deviations for the group CDCI=0 and the second row for CDCI$>$0. The first, second and third column summarize the results when logistic regression, random forest and XG-boost are implemented to estimate $\Ezero(Y_0\mid\X,0)$, respectively.}
		\label{fig:att_pro1}
	\end{center}
\end{figure}

\begin{figure}[h]
	\begin{center}
		\includegraphics[width=0.32\textwidth,height=0.18\textheight]{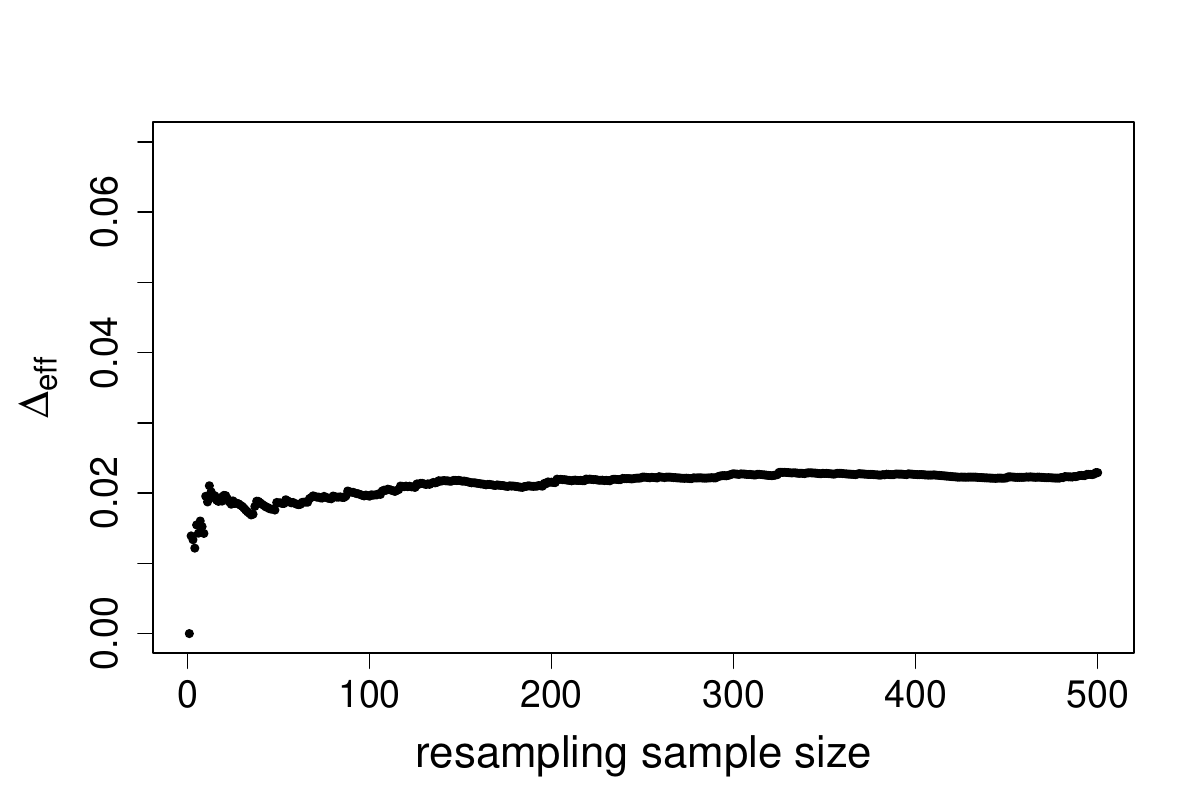}
		\includegraphics[width=0.32\textwidth,height=0.18\textheight]{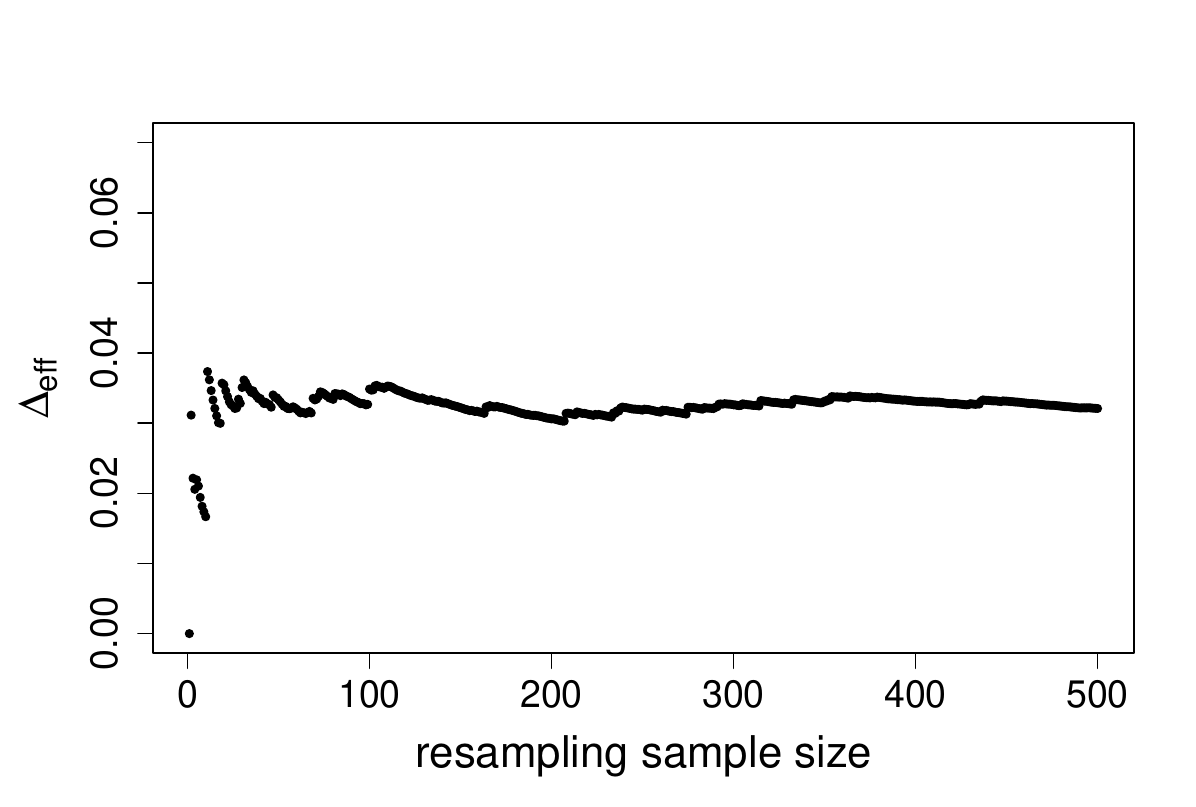}
		\includegraphics[width=0.32\textwidth,height=0.18\textheight]{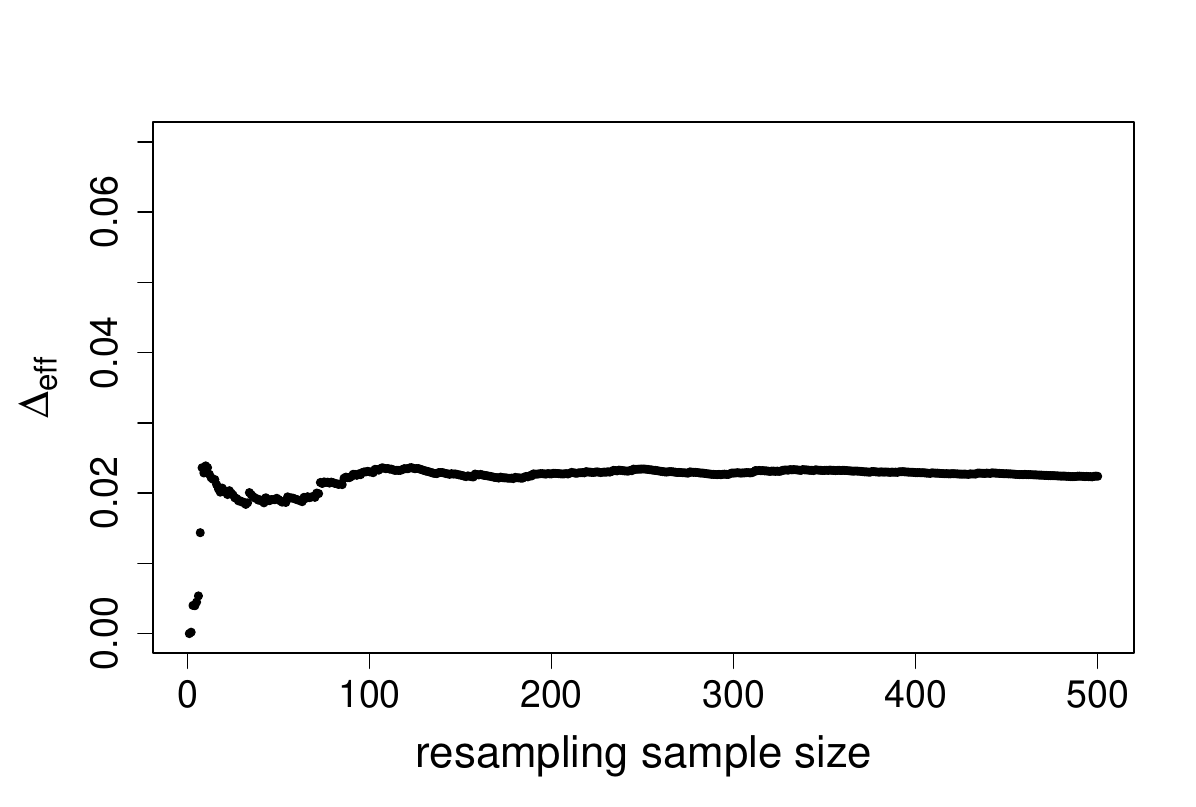}\\
		\includegraphics[width=0.32\textwidth,height=0.18\textheight]{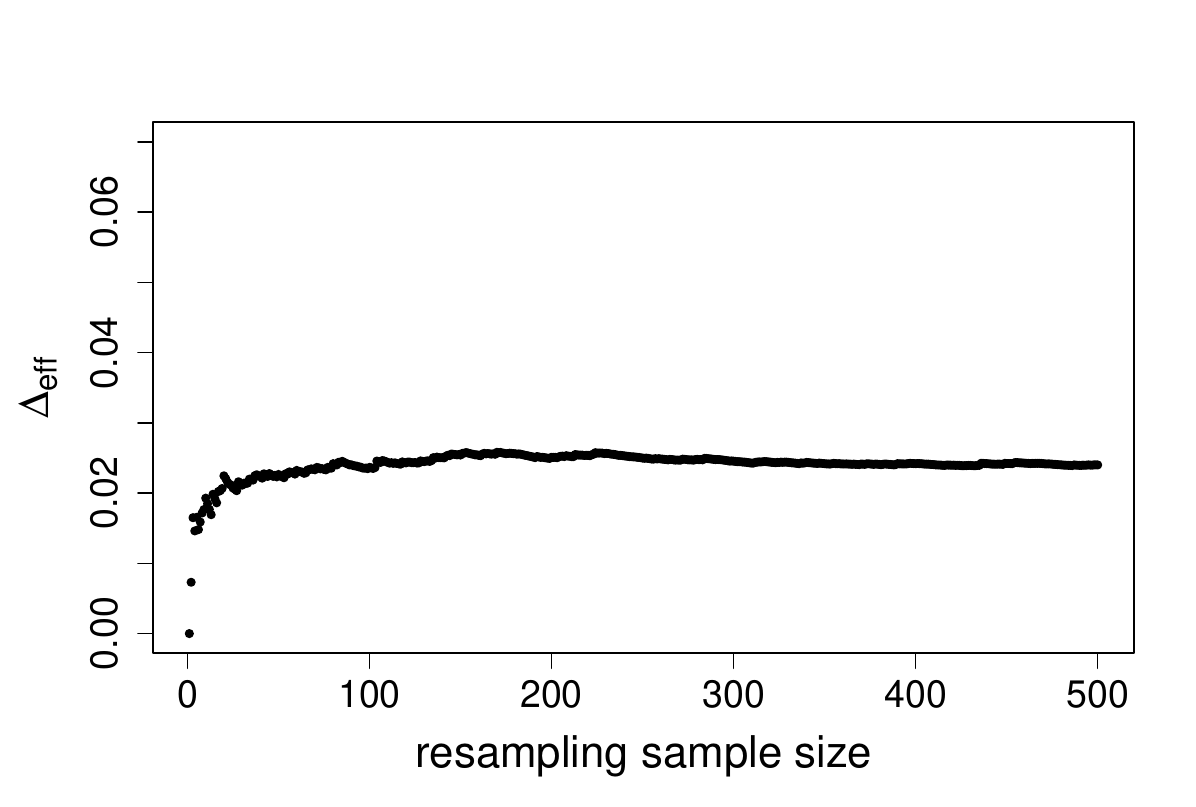}
		\includegraphics[width=0.32\textwidth,height=0.18\textheight]{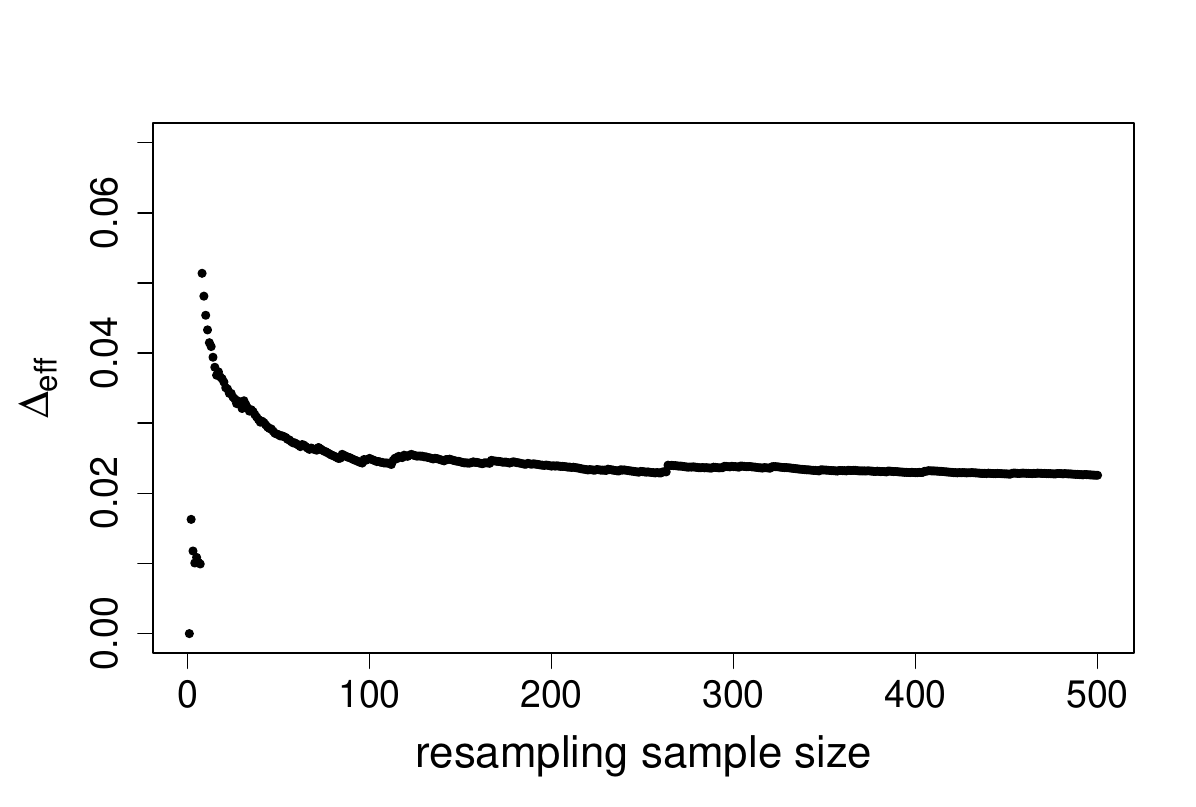}
		\includegraphics[width=0.32\textwidth,height=0.18\textheight]{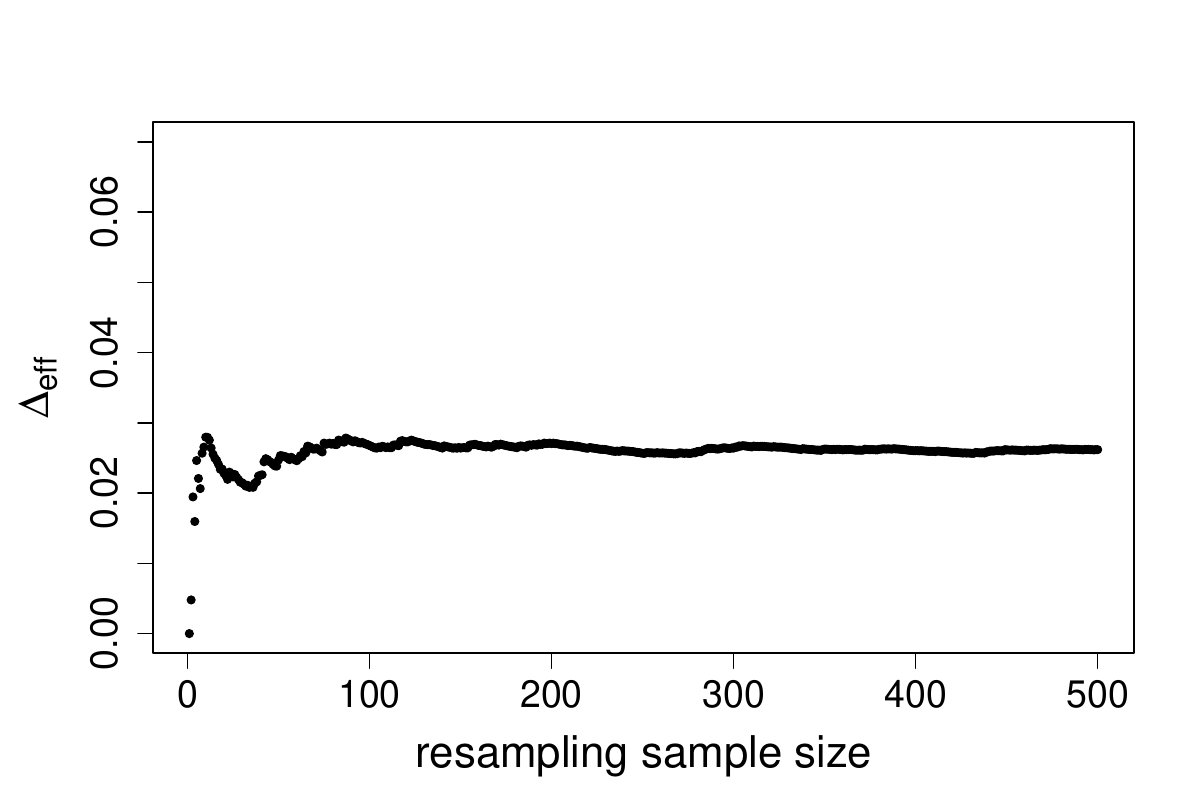}
		\caption{Real Data Application: standard deviations for $\wh\Delta\eff(\wh\bt\eff)$ based on 500 perturbed resamplings. The first row illustrates the standard deviations for the group CDCI=0 and the second row for CDCI$>$0. The first, second and third column summarize the results when logistic regression, random forest and XG-boost are implemented to estimate $\Ezero(Y_0\mid\X,0)$ and $\Eone(Y_1\mid\X,1)$, respectively.}
		\label{fig:att_pro2}
	\end{center}
\end{figure}

Finally, we primarily use perturbation resampling with 500 replications to estimate the standard deviation in this application, reported as $\wh\sd_p$ in the tables. We also validated these SD estimates using a separate nonparametric bootstrap with 500 bootstrap samples; the results were similar and are therefore omitted. In our experience, 500 resamples are sufficient for this application.
Below, in Figures~\ref{fig:att_pro1} and \ref{fig:att_pro2}, we provide the sensitivity plots of the number of subsamples in the perturbation resampling method for implementing $\wh\Delta\proone$ and $\wh\Delta\eff$ in Table~\ref{tab:comparison_ATT_cdcc}, respectively.

\begin{figure}[h]
	\begin{center}
		\includegraphics[width=0.32\textwidth,height=0.18\textheight]{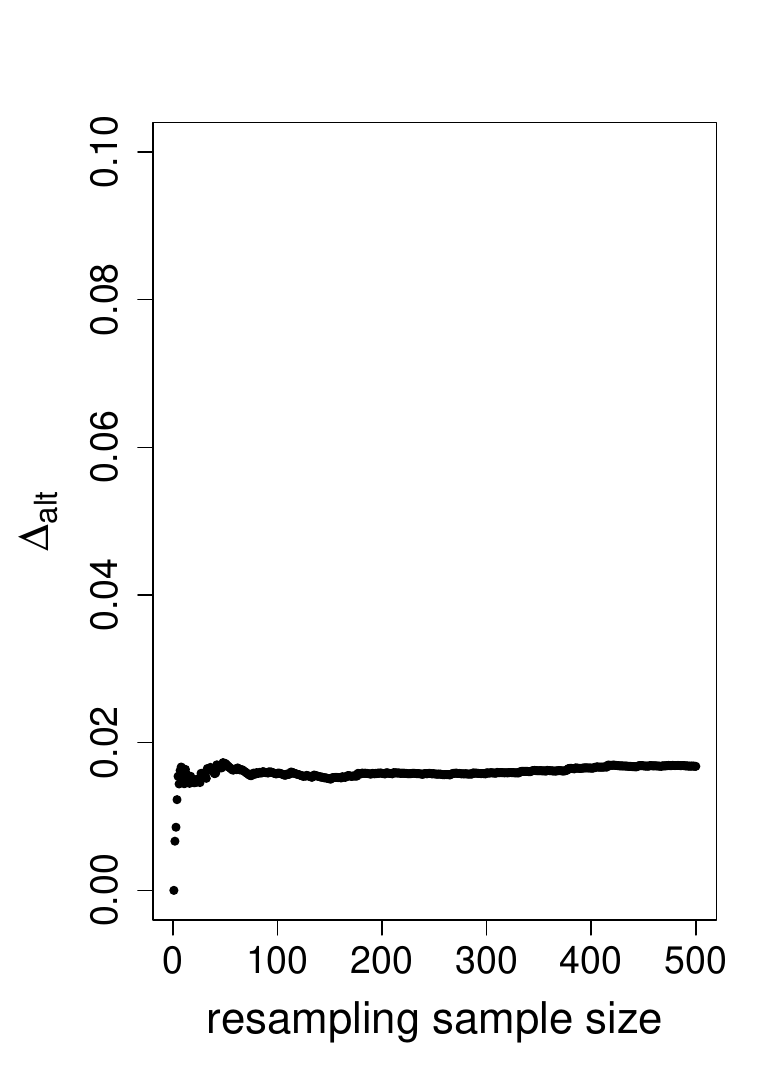}
		\includegraphics[width=0.32\textwidth,height=0.18\textheight]{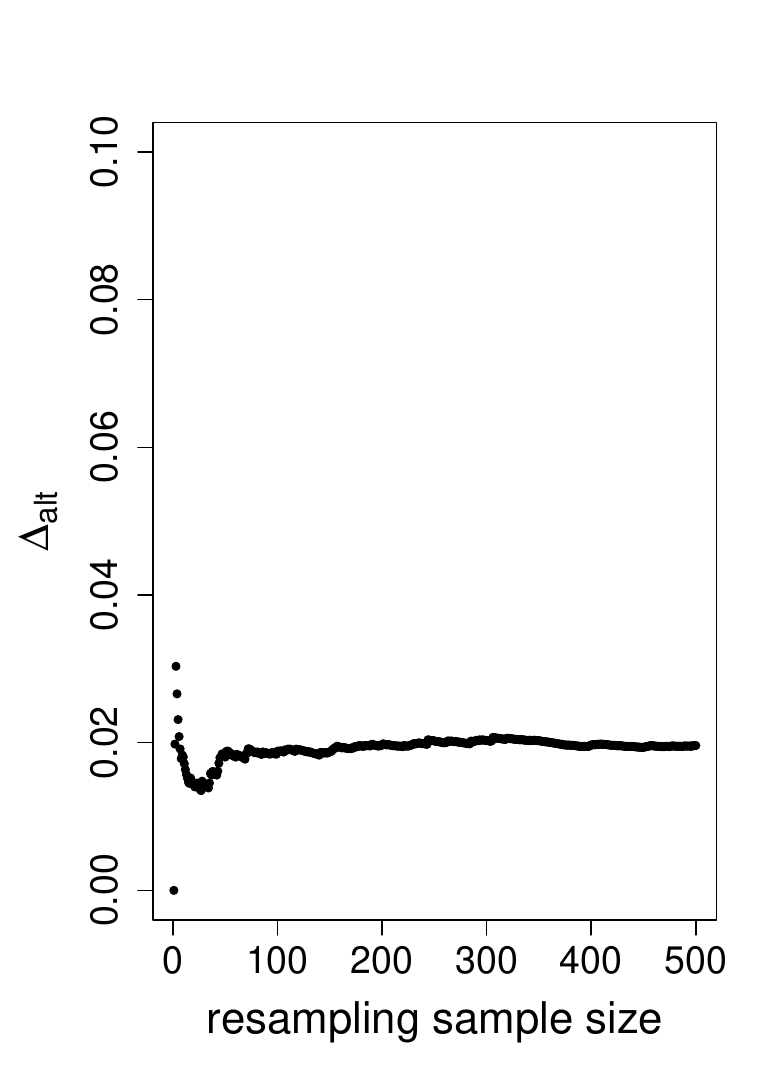}
		\includegraphics[width=0.32\textwidth,height=0.18\textheight]{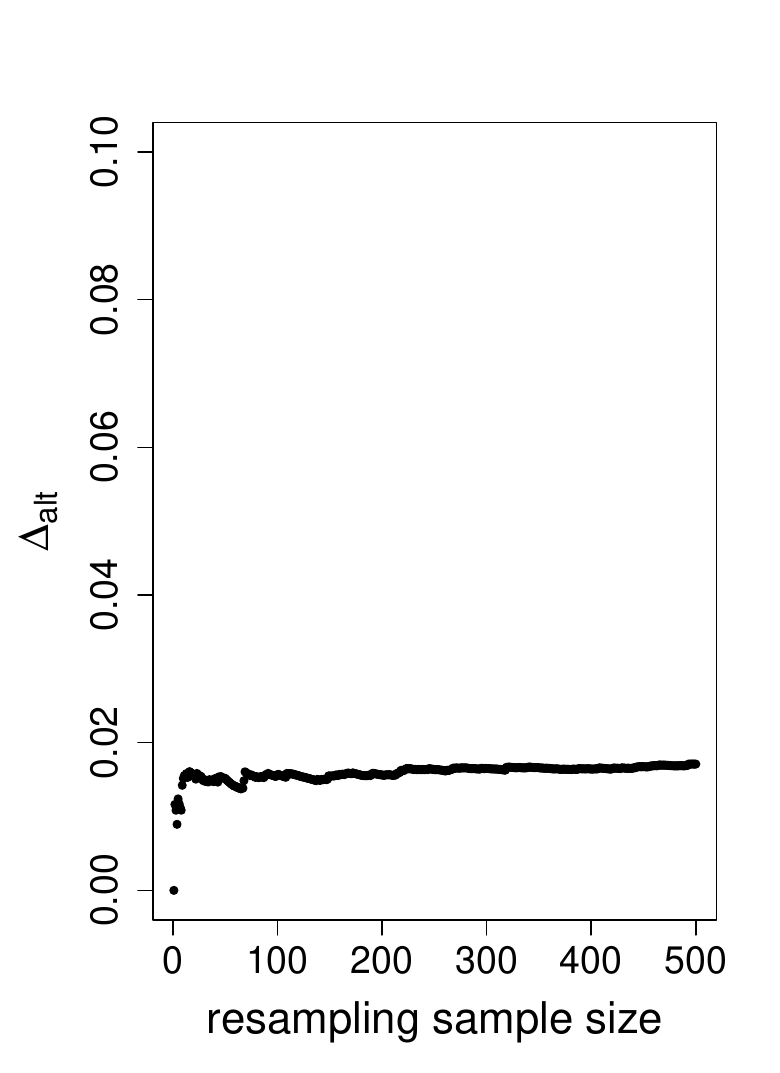}
		\caption{Real Data Application: standard deviations for $\wh\Delta\proone(\wh\bt\eff)$ based on 500 perturbed resamplings. The row illustrates the standard deviations based on the whole dataset. The first, second and third column summarize the results when logistic regression, random forest and XG-boost are implemented to estimate $\Ezero(Y_0\mid\X,0)$, respectively.}
		\label{fig:att_pro1_whole}
	\end{center}
\end{figure}

\begin{figure}[h]
	\begin{center}
		\includegraphics[width=0.32\textwidth,height=0.18\textheight]{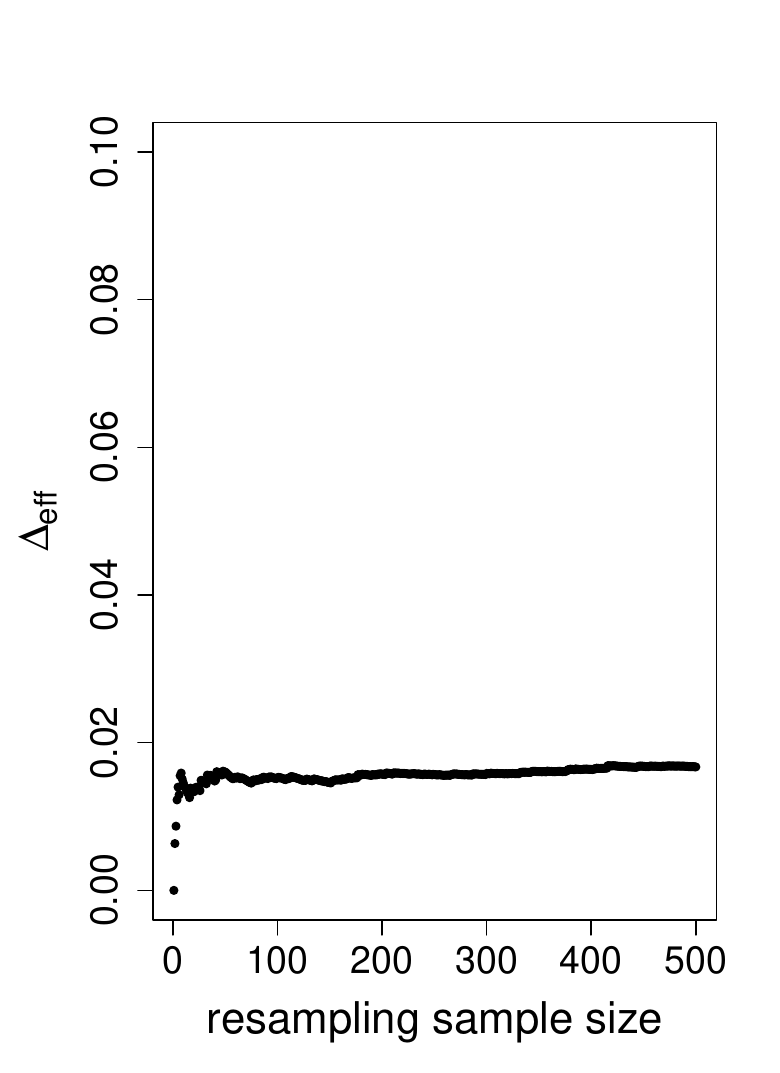}
		\includegraphics[width=0.32\textwidth,height=0.18\textheight]{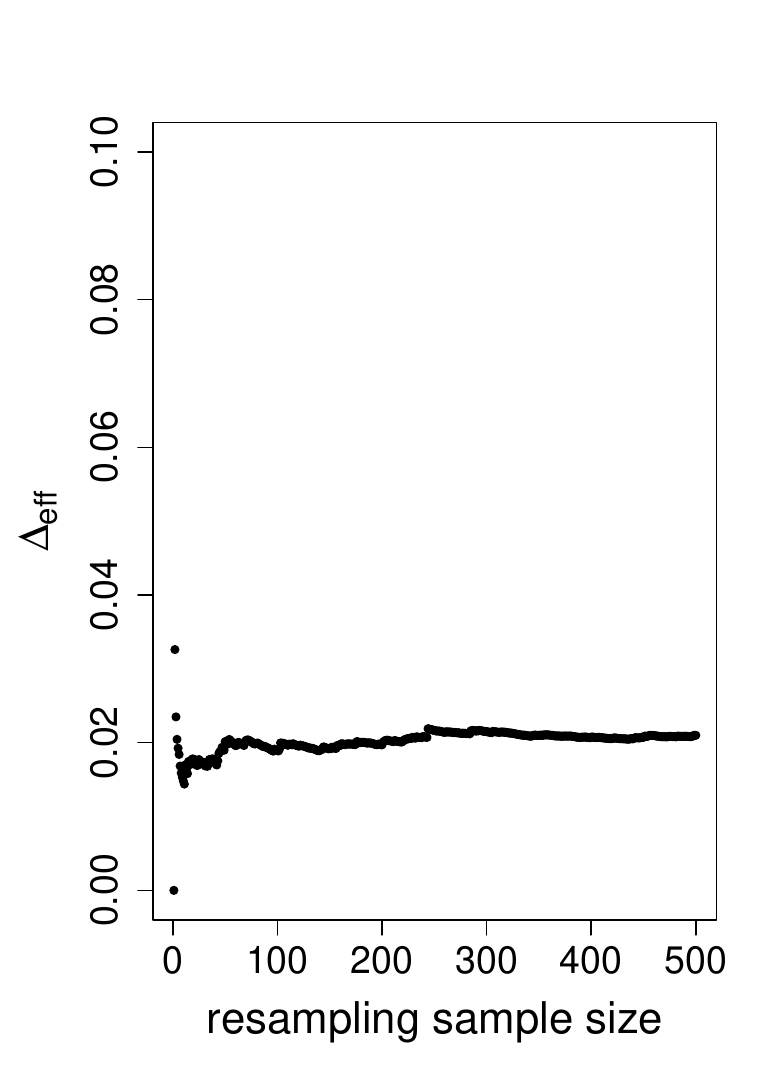}
		\includegraphics[width=0.32\textwidth,height=0.18\textheight]{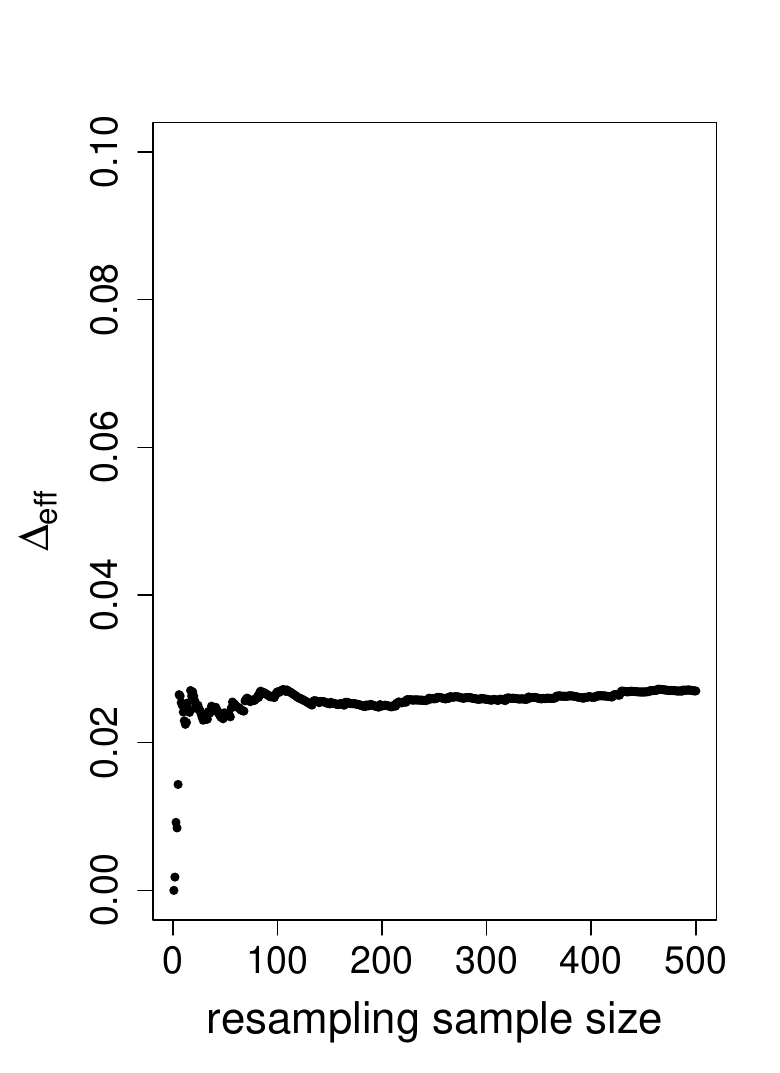}
		\caption{Real Data Application: standard deviations for $\wh\Delta\eff(\wh\bt\eff)$ based on 500 perturbed resamplings. The row illustrates the standard deviations based on the whole datasets. The first, second and third column summarize the results when logistic regression, random forest and XG-boost are implemented to estimate $\Ezero(Y_0\mid\X,0)$ and $\Eone(Y_1\mid\X,1)$, respectively.}
		\label{fig:att_pro2_whole}
	\end{center}
\end{figure}

\clearpage

\appendix

\setcounter{assump}{0}\renewcommand{\theassump}{A\arabic{assump}}
\setcounter{Lem}{0}\renewcommand{\theLem}{A\arabic{Lem}}
\setcounter{equation}{0}\renewcommand{\theequation}{A.\arabic{equation}}
\setcounter{subsection}{0}\renewcommand{\thesubsection}{S.\arabic{subsection}}

\section*{Appendix}
\section{Proof of Lemma~\ref{lemmaiden}}\label{sec:lemmaidenproof}

Assume we have two different sets of models $\fone(y_1\mid \x,1)$, $\fzero(y_0\mid \x,0)$, $g(\x)$, $\pi(y_0,\u)$ and $\wt\fone(y_1\mid \x,1)$, $\wt\fzero(y_0\mid \x,0)$, $\wt g(\x)$, $\wt\pi(y_0,\u)$, such that
\begin{eqnarray*}
&& \left[\int\frac{\fzero(y_0\mid \x,0)}{1-\pi(y_0,\u)}\dd y_0-1\right]^{t}
\left[\int\frac{\fzero(y_0\mid \x,0)}{1-\pi(y_0,\u)}\dd y_0\right]^{-1}
\fone(y_1\mid\x,1)^t \fzero(y_0\mid \x,0)^{1-t} g(\x)\\
&=& \left[\int\frac{\wt \fzero(y_0\mid \x,0)}{1-\wt\pi(y_0,\u)}\dd y_0-1\right]^{t}
\left[\int\frac{\wt \fzero(y_0\mid \x,0)}{1-\wt\pi(y_0,\u)}\dd y_0\right]^{-1}
\wt\fone(y_1\mid\x,1)^t \wt\fzero(y_0\mid \x,0)^{1-t} \wt g(\x).
\end{eqnarray*}
Taking $t=0$ and taking integration with respect to $Y_0$ on both sides of the above equation, one would have
\begin{eqnarray*}
\left[\int\frac{\fzero(y_0\mid \x,0)}{1-\pi(y_0,\u)}\dd y_0\right]^{-1} g(\x)=
\left[\int\frac{\wt \fzero(y_0\mid \x,0)}{1-\wt\pi(y_0,\u)}\dd y_0\right]^{-1} \wt g(\x).
\end{eqnarray*}
Taking $t=1$ and taking integration with respect to $Y_1$, one would obtain
\begin{eqnarray*}
&&\left[\int\frac{\fzero(y_0\mid \x,0)}{1-\pi(y_0,\u)}\dd y_0-1\right]
\left[\int\frac{\fzero(y_0\mid \x,0)}{1-\pi(y_0,\u)}\dd y_0\right]^{-1}
g(\x)\\
&=& \left[\int\frac{\wt \fzero(y_0\mid \x,0)}{1-\wt\pi(y_0,\u)}\dd y_0-1\right]
\left[\int\frac{\wt \fzero(y_0\mid \x,0)}{1-\wt\pi(y_0,\u)}\dd y_0\right]^{-1}
\wt g(\x).
\end{eqnarray*}
Hence, immediately, we have
\begin{eqnarray*}
\int\frac{\fzero(y_0\mid \x,0)}{1-\pi(y_0,\u)}\dd y_0 = \int\frac{\wt \fzero(y_0\mid \x,0)}{1-\wt\pi(y_0,\u)}\dd y_0, \mbox{ and } g(\x)=\wt g(\x).
\end{eqnarray*}
Then, by taking $t=0$ and $t=1$ respectively, we also obtain
\begin{eqnarray*}
\fzero(y_0\mid \x,0)=\wt\fzero(y_0\mid \x,0), \mbox{ and } \fone(y_1\mid \x,1)=\wt\fone(y_1\mid \x,1).
\end{eqnarray*}
Finally, from
\begin{eqnarray*}
\int\frac{\fzero(y_0\mid \x,0)}{1-\pi(y_0,\u)}\dd y_0 = \int\frac{\fzero(y_0\mid \x,0)}{1-\wt\pi(y_0,\u)}\dd y_0,
\end{eqnarray*}
and the completeness condition imposed in (\ref{eq:assumecompleteness}), it is immediate that
\begin{eqnarray*}
\pi(y_0,\u)=\wt \pi(y_0,\u),
\end{eqnarray*}
and this completes the proof.

\section{Proof of Proposition~\ref{prop:Hdecomp}}\label{sec:derivelambdaperp}

Firstly,
it can be easily verified, with the proofs omitted, that $\Lambda_0\perp\Lambda_1$, $\Lambda_0\perp\Lambda_2$ and $\Lambda_1\perp\Lambda_2$, hence the nuisance tangent space $\Lambda=\Lambda_0\oplus\Lambda_1\oplus\Lambda_2$.
Now we derive $\Lambda^\perp$. Denote a generic element as $t\g_1(y_1,\x)+(1-t)\g_0(y_0,\x)$. It is straightforward that
\begin{eqnarray*}
\Lambda_2^\perp &=& \left[ t\g_1(y_1,\x)+(1-t)\g_0(y_0,\x): \Eone(\g_1\mid\x,1)w(\x) + \Ezero(\g_0\mid\x,0)\{1-w(\x)\}=\0 \right],\\
\Lambda_1^\perp &=& \left\{ t\g_1(y_1,\x)+(1-t)\g_0(y_0,\x): \g_1(y_1,\x) = \Eone(\g_1\mid\x,1) \right\},\\
&=& \left\{ t\g(\x)+(1-t)\g_0(y_0,\x) \right\}.
\end{eqnarray*}
Hence
\begin{eqnarray*}
\Lambda_2^\perp \cap \Lambda_1^\perp =
\left[ t\g(\x)+(1-t)\g_0(y_0,\x): \Ezero(\g_0\mid\x,0) = - w(\x)\{1-w(\x)\}^{-1} \g(\x)
\right].
\end{eqnarray*}
Keeping this form in mind, we consider the element $t\g(\x)+(1-t)\g_0(y_0,\x)$ which satisfies that, for any $\c(y_0,\x)$ such that $\Ezero\{\c(y_0,\x)\mid\x,0\}=\0$, we have
\begin{eqnarray*}
0 &=& \E\left[\{t\g(\x)+(1-t)\g_0(y_0,\x)\}\trans \left\{(1-t)\c(y_0,\x)+\frac{t-w(\x)}{w(\x)}\Ezero(\c\mid\x)\right\}\right]\\
&=& \E\left[ \{1-w(\x)\}\g(\x)\trans \Ezero(\c\mid\x)+\{1-w(\x)\}\Ezero(\g_0\trans \c\mid\x,0) \right.\\
&& \left.- \{1-w(\x)\}\Ezero(\g_0\mid\x,0)\trans \Ezero(\c\mid\x)\right]\\
&=& \E\left[\g(\x)\trans \Ezero(\c\mid\x)+\{1-w(\x)\}\Ezero(\g_0\trans \c\mid\x,0)\right]\\
&=&
\E\left\{\Ezero\left(\{1-w(\x)\} [\{1-\pi(\bt^0)\}^{-1}\g(\x)+\g_0(y_0,\x)]\trans \c(y_0,\x)\mid\x,0\right)\right\}.
\end{eqnarray*}
Therefore, it must be the situation that $\{1-\pi(\bt^0)\}^{-1}\g(\x)+\g_0(y_0,\x)$ is a function of $\x$ only, say, $\v(\x)$:
\begin{eqnarray*}
\{1-\pi(\bt^0)\}^{-1}\g(\x)+\g_0(y_0,\x)=\v(\x).
\end{eqnarray*}
Taking conditional expectations on both sides, we can solve $\v(\x)=\g(\x)$, hence,
\begin{eqnarray*}
\{1-\pi(\bt^0)\}^{-1}\g(\x)+\g_0(y_0,\x)=\g(\x).
\end{eqnarray*}
Finally,
\begin{eqnarray*}
\Lambda^\perp &=& \left( [t-\pi(\bt^0)\{1-\pi(\bt^0)\}^{-1}(1-t)]\g(\x): \g(\x)\in {\mR^{d}} \right)\\
&=& \left\{\frac{t-\pi(\bt^0)}{1-\pi(\bt^0)}\g(\x): \g(\x)\in {\mR^{d}} \right\}.
\end{eqnarray*}

\section{Proof of Proposition~\ref{prop:Hphiprotwo}}
\label{sec:eifdelta}

The efficient influence function of $\Delta$, $\phi\eff$, must be of the form
\begin{eqnarray*}
\phi\eff = a_2(\x) + t b_2(y_1,\x) + (1-t)c_2(y_0,\x)+\frac{t-w(\x)}{w(\x)}\Ezero(c_2\mid\x)+ \H\trans \S\eff,
\end{eqnarray*}
where $\E(a_2)=\0$, $\Eone(b_2\mid\x,1)=\0$, $\Ezero(c_2\mid\x,0)=\0$ and $\H$ is a $d\times 1$ vector.

We consider parametric submodels $\fx(\x;\ba)$, $f_1(y_1\mid\x,1;\bb)$ and $f_0(y_0\mid\x,0;\bg)$, which coincide with the true models when $\ba=\ba^0$, $\bb=\bb^0$ and $\bg=\bg^0$, respectively. Then, the likelihood under the submodel is
\begin{eqnarray*}
w(\x;\bg,\bt)^t \{1-w(\x;\bg,\bt)\}^{1-t} \fzero(y\mid\x,0;\bg)^{1-t} \fone(y\mid\x,1;\bb)^t \fx(\x;\ba).
\end{eqnarray*}
We can derive
\begin{eqnarray*}
\S_\ba(\x) &=& \pdv{\log \fx(\x;\ba)}{\ba} \in \Lambda_2,\\
\S_\bb(t,y,\x) &=& t\pdv{\log f_1(y_1\mid\x,1;\bb)}{\bb} \in \Lambda_1,\\
\S_\bg(t,y,\x) &=& (1-t)\pdv{\log f_0(y_0\mid\x,0;\bg)}{\bg} \\
&&+ \frac{t-w(\x)}{w(\x)}\Ezero\left\{ \pdv{\log f_0(y_0\mid\x,0;\bg)}{\bg}\mid \x\right\}\in \Lambda_0,\\
\S_\bt(t,\x) &=& \frac{t-w(\x)}{w(\x)} \Ezero\left[ \{1-\pi(\bt^0)\}^{-1}\dot\bpi(\bt^0)\mid\x \right],
\end{eqnarray*}
where the partial derivatives are evaluated at the true values if not otherwise specified.
Next, we can derive
\begin{eqnarray*}
\E(\phi\eff\S_\ba) &=& \E\left\{a_2(\x)\pdv*{\log \fx}{\ba}\right\},\\
\E(\phi\eff\S_\bb) &=& \E\{\Eone(b_2 \pdv*{\log f_1}{\bb\trans}\mid\x,1)w(\x)\} = \E\{w(\x)b_2 \pdv*{\log f_1}{\bb\trans}\},\\
\E(\phi\eff\S_\bg) &=&
\E\left( \left[ \{1-\pi(\bt^0)\}c_2 + \frac{1-w(\x)}{w(\x)}\Ezero(c_2\mid\x) \right] \pdv*{\log f_0}{\bg}\right),\\
\E(\phi\eff\S_\bt\trans)&=&\E\left[ \frac{1-w(\x)}{w(\x)}\E_0\left(c_2\mid\x\right)\Ezero\left\{\frac{\dot\bpi(\bt^0)\trans}{1-\pi(\bt^0)}\mid\x\right\}\right] + \H\trans\M.
\end{eqnarray*}

Now, we rewrite
\begin{eqnarray*}
\Delta = \frac{A(\ba,\bb,\bg,\bt)-B(\ba,\bg,\bt)}{C(\ba,\bg,\bt)},
\end{eqnarray*}
where
\begin{eqnarray*}
A(\ba,\bb,\bg,\bt) &=& \iint y_1 w(\x;\bg,\bt)f_1(y_1\mid\x,1;\bb)\fx(\x;\ba)\dd y_1 \dd \x = \E(T Y_1),\\
B(\ba,\bg,\bt) &=& \iint y_0 \pi(\bt)\{1-\pi(\bt)\}^{-1} \{1-w(\x;\bg,\bt)\}f_0(y_0\mid\x,0;\bg)\fx(\x;\ba)\dd y_0 \dd \x \\
&=& \E(T Y_0),\\
C(\ba,\bg,\bt) &=& \int w(\x;\bg,\bt)\fx(\x;\ba)\dd \x = \E(T) = p.
\end{eqnarray*}
For $\pdv*{\Delta}{\ba}$, we have
\begin{eqnarray*}
\pdv{A}{\ba}=\E\left\{T Y_1 \pdv{\log \fx}{\ba}\right\}, \pdv{B}{\ba}=\E\left\{T Y_0 \pdv{\log \fx}{\ba}\right\}, \pdv{C}{\ba}=\E\left\{T \pdv{\log \fx}{\ba}\right\}.
\end{eqnarray*}
Hence
\begin{eqnarray*}
\pdv{\Delta}{\ba} = \E\left[ \left\{\frac{T}{p}(Y_1-Y_0)-\frac{T}{p}\Delta^0\right\}\pdv{\log \fx}{\ba}\right].
\end{eqnarray*}
This implies
\begin{eqnarray*}
a_2(\x) &=& \E\left\{ \frac{T}{p}(Y_1-Y_0)-\frac{T}{p}\Delta^0\mid \x \right\} \\
&=& \frac1p \Eone(Y_1\mid \x,1)w(\x) -\frac1p \Ezero\left\{\frac{\pi(\bt^0)}{1-\pi(\bt^0)}Y_0\mid \x,0\right\}\{1-w(\x)\}-\frac{\Delta^0}{p}w(\x),\\
&=&\frac1p \Eone(Y_1\mid \x,1)w(\x) -\frac1p \Ezero\{\pi(\bt^0)Y_0\mid \x\}-\frac{\Delta^0}{p}w(\x).
\end{eqnarray*}
For $\pdv*{\Delta}{\bb}$, we have
\begin{eqnarray*}
\pdv{\Delta}{\bb}=\E\left\{ \frac{w(\X)}{p}Y_1 \pdv{\log \fx}{\bb}\right\},
\end{eqnarray*}
hence $b_2 - y_1/p$ must be a function of $\x$ alone. From this, we can solve
\begin{eqnarray*}
b_2 = \frac1p y_1 - \frac1p \Eone(Y_1\mid \x,1).
\end{eqnarray*}
For $\pdv*{\Delta}{\bg}$, we can derive
\begin{eqnarray*}
\pdv{A}{\bg} &=& \E[\{1-w(\X)\}\Eone(Y_1\mid\X,1)\pdv*{\log f_0}{\bg}],\\
\pdv{B}{\bg} &=& \E\left([\pi(\bt^0)Y_0-\Ezero\{\pi(\bt^0)Y_0\mid\X\}]\pdv*{\log f_0}{\bg}\right),\\
\pdv{p}{\bg} &=& \E[\{1-w(\X)\}\pdv*{\log f_0}{\bg}].
\end{eqnarray*}
Since $\pdv*{\log f_0}{\bg}$ satisfies $\Ezero[\{1-\pi(\bt^0)\}\pdv*{\log f_0}{\bg}\mid\x]=\0$, it must be the situation that
\begin{eqnarray*}
&&\frac{1-w(\x)}{1-\pi(\bt^0)}\Eone(Y_1\mid\x,1) - \frac{\pi(\bt^0)}{1-\pi(\bt^0)}Y_0 + \frac{\Ezero\{\pi(\bt^0)Y_0\mid\x\}}{1-\pi(\bt^0)} - \frac{1-w(\x)}{1-\pi(\bt^0)}\Delta^0- p c_2 \\
&&- \frac{p}{1-\pi(\bt^0)}\frac{1-w(\x)}{w(\x)}\Ezero(c_2\mid\x)
\end{eqnarray*}
is a function of $\x$ only, say, $e(\x)$, then taking conditional expectations on both sides, we must have
\begin{eqnarray*}
e(\x) = \Eone(Y_1\mid\x,1) - \Delta^0- \frac{p}{w(\x)}\Ezero(c_2\mid\x).
\end{eqnarray*}
Then taking expectations on both sides again, and after a little bit tedious calculations, we will have
\begin{eqnarray*}
\Ezero(c_2\mid\x) &=& \frac{w(\x)}{1-w(\x)}\frac1{B(\x)}\frac1p [\{1-w(\x)\}B(\x)-w(\x)]\{\Eone(Y_1\mid\x,1)-\Delta^0\}\\
&& -\frac{w(\x)}{1-w(\x)}\frac1{B(\x)}\frac1p\Ezero\left\{\frac{\pi(\bt^0)}{1-\pi(\bt^0)}Y_0\mid\x\right\}\\
&& + \frac{w(\x)}{1-w(\x)}\frac1{B(\x)}\frac1p\{1+B(\x)\}\Ezero\{\pi(\bt)Y_0\mid\x\}, \mbox{ and }\\
c_2 &=& \frac1p \frac{\pi(\bt^0)-w(\x)}{1-\pi(\bt^0)}\frac{w(\x)}{1-w(\x)}\frac1{B(\x)}\{\Eone(Y_1\mid\x,1)-\Delta^0\}\\
&& + \frac1p \Ezero\{\pi(\bt^0)Y_0\mid\x\}\left\{\frac1{1-w(\x)}-\frac{\pi(\bt^0)-w(\x)}{\{1-\pi(\bt^0)\}\{1-w(\x)\}B(\x)}\right\}\\
&& - \frac1p \frac{\pi(\bt^0)}{1-\pi(\bt^0)}Y_0 +\frac1p \frac{\pi(\bt^0)-w(\x)}{\{1-\pi(\bt^0)\}\{1-w(\x)\}B(\x)}\Ezero\left\{\frac{\pi(\bt^0)}{1-\pi(\bt^0)}Y_0\mid\x\right\}.
\end{eqnarray*}
Now we put the first three terms in $\phi\eff$, $a_2(\x) + t b_2(y_1,\x) + (1-t)c_2(y_0,\x)+\frac{t-w(\x)}{w(\x)}\Ezero(c_2\mid\x)$, together, it equals
\begin{align*}
&\frac1p \frac{t-\pi(\bt^0)}{1-\pi(\bt^0)}\left(y - \frac1{B(\x)}\left[w(\x)\Eone(Y_1\mid\x,1)+\Ezero\left\{\frac{\pi(\bt^0)^2}{1-\pi(\bt^0)}Y_0\mid\x\right\}\right]\right) \\
&-\frac{\Delta^0}p\left\{t-\frac{t-\pi(\bt^0)}{1-\pi(\bt^0)}\frac{w(\x)}{B(\x)}\right\}\\
=&~\frac1p \frac{t-\pi(\bt^0)}{1-\pi(\bt^0)}\bigg(y 
- \frac1{B(\x)}\left[w(\x)\Eone(Y_1\mid\x,1)+\{1-w(\x)\}\Ezero\left\{\frac{\pi(\bt^0)^2}{\{1-\pi(\bt^0)\}^2}Y_0\mid\x,0\right\}\right]\bigg)\\
&-\frac{\Delta^0}p \left\{t-\frac{t-\pi(\bt^0)}{1-\pi(\bt^0)}\frac{w(\x)}{B(\x)}\right\}\\
=&~ \phi(\w;\bt^0,\Delta^0).
\end{align*}
For $\pdv*{\Delta}{\bt}$, we can derive
\begin{eqnarray*}
\pdv{A}{\bt} &=& \E\left[\left\{1-w(\x)\right\}\Eone(Y_1\mid\x,1)\frac{\dot\bpi(\bt^0)}{1-\pi(\bt^0)}\right],\\
\pdv{B}{\bt}&=& \E\left\{Y_0\frac{\dot\bpi(\bt^0)}{1-\pi(\bt^0)}\right\} - \E\left[\Ezero\left\{Y_0\pi(\bt^0)\mid\x\right\}\frac{\dot\bpi(\bt^0)}{1-\pi(\bt^0)}\right],\\
\pdv{p}{\bt}&=& \E\left[\frac{\dot\bpi(\bt^0)}{1-\pi(\bt^0)}\left\{1-w(\x)\right\}\right].
\end{eqnarray*}
Hence,
\begin{eqnarray*}
\pdv{\Delta}{\bt}=p^{-1}\E\left(\left[\left\{1-w(\x)\right\}\left\{\Eone(Y_1\mid\x,1)-\Delta^0\right\}-Y_0+\Ezero\left\{Y_0\pi(\bt^0)\mid\x\right\}\right]\frac{\dot\bpi(\bt^0)}{1-\pi(\bt^0)}\right).
\end{eqnarray*}
Then plugging into $\Ezero(c_2\mid\x)$ in $\E(\phi\eff\S_\bt\trans)=\pdv*{\Delta}{\bt\trans}$, we get
\begin{eqnarray*}
&&\E\left\{\left(\frac1{B(\x)}\frac1p\left[\left\{1-w(\x)\right\}B(\x)-w(\x)\right]\left\{\Eone(Y_1\mid\x,1)-\Delta^0\right\}-\frac1{B(\x)}\frac1p\Ezero\left\{\frac{\pi(\bt^0)}{1-\pi(\bt^0)}Y_0\mid \x\right\}  \right.\right.\\
&&\left.\left.+\frac1{B(\x)}\frac1p\left\{1+B(\x)\right\}\Ezero\left\{\pi(\bt^0)Y_0\mid\x\right\}\right)\A(\x)\trans\right\} + \H\trans\M \\
&=& \frac1p\E\left(\left[\left\{1-w(x)\right\}\left\{\Eone(Y_1\mid\x,1)-\Delta^0\right\}-Y_0+\Ezero\left\{Y_0\pi(\bt^0)\mid\x\right\}\right]\frac{\dot\bpi(\bt^0)\trans}{1-\pi(\bt^0)}\right).
\end{eqnarray*}
After simplifying the above equation, we get $p\H\trans\M = (\D-\Q)\trans$, i.e., $p\H\trans=(\D-\Q)\trans\M^{-1}$.

\section{Proof of Theorem \ref{th:theta.ATT2}}\label{sec:proof.theta.ATT2}

\begin{Lem}\label{lem:Neyman_orth}
	 The score function $\G$ satisfies the Neyman orthogonality condition, i.e.,
	\begin{eqnarray*}
		\partial_r \E\left[ \G\{\W;\bpsi^0,\boeta^0+r(\boeta-\boeta^0)\}\right] |_{r=0}=\0, ~~ \mbox{for any }\boeta\in\calT.
	\end{eqnarray*}
\end{Lem}

\noindent\emph{Proof of Lemma~\ref{lem:Neyman_orth}.}
Let 
\begin{align*}
		\eta_1^0(\x) =&~  \Eone(Y_1\mid \x,1),\\
		\eta_2^0(\x) =&~ \Ezero[\{1-\pi(\bt^0)\}^{-1}\mid \x,0], \\
		\eta_3^0(\x) =&~ \Ezero[\{1-\pi(\bt^0)\}^{-2}\pi(\bt^0)\mid \x,0],\\
		\eta_4^0(\x) =&~ \Ezero[Y_0\{1-\pi(\bt^0)\}^{-2}\pi(\bt^0)^2\mid \x,0],\\
		\boeta_5^0(\x) =&~ \Ezero[\{1-\pi(\bt^0)\}^{-2}\dot\bpi(\bt^0)\mid \x,0].
\end{align*}
Note that 
\begin{align*}
\S\eff(\w;\bt,\boeta) = 
\frac{t-\pi(\bt)}{1-\pi(\bt)} 
\frac{\boeta_5(\x)}{\eta_3(\x)}.
\end{align*}
We have 
\begin{align*}
& \partial_r \E\left[\S\eff(\W;\bt,\boeta^0+r(\boeta-\boeta^0))\right]|_{r=0} \\ 
=&~ \E\Bigg[\frac{T-\pi(\bt^0)}{1-\pi(\bt^0)} 
  \Bigg(\frac{\boeta_5(\X)-\boeta_5^0(\X)}{\eta_3^0(\X)}  
  + 
\frac{\boeta_5^0(\X)}{\{\eta_3^0(\X)\}^2}\left\{\eta_3(\X)-\eta_3^0(\X)\right\}\Bigg) \Bigg] \\ 
=&~  \E\Bigg[\frac{\pr(T=1\mid\X,Y_0)-\pi(\bt^0)}{1-\pi(\bt^0)} 
  \Bigg(\frac{\boeta_5(\X)-\boeta_5^0(\X)}{\eta_3^0(\X)}  
  + 
\frac{\boeta_5^0(\X)}{\{\eta_3^0(\X)\}^2}\left\{\eta_3(\X)-\eta_3^0(\X)\right\}\Bigg) \Bigg]\\ 
=&~  \0,
\end{align*}
for any $\boeta\in\calT$,
where the second equality is by the law of total expectation, and the last equality is by \eqref{eq:assumeshadow}. We can also rewrite $\phi$ as 
\begin{align*}
\phi(\w;\bpsi,\boeta) = p^{-1} \left[\frac{t-\pi(\bt)}{1-\pi(\bt)} \left(Y-\frac{\{\eta_1(\x)-\Delta\}\{\eta_2(\x)-1\}+\eta_4(\x)}{\eta_3(\x)} \right)
		-\Delta t \right].
\end{align*}
Then, we have 
\begin{align*}
& \partial_r \E\left[\phi(\W;\bpsi^0,\boeta^0+r(\boeta-\boeta^0))\right]|_{r=0} \\ 
=&~ \E\Bigg[\frac{T-\pi(\bt^0)}{1-\pi(\bt^0)} 
  \Bigg(
	-\frac{\eta_2^0(\X)-1}{\eta_3^0(\X)} \left\{\eta_1(\X)-\eta_1^0(\X)\right\}
	-\frac{\eta_1^0(\X)-\Delta^0}{\eta_3^0(\X)}\left\{\eta_2(\X)-\eta_2^0(\X)\right\}\\
	&\qquad+\frac{\{\eta_1(\x)-\Delta^0\}\{\eta_2(\x)-1\}+\eta_4(\x)}{\eta_3^0(\X)^2}\left\{\eta_3(\X)-\eta_3^0(\X)\right\}
	-\frac{1}{\eta_3^0(\X)}\left\{\eta_4(\X)-\eta_4^0(\X)\right\}\Bigg) \Bigg] \\ 
=&~  0,
\end{align*}
for any $\boeta\in\calT$, where the last equality is due to the fact 
\begin{align*}
\E\left\{\frac{T-\pi(\bt^0)}{1-\pi(\bt^0)}\mid\X\right\}
= 1-\E\left\{\frac{1-T}{1-\pi(\bt^0)}\mid\X\right\} = 0.
\end{align*}
This completes the proof.

\hfill$\square$

	We also introduce the following high-level regularity conditions, parallel to Assumptions 3.3 and 3.4 in Section 3.3 of \cite{chernozhukov2018double}. Specifically, Assumption~\ref{assump:DML1} is mild and fairly standard in problems with moment conditions. It concerns the smoothness of the estimating equations and the unique identification of the parameter, and can be viewed as a restatement and strengthening of Assumption~\ref{assump:consistency.para} for the purpose of inference. Assumption~\ref{assump:DML2} provides a high-level characterization of the quality of the nuisance estimation, in that it is formulated in terms of the intrinsic rates most closely tied to the statistical problem at hand, rather than directly in terms of the norm $\|\cdot\|_{P,2}$ for $\wh\boeta$. However, as remarked by \cite{chernozhukov2018double}, the rate conditions in (A1)-(A3) can generally be implied from the crude but more intuitive requirement on the convergence rate of $\wh\boeta$, namely that given in Assumption~\ref{assump:quality.nuisance.est}. As discussed there, such convergence rate is satisfied by many commonly used machine learning methods.

	\begin{assump}\label{assump:DML1}(Non-linear moment condition problem with Neyman orthogonality) It holds that :
		\begin{enumerate}
			\item[(a)] $\E\left\{\G(\W;\bpsi^0,\boeta^0)\right\}=\0$ and $\calB$ contains a ball of radius $\Theta(n^{-1/2}\log n)$ centered at $\bpsi^0$.
			\item[(b)] The map $(\bpsi,\boeta) \rightarrow \E\left\{\G(\W;\bpsi,\boeta)\right\}$ is twice continuously Gateaux-differentiable on $\calB\times\calT$.
			\item[(c)] For all $\bpsi\in \calB$, $2\Vert \E\left\{\G(W;\bpsi,\boeta^0)\right\} \Vert \geq \Vert \J (\bpsi-\bpsi^0)\Vert \wedge c_0$ is satisfied, and $$\J=\partial_{\bpsi\trans}\E\left\{\G(\W;\bpsi,\boeta^0)\right\} \mid_{\bpsi=\bpsi^0}$$ has singular values between $c_0$ and $c_1$. Here $c_0$ and $c_1$ are some finite positive constants.
		\end{enumerate}
	\end{assump}
	
	\begin{assump}\label{assump:DML2}(Score regularity and requirements on the quality of estimation of nuisance parameters) It holds that:
		\begin{enumerate}
			\item[(a)]\label{con:C1} $\wh\boeta^{[-k]}$ belongs to the realization set $\calT_{n}$ for each $k\in\{1,2,\ldots,K\}$, with probability tending to 1 where $\calT_{n}$ satisfies $\boeta^0\in \calT_{n}$ and conditions provided as below.
			\item[(b)]\label{con:C2} The space of $\bpsi$, $\calB$ is bounded and for each $\boeta\in\calT_{n}$, the function class $\calF_{1,\boeta}=\{G_j(\cdot,\bpsi,\boeta): j= 1,2,\ldots,(d+1); \bpsi\in\calB\}$ is suitably measurable and its uniform covering entropy obeys
			\begin{eqnarray*}
			\underset{Q}{\sup} \log N(\epsilon\Vert \F_{1,\boeta} \Vert_{Q,2}, \calF_{1,\boeta},\Vert\cdot\Vert_{Q,2}) \leq \nu \log(a/\epsilon), \quad \text{for all} \quad 0\leq\epsilon\leq1,
			\end{eqnarray*}
			where $\F_{1,\boeta}$ is a measurable envelope for $\calF_{1,\boeta}$  satisfying $\Vert \F_{1,\boeta}(\cdot)\Vert_{P,q}\leq c_1$. Here $q>2$, $c_1>0$, $a>1$ and $\nu>0$ are some finite constants.
			\item[(c)]\label{con:C3} There exist sequences $\tau_n>0$ and $\delta_n>0$ that converge to zero such that:
			\begin{align}
			& \underset{\boeta\in\calT_{n},\bpsi\in\calB}{\sup} \Vert \E\left\{\G(\W;\bpsi,\boeta)  - \E\G(\W;\bpsi,\boeta^0)\right\}\Vert \le \tau_n\delta_n,\label{eq:assump3.4c.1}\\
			& \underset{\boeta\in\calT_{n},\Vert \bpsi-\bpsi^0\Vert\leq\tau_n}{\sup}  \left[\E\left\{\Vert \G(\W;\bpsi,\boeta)-\G(\W;\bpsi^0,\boeta^0)\Vert^2\right\}\right]^{1/2}\le \delta_n,
			\label{eq:assump3.4c.2} \\
			& \underset{r\in(0,1),\boeta\in\calT_{n},\Vert \bpsi-\bpsi^0\Vert\leq\tau_n}{\sup} \Vert  \partial_r^2 \E\left[\G\left\{\W;\bpsi^0+r(\bpsi-\bpsi^0),\boeta^0+r(\boeta-\boeta^0)\right\}\right]\Vert \le \delta_n n^{-1/2}.
			\label{eq:assump3.4c.3}
			\end{align}
			\item[(d)]\label{con:C4} The variance of the score is non-degenerate. All eigenvalues of the matrix  $$\E\left\{\G(\W;\bpsi^0,\boeta^0)\G(\W;\bpsi^0,\boeta^0)\trans\right\}$$ are bounded from below by $c_0$.
		\end{enumerate}
	    \end{assump}

\noindent\emph{Proof of Theorem~\ref{th:theta.ATT2}.}
	Under Assumptions \ref{assump:consistency.para}, \ref{assump:quality.nuisance.est}, \ref{assump:DML1}, and \ref{assump:DML2}, together with Lemma~\ref{lem:Neyman_orth},  Theorem 3.3 and Lemma 6.3 of \cite{chernozhukov2018double} show that $\sqrt{n}(\wh\bpsi - \bpsi^0)$ is asymptotically normal with mean $\0$ and covariance matrix $\V = \J^{-1} \E\{ \G(\W;\bpsi^0,\boeta^0)\G(\W;\bpsi^0,\boeta^0)\trans\} (\J^{-1})\trans$, where $\J = \partial_{\bpsi\trans} \E\{\G(\W;\bpsi,\boeta^0)\} \mid_{\bpsi=\bpsi^0}$. Specifically,
	\begin{eqnarray*}
	\J &=& \begin{bmatrix}
		\partial_{\bt\trans} \E\{\S\eff(\W;\bpsi,\boeta^0)\} & \0 \\
		\partial_{\bt\trans} \E\{\phi(\W;\bpsi,\boeta^0)\} & \partial_{\Delta} \E\{\phi(\W;\bpsi,\boeta^0)\}
	\end{bmatrix} \Bigg\vert_{\bpsi=\bpsi^0} \\
	&=& \begin{bmatrix}
		\J_{11} & \0 \\
		\J_{21}\trans & J_{22}
	\end{bmatrix},
	\end{eqnarray*}
	where, $\J_{11} = -\M$, $\J_{21}=p^{-1}(\D-\Q)$
	and
	\begin{eqnarray*}
	J_{22} &=& \E\left\{-\frac1p T + \frac1p \frac{T-\pi(\bt)}{1-\pi(\bt)} \frac{w(\X)}{B(\X)}  \right\} \bigg\vert_{\bpsi=\bpsi_0} = -1.
	\end{eqnarray*}
	Further,
	\begin{eqnarray*}
	\J^{-1} &=& \begin{pmatrix}
		\J_{11}^{-1} & \0 \\
		\0\trans & -1
	\end{pmatrix} \begin{pmatrix}
		\I & \0 \\
		-\J_{21}\trans\J_{11}^{-1} & 1
	\end{pmatrix} = \begin{pmatrix}
		\J_{11}^{-1} & \0 \\
		\J_{21}\trans\J_{11}^{-1} & -1
	\end{pmatrix} = \begin{pmatrix}
		-\M^{-1} & \0 \\
		-\J_{21}\trans\M^{-1} & -1
	\end{pmatrix}
	\end{eqnarray*}
	and for simplicity, we can write
	\begin{eqnarray*}
	&& \E\left\{ \G(\W;\bpsi^0,\boeta^0) \G(\W;\bpsi^0,\boeta^0)\trans \right\} \\
	&=& \begin{bmatrix}
		\E\{\S\eff(\W;\bpsi^0,\boeta^0)\S\eff(\W;\bpsi^0,\boeta^0)\trans\} &  \E\{\S\eff(\W;\bpsi^0,\boeta^0)\phi(\W;\bpsi^0,\boeta^0)\} \\
		\E\{\phi(\W;\bpsi^0,\boeta^0)\S\eff(\W;\bpsi^0,\boeta^0)\trans\} & \E\{\phi(\W;\bpsi^0,\boeta^0)^2\}
	\end{bmatrix} \\
	&=& \begin{pmatrix}
		\M & \A_{12} \\
		\A_{12}\trans & A_{22}
	\end{pmatrix} = \begin{pmatrix}
		\M & \0 \\
		\0\trans & A_{22}
	\end{pmatrix}.
	\end{eqnarray*}
	Here $\A_{12} = \0$ as $\S\eff(\W;\bpsi^0,\boeta^0)\in \Lambda^\perp$ and $\phi(\W;\bpsi^0,\boeta^0) \in \Lambda$.
	Thus, we get
	\begin{eqnarray*}
	\V &=& \left(\begin{matrix}
		\M^{-1} & \M^{-1} \J_{21} \\
		\J_{21}\trans \M^{-1} & \J_{21}\trans\M^{-1}\J_{21}+A_{22}
	\end{matrix}\right).
	\end{eqnarray*}
	This completes the proof.

	\hfill$\square$

\section{Nonparametric Estimation} \label{sec:nonparametric}
As shown in Appendix \ref{sec:lemmaidenproof}, under assumptions (\ref{eq:assumecompleteness}) and (\ref{eq:assumeshadow}), $\pi(y_0,\u)$ is nonparametrically identified by the integral equation \eqref{eq:iden_pi}. 
In this section, we propose a nonparametric approach to estimating $\pi(\cdot)$, and subsequently $\Delta$, to mitigate the risk of model misspecification. For a square-integrable function $g(y_0,\u)$, we define $m(\x;g)= \E\{(1-T)g(\Y_0,\U)\mid \x\}$. It follows from equation \eqref{eq:iden_pi} that $\pi(\cdot)$ satisfies $m\{\X;(1-\pi)^{-1}\} = 1$ almost surely. Consequently, $\pi(\cdot)$ minimizes
\begin{align}\label{eq:app_SMD_exp}
\inf_{\tilde\pi\in\Pi} \E\left[m\{\X;(1-\tilde\pi)^{-1}\}-1\right]^2,
\end{align} 
where $\Pi$ denotes a function space of $(y_0,\u)$.
This motivates estimating $\pi$ by solving the empirical analogue of \eqref{eq:app_SMD_exp}. 

We first estimate $m$ using a series estimator \citep{Newey1997}. Specifically, let $\{p_j(\cdot)\}_{j=1}^\infty$ denote a sequence of known basis functions (such as power series, splines, Fourier series, etc.), with the property that its linear combination can approximate any squared integral real-valued function of $\x$ well. 
For a detailed discussion on the construction of basis functions, we refer readers to \cite{Chen2007Handbook,BelloniChernozhukovChetverikovKato2015}.
Denote $\p^{k_n}(\x)= \big(p_1(\x),\ldots,p_{k_n}(\x)\big)\trans$, and $\P= \big(\p^{k_n}(\x_1),\ldots,\p^{k_n}(\x_n)\big)\trans\in\mR^{n\times k_n}$, where $k_n$ could increase with the sample size $n$. We then estimate $m$ by 
\begin{align*}
\wh m(\x;g) = \p^{k_n}(\x)\trans\big(\P\trans\P\big)^{-1}\sum_{j=1}^n  (1-t_j)g(y_{0j},\u_j) \p^{k_n}(\x_j).
\end{align*}
Even after substituting $\widehat m$, directly solving \eqref{eq:app_SMD_exp} remains infeasible because $\Pi$ is infinite-dimensional without further structural constraints on $\pi$.
To address this, we approximate $\Pi$ using a sieve space $\Pi_{l_n}$, which is finite-dimensional for each $n$ and becomes dense in $\Pi$ as $n$ grows. Specifically, let $\q^{l_n}(y_0,u)$ be an $l_n$-vector of basis functions analogous to $\p^{k_n}$. We construct
\begin{align*}
\Pi_{l_n} = \left\{\pi(y_0,\u)=\expit\{\q^{l_n}(y_0,\u)\trans\bb\}:\bb\in\mR^{l_n}\right\}.
\end{align*}
We then estimate $\pi$ by solving 
\begin{align}\label{eq:app_pi_SMD}
	\wh\pi\np = \argmin_{\tilde\pi\in\Pi_{l_n}} \frac{1}{n}\sum_{i=1}^{n} \left[\wh m\{\x_i;(1-\tilde\pi)^{-1}\}-1\right]^2.
\end{align}
As a result, an estimator of $\Delta$ is given by
\begin{align*}
\wh\Delta\np = \sumi \left\{t_i y_{1i} - (1-t_i)\frac{\wh\pi\np(y_{0i},\u_i)}{1-\wh\pi\np(y_{0i},\u_i)}y_{0i}\right\}\Big /\sumi t_i.
\end{align*}

We now comment on some potential issues with the nonparametric strategy. First, nonparametric estimation is well known to suffer from the curse of dimensionality. This issue mainly arises because we approximate $m$ and $\pi$ using expanding sieves $\p^{k_n}$ and $\q^{l_n}$. In general, approximation error decreases with the smoothness of the estimand and increases with the dimension of the arguments. Hence, when $\x$ is of moderate or high dimension, both approximation quality and finite-sample performance can deteriorate. Second, the identification equation \eqref{eq:iden_pi} is a Fredholm integral equation of the first kind and is therefore ill-posed. As a result, the estimator $\wh\pi\np$ in \eqref{eq:app_pi_SMD} may converge very slowly \citep{AiChen2003,newey2003instrumental}. In practice, it is often necessary to regularize the sieve space $\Pi_{l_n}$, e.g., by imposing boundedness restrictions on the parameters indexing $\Pi_{l_n}$, to ensure well-behaved asymptotic properties. Third, inference based on $\wh\Delta\np$ is challenging because a convenient linear representation is difficult to obtain using standard techniques such as Taylor expansion. This difficulty stems from the ill-posed nature of $\wh\pi\np$ so establishing asymptotic normality typically requires additional conditions and more delicate arguments. Accordingly, we omit a theoretical analysis of the nonparametric estimators $\wh\pi\np$ and $\wh\Delta\np$, as it is beyond the scope of this paper. We refer readers to \citet{newey2003instrumental,ChenPouzo2008,DarollesFanFlorensRenault2011} for discussions of regularization and the asymptotic behavior of ill-posed estimators, and to \citet{AiChen2012,ShanLiAi2025} for asymptotic distribution theory for plug-in estimators in ill-posed settings.

\vskip 0.2in
\bibliographystyle{apalike}
\bibliography{reference}

\begin{thebibliography}{}

\bibitem[Abadie and Imbens, 2002]{abadie2002simple}
Abadie, A. and Imbens, G. (2002).
\newblock Simple and bias-corrected matching estimators for average treatment
  effects.

\bibitem[Abadie and Imbens, 2006]{abadie2006large}
Abadie, A. and Imbens, G.~W. (2006).
\newblock Large sample properties of matching estimators for average treatment
  effects.
\newblock {\em Econometrica}, 74(1):235--267.

\bibitem[Abadie and Imbens, 2011]{abadie2011bias}
Abadie, A. and Imbens, G.~W. (2011).
\newblock Bias-corrected matching estimators for average treatment effects.
\newblock {\em Journal of Business \& Economic Statistics}, 29(1):1--11.

\bibitem[Ai and Chen, 2003]{AiChen2003}
Ai, C. and Chen, X. (2003).
\newblock Efficient estimation of models with conditional moment restrictions
  containing unknown functions.
\newblock {\em Econometrica}, 71(6):1795--1843.

\bibitem[Ai and Chen, 2012]{AiChen2012}
Ai, C. and Chen, X. (2012).
\newblock The semiparametric efficiency bound for models of sequential moment
  restrictions containing unknown functions.
\newblock {\em Journal of Econometrics}, 170(2):442--457.

\bibitem[Angrist et~al., 1996]{AngristImbensRubin1996}
Angrist, J.~D., Imbens, G.~W., and Rubin, D.~B. (1996).
\newblock Identification of causal effects using instrumental variables.
\newblock {\em Journal of the American Statistical Association},
  91(434):444--455.

\bibitem[Angrist and Krueger, 2001]{angrist2001instrumental}
Angrist, J.~D. and Krueger, A.~B. (2001).
\newblock Instrumental variables and the search for identification: From supply
  and demand to natural experiments.
\newblock {\em Journal of Economic Perspectives}, 15(4):69--85.

\bibitem[Arruda et~al., 2022]{SBRT2022}
Arruda, G.~V., Lourencao, M., Caldeira~de Oliveira, J.~H., Galendi, J. S.~C.,
  and Jacinto, A.~A. (2022).
\newblock Cost-effectiveness of stereotactic body radiotherapy versus
  conventional radiotherapy for the treatment of surgically ineligible stage i
  non-small cell lung cancer in the brazilian public health system.
\newblock {\em The Lancet Regional Health -- Americas}, 14:100329.

\bibitem[Baiocchi et~al., 2014]{IVSIMreview2014}
Baiocchi, M., Cheng, J., and Small, D.~S. (2014).
\newblock Instrumental variable methods for causal inference.
\newblock {\em Statistics in Medicine}, 33(13):2297--2340.

\bibitem[Belloni et~al., 2015]{BelloniChernozhukovChetverikovKato2015}
Belloni, A., Chernozhukov, V., Chetverikov, D., and Kato, K. (2015).
\newblock Some new asymptotic theory for least squares series: {{Pointwise}}
  and uniform results.
\newblock {\em Journal of Econometrics}, 186(2):345--366.

\bibitem[Bickel, 1982]{bickel1982adaptive}
Bickel, P.~J. (1982).
\newblock On adaptive estimation.
\newblock {\em The Annals of Statistics}, pages 647--671.

\bibitem[Bickel et~al., 1993]{bickel1993efficient}
Bickel, P.~J., Klaassen, J., Ritov, Y., and Wellner, J.~A. (1993).
\newblock {\em Efficient and Adaptive Estimation for Semiparametric Models}.
\newblock Johns Hopkins University Press Baltimore.

\bibitem[Bickel et~al., 2009]{BickelRitovTsybakov2009}
Bickel, P.~J., Ritov, Y., and Tsybakov, A.~B. (2009).
\newblock Simultaneous analysis of lasso and dantzig selector.
\newblock {\em The Annals of Statistics}, 37(4):1705--1732.

\bibitem[Chen, 2007]{Chen2007Handbook}
Chen, X. (2007).
\newblock Large sample sieve estimation of semi-nonparametric models.
\newblock In Heckman, J.~J. and Leamer, E.~E., editors, {\em Handbook of
  Econometrics}, volume~6, pages 5549--5632. Elsevier.

\bibitem[Chen and Pouzo, 2008]{ChenPouzo2008}
Chen, X. and Pouzo, D. (2008).
\newblock Efficient estimation of semiparametric conditional moment models with
  possibly nonsmooth residuals.
\newblock (1095779).

\bibitem[Chen and Briesacher, 2011]{ChenJCE2011}
Chen, Y. and Briesacher, B.~A. (2011).
\newblock Use of instrumental variable in prescription drug research with
  observational data: a systematic review.
\newblock {\em Journal of Clinical Epidemiology}, 64:687--700.

\bibitem[Chernozhukov et~al., 2018]{chernozhukov2018double}
Chernozhukov, V., Chetverikov, D., Demirer, M., Duflo, E., Hansen, C., Newey,
  W., and Robins, J. (2018).
\newblock Double/debiased machine learning for treatment and structural
  parameters.

\bibitem[Clarke and Windmeijer, 2012]{clarke2012instrumental}
Clarke, P.~S. and Windmeijer, F. (2012).
\newblock Instrumental variable estimators for binary outcomes.
\newblock {\em Journal of the American Statistical Association},
  107(500):1638--1652.

\bibitem[Cochran and Rubin, 1973]{cochran1973controlling}
Cochran, W.~G. and Rubin, D.~B. (1973).
\newblock Controlling bias in observational studies: A review.
\newblock {\em Sankhy{\=a}: The Indian Journal of Statistics, Series A}, pages
  417--446.

\bibitem[Darolles et~al., 2011]{DarollesFanFlorensRenault2011}
Darolles, S., Fan, Y., Florens, J.~P., and Renault, E. (2011).
\newblock Nonparametric instrumental regression.
\newblock {\em Econometrica}, 79(5):1541--1565.

\bibitem[Dehejia and Wahba, 1999]{dehejia1999causal}
Dehejia, R.~H. and Wahba, S. (1999).
\newblock Causal effects in nonexperimental studies: Reevaluating the
  evaluation of training programs.
\newblock {\em Journal of the American statistical Association},
  94(448):1053--1062.

\bibitem[d'Haultfoeuille, 2011]{d2011completeness}
d'Haultfoeuille, X. (2011).
\newblock On the completeness condition in nonparametric instrumental problems.
\newblock {\em Econometric Theory}, 27(3):460--471.

\bibitem[Farrell et~al., 2021]{FarrellLiangMisra2021}
Farrell, M.~H., Liang, T., and Misra, S. (2021).
\newblock Deep neural networks for estimation and inference.
\newblock {\em Econometrica}, 89(1):181--213.

\bibitem[Fr{\"o}lich and Melly, 2013]{frolich2013identification}
Fr{\"o}lich, M. and Melly, B. (2013).
\newblock Identification of treatment effects on the treated with one-sided
  non-compliance.
\newblock {\em Econometric Reviews}, 32(3):384--414.

\bibitem[Hahn, 1998]{hahn1998role}
Hahn, J. (1998).
\newblock On the role of the propensity score in efficient semiparametric
  estimation of average treatment effects.
\newblock {\em Econometrica}, pages 315--331.

\bibitem[Hainmueller, 2012]{hainmueller2012entropy}
Hainmueller, J. (2012).
\newblock Entropy balancing for causal effects: A multivariate reweighting
  method to produce balanced samples in observational studies.
\newblock {\em Political Analysis}, 20(1):25--46.

\bibitem[Hartman et~al., 2015]{hartman2015sample}
Hartman, E., Grieve, R., Ramsahai, R., and Sekhon, J.~S. (2015).
\newblock From sample average treatment effect to population average treatment
  effect on the treated: combining experimental with observational studies to
  estimate population treatment effects.
\newblock {\em Journal of the Royal Statistical Society: Series A (Statistics
  in Society)}, 178(3):757--778.

\bibitem[Heckman et~al., 1998]{heckman1998matching}
Heckman, J.~J., Ichimura, H., and Todd, P. (1998).
\newblock Matching as an econometric evaluation estimator.
\newblock {\em The Review of Economic Studies}, 65(2):261--294.

\bibitem[Heckman et~al., 1997]{heckman1997matching}
Heckman, J.~J., Ichimura, H., and Todd, P.~E. (1997).
\newblock Matching as an econometric evaluation estimator: Evidence from
  evaluating a job training programme.
\newblock {\em The Review of Economic Studies}, 64(4):605--654.

\bibitem[Heckman and Smith, 1995]{heckmansmith1995}
Heckman, J.~J. and Smith, J.~A. (1995).
\newblock Assessing the case for social experiments.
\newblock {\em Journal of Economic Perspectives}, 9(2):85--110.

\bibitem[Heckman and Vytlacil, 2001]{heckman2001policy}
Heckman, J.~J. and Vytlacil, E. (2001).
\newblock Policy-relevant treatment effects.
\newblock {\em American Economic Review}, 91(2):107--111.

\bibitem[Hern{\'a}n and Robins, 2006]{hernan2006instruments}
Hern{\'a}n, M.~A. and Robins, J.~M. (2006).
\newblock Instruments for causal inference: an epidemiologist's dream?
\newblock {\em Epidemiology}, pages 360--372.

\bibitem[Hernan and Robins, 2022]{hernan_robins_2022}
Hernan, M.~A. and Robins, J.~M. (2022).
\newblock {\em Causal Inference: What If}.
\newblock Taylor \& Francis.

\bibitem[Hirano et~al., 2003]{hirano2003efficient}
Hirano, K., Imbens, G.~W., and Ridder, G. (2003).
\newblock Efficient estimation of average treatment effects using the estimated
  propensity score.
\newblock {\em Econometrica}, 71(4):1161--1189.

\bibitem[Hu and Shiu, 2018]{hu2018nonparametric}
Hu, Y. and Shiu, J.-L. (2018).
\newblock Nonparametric identification using instrumental variables: sufficient
  conditions for completeness.
\newblock {\em Econometric Theory}, 34(3):659--693.

\bibitem[Imbens, 2004]{imbens2004nonparametric}
Imbens, G.~W. (2004).
\newblock Nonparametric estimation of average treatment effects under
  exogeneity: A review.
\newblock {\em Review of Economics and Statistics}, 86(1):4--29.

\bibitem[Imbens et~al., 2005]{imbens2005mean}
Imbens, G.~W., Newey, W.~K., and Ridder, G. (2005).
\newblock Mean-square-error calculations for average treatment effects.

\bibitem[Imbens and Rubin, 2015]{imbens2015causal}
Imbens, G.~W. and Rubin, D.~B. (2015).
\newblock {\em Causal Inference in Statistics, Social, and Biomedical
  Sciences}.
\newblock Cambridge University Press.

\bibitem[Jin et~al., 2001]{jin2001simple}
Jin, Z., Ying, Z., and Wei, L. (2001).
\newblock A simple resampling method by perturbing the minimand.
\newblock {\em Biometrika}, 88(2):381--390.

\bibitem[Kress, 1989]{Kress_LinearIntegralEquations}
Kress, R. (1989).
\newblock {\em Linear Integral Equations}, volume~82 of {\em Applied
  {{Mathematical Sciences}}}.
\newblock Springer New York, New York.

\bibitem[Lechner, 1999]{lechner1999earnings}
Lechner, M. (1999).
\newblock Earnings and employment effects of continuous gff-the-job training in
  east germany after unification.
\newblock {\em Journal of Business \& Economic Statistics}, 17(1):74--90.

\bibitem[Liu et~al., 2020]{liu2020identification}
Liu, L., Miao, W., Sun, B., Robins, J., and Tchetgen, E.~T. (2020).
\newblock Identification and inference for marginal average treatment effect on
  the treated with an instrumental variable.
\newblock {\em Statistica Sinica}, 30(3):1517.

\bibitem[Martens et~al., 2006]{martens2006instrumental}
Martens, E.~P., Pestman, W.~R., de~Boer, A., Belitser, S.~V., and Klungel,
  O.~H. (2006).
\newblock Instrumental variables: application and limitations.
\newblock {\em Epidemiology}, pages 260--267.

\bibitem[McCaffrey et~al., 2004]{mccaffrey2004propensity}
McCaffrey, D.~F., Ridgeway, G., and Morral, A.~R. (2004).
\newblock Propensity score estimation with boosted regression for evaluating
  causal effects in observational studies.
\newblock {\em Psychological Methods}, 9(4):403.

\bibitem[Mebane and Poast, 2013]{mebane2013causal}
Mebane, W.~R. and Poast, P. (2013).
\newblock Causal inference without ignorability: Identification with nonrandom
  assignment and missing treatment data.
\newblock {\em Political Analysis}, 21(2):233--251.

\bibitem[Miao et~al., 2016]{miao2016identifiability}
Miao, W., Ding, P., and Geng, Z. (2016).
\newblock Identifiability of normal and normal mixture models with nonignorable
  missing data.
\newblock {\em Journal of the American Statistical Association},
  111(516):1673--1683.

\bibitem[Miao et~al., 2018]{MiaoGengTchetgenTchetgen2018}
Miao, W., Geng, Z., and Tchetgen~Tchetgen, E. (2018).
\newblock Identifying causal effects with proxy variables of an unmeasured
  confounder.
\newblock {\em Biometrika}, 105(4):987--993.

\bibitem[Moodie et~al., 2018]{moodie2018doubly}
Moodie, E.~E., Saarela, O., and Stephens, D.~A. (2018).
\newblock A doubly robust weighting estimator of the average treatment effect
  on the treated.
\newblock {\em Stat}, 7(1):e205.

\bibitem[Morgan and Winship, 2014]{Morgan_Winship_2014}
Morgan, S.~L. and Winship, C. (2014).
\newblock {\em Counterfactuals and Causal Inference}.
\newblock Cambridge University Press, 2nd edition.

\bibitem[Newey, 1997]{Newey1997}
Newey, W.~K. (1997).
\newblock Convergence rates and asymptotic normality for series estimators.
\newblock {\em Journal of Econometrics}, 79(1):147--168.

\bibitem[Newey and McFadden, 1994]{NeweyMcFadden1994}
Newey, W.~K. and McFadden, D. (1994).
\newblock Large sample estimation and hypothesis testing.
\newblock In {\em Handbook of {{Econometrics}}}, volume~4, pages 2111--2245.
  Elsevier.

\bibitem[Newey and Powell, 2003]{newey2003instrumental}
Newey, W.~K. and Powell, J.~L. (2003).
\newblock Instrumental variable estimation of nonparametric models.
\newblock {\em Econometrica}, 71(5):1565--1578.

\bibitem[Newhouse and McClellan, 1998]{newhouse1998econometrics}
Newhouse, J.~P. and McClellan, M. (1998).
\newblock Econometrics in outcomes research: the use of instrumental variables.
\newblock {\em Annual Review of Public Health}, 19(1):17--34.

\bibitem[Neyman, 1923]{neyman1923applications}
Neyman, J. (1923).
\newblock Sur les applications de la thar des probabilities aux experiences
  agaricales: Essay des principle. excerpts reprinted (1990) in english.
\newblock {\em Statistical Science}, 5(463-472):4.

\bibitem[Neyman, 1959]{neyman1959optimal}
Neyman, J. (1959).
\newblock Optimal asymptotic tests of composite hypotheses.
\newblock {\em Probability and Statsitics}, pages 213--234.

\bibitem[Neyman, 1979]{neyman1979c}
Neyman, J. (1979).
\newblock C ($\alpha$) tests and their use.
\newblock {\em Sankhy{\=a}: The Indian Journal of Statistics, Series A}, pages
  1--21.

\bibitem[Pearl, 1995]{PearlBiometrika95}
Pearl, J. (1995).
\newblock Causal diagrams for empirical research.
\newblock {\em Biometrika}, 82(4):669--688.

\bibitem[Pearl, 2009]{Pearl_2009}
Pearl, J. (2009).
\newblock {\em Causality: Models, Reasoning and Inference}.
\newblock Cambridge University Press.

\bibitem[Pirracchio and Carone, 2018]{pirracchio2018balance}
Pirracchio, R. and Carone, M. (2018).
\newblock The balance super learner: A robust adaptation of the super learner
  to improve estimation of the average treatment effect in the treated based on
  propensity score matching.
\newblock {\em Statistical Methods in Medical Research}, 27(8):2504--2518.

\bibitem[Pirracchio et~al., 2016]{pirracchio2016propensity}
Pirracchio, R., Carone, M., Rigon, M.~R., Caruana, E., Mebazaa, A., and
  Chevret, S. (2016).
\newblock Propensity score estimators for the average treatment effect and the
  average treatment effect on the treated may yield very different estimates.
\newblock {\em Statistical Methods in Medical Research}, 25(5):1938--1954.

\bibitem[Robins, 1994]{robins1994correcting}
Robins, J.~M. (1994).
\newblock Correcting for non-compliance in randomized trials using structural
  nested mean models.
\newblock {\em Communications in Statistics-Theory and methods},
  23(8):2379--2412.

\bibitem[Rosenbaum and Rubin, 1983]{rosenbaum1983central}
Rosenbaum, P.~R. and Rubin, D.~B. (1983).
\newblock The central role of the propensity score in observational studies for
  causal effects.
\newblock {\em Biometrika}, 70(1):41--55.

\bibitem[Rubin, 1973a]{rubin1973matching}
Rubin, D.~B. (1973a).
\newblock Matching to remove bias in observational studies.
\newblock {\em Biometrics}, pages 159--183.

\bibitem[Rubin, 1973b]{rubin1973use}
Rubin, D.~B. (1973b).
\newblock The use of matched sampling and regression adjustment to remove bias
  in observational studies.
\newblock {\em Biometrics}, pages 185--203.

\bibitem[Rubin, 1974]{rubin1974estimating}
Rubin, D.~B. (1974).
\newblock Estimating causal effects of treatments in randomized and
  nonrandomized studies.
\newblock {\em Journal of Educational Psychology}, 66(5):688.

\bibitem[Rubin, 1990]{rubin1990comment}
Rubin, D.~B. (1990).
\newblock Comment: Neyman (1923) and causal inference in experiments and
  observational studies.
\newblock {\em Statistical Science}, 5(4):472--480.

\bibitem[Schick, 1986]{schick1986asymptotically}
Schick, A. (1986).
\newblock On asymptotically efficient estimation in semiparametric models.
\newblock {\em The Annals of Statistics}, pages 1139--1151.

\bibitem[Shah et~al., 2013]{ShahHahnStetsonFriedbergEtAl2013}
Shah, A., Hahn, S.~M., Stetson, R.~L., Friedberg, J.~S., Pechet, T. T.~V., and
  Sher, D.~J. (2013).
\newblock Cost-effectiveness of stereotactic body radiation therapy versus
  surgical resection for stage {{I}} non--small cell lung cancer.
\newblock {\em Cancer}, 119(17):3123--3132.

\bibitem[Shan et~al., 2026]{ShanLiAi2025}
Shan, J., Li, W., and Ai, C. (2026).
\newblock Efficient nonparametric inference for mediation analysis with
  nonignorable missing confounders.
\newblock {\em Journal of the American Statistical Association}.

\bibitem[Tchetgen~Tchetgen et~al., 2024]{TchetgenTchetgenYingCuiShiEtAl2024}
Tchetgen~Tchetgen, E.~J., Ying, A., Cui, Y., Shi, X., and Miao, W. (2024).
\newblock An introduction to proximal causal inference.
\newblock {\em Statistical Science}, in press.

\bibitem[Tsiatis, 2006]{tsiatis2006semiparametric}
Tsiatis, A.~A. (2006).
\newblock {\em Semiparametric Theory and Missing Data}.
\newblock New York: Springer.

\bibitem[Wager and Athey, 2018]{WagerAthey2018}
Wager, S. and Athey, S. (2018).
\newblock Estimation and inference of heterogeneous treatment effects using
  random forests.
\newblock {\em Journal of the American Statistical Association},
  113(523):1228--1242.

\bibitem[Wang et~al., 2017]{wang2017g}
Wang, A., Nianogo, R.~A., and Arah, O.~A. (2017).
\newblock G-computation of average treatment effects on the treated and the
  untreated.
\newblock {\em BMC Medical Research Methodology}, 17(1):1--5.

\bibitem[Wang and Tchetgen~Tchetgen, 2018]{WangTchetgenTchetgen2018a}
Wang, L. and Tchetgen~Tchetgen, E. (2018).
\newblock Bounded, efficient and multiply robust estimation of average
  treatment effects using instrumental variables.
\newblock {\em Journal of the Royal Statistical Society Series B: Statistical
  Methodology}, 80(3):531--550.

\bibitem[Wang et~al., 2014]{wang2014instrumental}
Wang, S., Shao, J., and Kim, J.~K. (2014).
\newblock An instrumental variable approach for identification and estimation
  with nonignorable nonresponse.
\newblock {\em Statistica Sinica}, pages 1097--1116.

\bibitem[Wright, 1928]{wright1928tariff}
Wright, P.~G. (1928).
\newblock {\em Tariff on animal and vegetable oils}.
\newblock Macmillan Company, New York.

\bibitem[Zhao and Ma, 2022]{zhao2022versatile}
Zhao, J. and Ma, Y. (2022).
\newblock A versatile estimation procedure without estimating the nonignorable
  missingness mechanism.
\newblock {\em Journal of the American Statistical Association},
  117(540):1916--1930.

\bibitem[Zhao and Shao, 2015]{zhao2015semiparametric}
Zhao, J. and Shao, J. (2015).
\newblock Semiparametric pseudo-likelihoods in generalized linear models with
  nonignorable missing data.
\newblock {\em Journal of the American Statistical Association},
  110(512):1577--1590.

\bibitem[Zhao and Percival, 2017]{zhao2017entropy}
Zhao, Q. and Percival, D. (2017).
\newblock Entropy balancing is doubly robust.
\newblock {\em Journal of Causal Inference}, 5(1).

\end{thebibliography}

\end{document}